\documentclass[]{aa}
\usepackage{amsmath}
\usepackage{graphicx}
\DeclareGraphicsExtensions{.eps,.ps,.pdf}
\usepackage{txfonts}
\usepackage{subfigure}
\usepackage{float}
\usepackage{natbib}
\bibpunct{(}{)}{;}{a}{}{,}
\usepackage{hyperref}
\usepackage{xspace}
\usepackage{balance}
\usepackage{upgreek}

\begin{document}

\newcommand{\XMM}{XMM-\textit{Newton}\xspace}
\newcommand{\IRAS}{\textit{IRAS}\xspace}
\newcommand{\ISO}{\textit{ISO}\xspace}
\newcommand{\Spitzer}{\textit{Spitzer}\xspace}
\newcommand{\micr}{$\upmu$m\xspace}

\newcommand{\alfa}{\alpha_{6} }
\newcommand{\Fint}{f_{6}^{\mathrm{int}} }
\newcommand{\Fobs}{f_{\lambda}^{\mathrm{obs}} }
\newcommand{\tempAGN}{u_{\lambda}^{\mathrm{AGN}} }
\newcommand{\tempSB}{u_{\lambda}^{\mathrm{SB}} }
\newcommand{\Lsun}{L_\mathrm{\sun}}
\newcommand{\Msun}{M_\mathrm{\sun}}
\newcommand{\RAGN}{R^\mathrm{AGN}}
\newcommand{\RSB}{R^\mathrm{SB}}
   \title{Analysis of \textit{Spitzer}-IRS spectra of hyperluminous infrared galaxies}


   \author{A. Ruiz
          \inst{1,2}
	  \and
	  G. Risaliti
          \inst{3,4}
          \and
          E. Nardini
          \inst{4}
	  \and
	  F. Panessa
          \inst{5}
          \and
	  F.J. Carrera
          \inst{6}
         }

   \institute{
	     Istituto Nazionale di Astrofisica (INAF), Osservatorio Astronomico di Brera,
	     via Brera 21, 20121 Milano, Italy
          \and
             Inter-University Centre for Astronomy and Astrophysics (IUCAA),
             Post Bag 4, Ganeshkhind, Pune 411 007, India
	  \and
             INAF - Osservatorio Astrofisico di Arcetri,
             L.go E. Fermi 5, 50125 Firenze, Italy
          \and
             Harvard-Smithsonian Center for Astrophysics, 
             60 Garden Street, Cambridge, MA 02138, USA
	  \and
            INAF - Istituto di Astrofisica e Planetologia Spaziali di Roma (IAPS)
            Via Fosso del Cavaliere 100, I-00133 Rome, Italy
          \and
            Instituto de F\'isica de Cantabria (IFCA), CSIC-UC,
            Avda. de los Castros, 39005 Santander, Spain \\
          \email{ruizca@iucaa.ernet.in}
             }

  \date{Received June 6, 2010; accepted October 13, 2012}


  \abstract 
   {Hyperluminous infrared galaxies (HLIRG) are the most luminous
    persistent objects in the Universe. They exhibit extremely high
    star formation rates, and most of them seem to harbour an Active
    Galactic Nucleus (AGN). They are unique laboratories to
    investigate the most extreme star formation, and its connection to
    super-massive black hole growth.}
   {The AGN and starburst (SB) relative contributions to the total output
    in these objects is still debated. Our aim is to disentangle the AGN
    and SB emission of a sample of thirteen HLIRG.}
   {We have studied the MIR low resolution spectra of a sample of thirteen HLIRG
   obtained with the Infrared Spectrograph on board \Spitzer. The $5-8$~\micr
   range is an optimal window to detect AGN activity even in a heavily
   obscured environment. We performed a SB/AGN decomposition of the continuum
   using templates, which has been successfully applied for ULIRG in previous
   works.}
   {The MIR spectra of all sources is largely dominated by AGN emission.
   Converting the 6~\micr luminosity into IR luminosity, we found that $\sim80\%$
   of the sample shows an IR output dominated by the AGN emission. However, the SB
   activity is significant in all sources (mean SB contribution $\sim30\%$),
   showing star formation rates $\sim300-3000~\Msun\mathrm{yr}^{-1}$. Using
   X-ray and MIR data we estimated the dust covering factor (CF) of these HLIRG,
   finding that a significant fraction presents a CF consistent with unity. Along with
   the high X-ray absorption shown by these sources, this suggests that large
   amounts of dust and gas enshroud the nucleus of these HLIRG, as also
   observed in ULIRG.}
   {Our results are in agreement with previous studies of the IR SED of HLIRG
   using radiative transfer models, and we find strong evidence that all HLIRG 
   harbour an AGN. Moreover, this work provides further support to the idea 
   that AGN and SB are both crucial to understand the properties of HLIRG. Our 
   study of the CF supports the hypothesis that HLIRG can be divided in two different
   populations.}


   \maketitle

\section{Introduction}
\label{sec:intro}
One of the most important results brought by the Infrared Astronomical
Satellite (\IRAS) was the detection of a new class of galaxies where the bulk 
of their bolometric emission lies in the infrared range \citep{Soifer84}. This
population, named luminous infrared galaxies (LIRG), becomes the dominant
extragalactic population at IR luminosities above $10^{11}\Lsun$, with a space
density higher than all other classes of galaxies of comparable bolometric
luminosity (see \citealt{Sanders96} for a complete review of these objects). At
the brightest end of this population distribution lie the ultraluminous (ULIRG;
$L_\mathrm{IR} > 10^{12}~\Lsun$) and the hyperluminous infrared galaxies (HLIRG;
$L_\mathrm{IR} > 10^{13}~\Lsun$).

Ever since the discovery of LIRG two possible origins have been
suggested for the observed luminosities: Active Galactic Nuclei (AGN) and/or
stellar formation in starburst (SB) episodes
\citep{Rieke80,Soifer84b,Wright84,Soifer87b,Carico90,Condon91}. The AGN/SB
relative contributions to the IR output of this kind of sources shows a clear
trend with luminosity: while LIRG and low-luminosity ULIRG are SB-dominated
sources, the AGN emission is essential to explain the bolometric luminosity in
high-luminosity ULIRG and HLIRG. The fraction of sources harbouring an AGN and
its relative contribution increases with IR luminosity
\citep{Veilleux99,Rowan00,Tran01,Farrah02submm,Veilleux02,Farrah06,Nardini10}.

Understanding the ultimate physical mechanism which triggers those
phenomena and, particularly in those sources with extreme IR luminosities, their
interplay, is crucial to obtain a complete view of galaxy evolution. Over the
last decade comprehensive observations from X-rays to radio have
produced a consistent paradigm for local ULIRG. Now is broadly accepted that
these objects are dusty galaxies where fierce star formation processes have been
triggered by mergers or interactions between gas-rich galaxies
\citep{Mihos94,Genzel01,Dasyra06}. Only half of them harbour an AGN, which usually
is a minor contributor to the total IR emission \citep[cf.][for a complete
review]{Lonsdale06}.

ULIRG are rare in the Local Universe \citep{Soifer87}, but a large number has
been detected instead in deep-IR surveys, being a fundamental constituent of the
high-redshift galaxy population
\citep{Fran01,Elbaz02,Lonsdale04,Daddi05,Farrah06,Caputi07,Magdis10,Magdis11}.
It has been proposed that ULIRG at high redshift could be at the origin of
massive elliptical and S0 galaxies \citep{Fran94,Lilly99,Genzel00,Genzel01}. A
large fraction of stars in present-day galaxies would have been formed during
these evolutionary phases. However, the paradigm described above is not so
well-grounded for the high-$z$ ULIRG population
\citep{Rodighiero11,Draper12,Niemi12}, neither for higher luminosity objects
like HLIRG.

HLIRG seem to be composite sources, i.e. AGN and SB phenomena are both needed to
fully explain their IR emission
\citep{Rowan00,Farrah02submm,Ruiz10,Rowan10,Han12}, and only about a third has
been found in interacting systems \citep{Farrah02hst}. This suggests that most
HLIRG are not triggered by mergers, as also found for a significant number of
high-$z$ ULIRG \citep{Sturm10,Draper12,Niemi12}. If HLIRG and high-$z$ ULIRG
share other properties, local HLIRG, since they are brighter and easier to study
with the current astronomical tools, could be key objects to understand the
high-$z$ population of ULIRG. Moreover, as HLIRG could represent the most
vigorous stage of galaxy formation, with SFR~$>1000~\Msun \mathrm{yr}^{-1}$,
they are unique laboratories to investigate extremely high stellar formation,
and its connection to super-massive black hole (SMBH) growth.

The relative contributions of AGN and SB to the bolometric luminosity of HLIRG is
still in debate. Early studies of small samples of HLIRG found contradictory
results, with some authors suggesting that the IR emission arises
predominantly from a SB \citep{Frayer98,Frayer99} with star formation rates
(SFR) of the order $1000~\mathrm{\Msun yr}^{-1}$, and other authors suggesting
that HLIRG are powered by a dusty AGN \citep{Granato96,Evans98,Yun98}. The
analysis of the IR Spectral Energy Distributions (SED) of larger samples of HLIRG
revealed that about half of these sources are AGN-dominated
\citep{Rowan00,Verma02,Rowan10}. Our analysis of the HLIRG's broadband (from
radio to X-rays) SED \citep{Ruiz10} obtained consistent results, showing that
both AGN and SB components are needed to reproduce the full emission of these
objects, being the bolometric luminosity of the AGN dominant, or at least
comparable to the SB luminosity, in $\sim70\%$ of the studied HLIRG. These
studies were however based on samples biased toward AGN.

\citealt{Farrah02submm} (hereafter F02) built a sample of HLIRG with a
negligible bias toward AGN and it was observed with the Submillimetre
Common-User Bolometer Array \citep[SCUBA;][]{Scuba}. Submillimetre data
introduced tight constraints on SB luminosities in the IR SED analysis. They
found most HLIRG being AGN-dominated sources, but with a significant
contribution ($\gtrsim 20\%$) due to star formation in all objects.

In \citet{Ruiz07} we tried to disentangle the AGN and SB emission of HLIRG
following a different approach. Since the 2-10~keV X-ray emission above
$\sim10^{42}~\mathrm{erg~s^{-1}}$ is the ``smoking gun'' of AGN activity
\citep{Mushotzky04b}, we studied the X-ray spectra of a sample of fourteen HLIRG
observed by \XMM. We found that all the detected sources (ten out of fourteen)
were AGN-dominated in the X-ray band, but most were underluminous in X-rays with
respect to the predicted luminosity by the standard AGN SED from
\citet{Risaliti04}, given their IR luminosities. We also found strong evidence
that a significant fraction of these objects is heavily obscured, reaching the
Compton Thick level (hydrogen column, $N_\mathrm{H}$
$>10^{24}~\mathrm{cm}^{-2}$), as suggested by previous studies
\citep{Wilman03,Iwasawa05}. Such absorption level allows, in the best cases,
only the detection of reflected X-ray emission. To avoid the strong obscuration
effects and in order to obtain a more complete and independent view of these
sources, we need to move to another spectral window.

In this paper we present a study of a sample of HLIRG in the Mid-Infrared (MIR)
band, $\lambda = 5-8$~\micr, a spectral range very efficient in detecting AGN
emission and less affected by absorption than X-rays. Several diagnostic methods
are available to unravel the AGN and SB activity through the MIR spectra, e.g.
the study of high-ionization emission lines and the polycyclic aromatic
hydrocarbon (PAH) features \citep{Laurent00,Armus07,Spoon07,Farrah07} or the
analysis of the continuum around 4~\micr \citep{Risaliti06}. Blind statistical
techniques, like principal component analysis, have been also successfully
employed with MIR spectra \citep{Wang11,Hurley12}. However, given the limited
size of our sample, we cannot apply such statistical methods.

The MIR continuum of pure AGN and pure SB show small dispersion below 10~\micr
\citep{Netzer07,Brandl06}, allowing the use of universal templates to reproduce
the AGN and SB emission in sources where both physical processes are present.
Hence, we can unravel the AGN and SB components of composite sources modelling
their MIR spectra with such templates. The key reason for using the continuum
emission at $\lambda \simeq 5-8$~\micr as a diagnostic is the difference of the
6~\micr-to-bolometric ratios between AGN and SB (approximately two orders of
magnitude larger in the former). This makes the detection of the AGN component
possible even when the AGN is heavily obscured and/or bolometrically weak
compared to the SB.

SB/AGN continuum spectral decomposition has been successfully applied in
ULIRG \citep{Nardini08,Nardini09,Nardini10,Risaliti10} to disentangle the
emission of both components. A significant number of HLIRG has been observed
with the Infrared Spectrograph \citep[IRS,][]{Spitzer-IRS} on board \Spitzer
\citep{Spitzer}, obtaining good quality spectra to apply this diagnostic
technique. 

The outline of the paper is as follows: Section~\ref{sec:sample} describes the selected
sample of HLIRG. The IRS data reduction is explained in
Sect.~\ref{sec:reduction}. In Sect.~\ref{sec:decomp} we explain the
decomposition process, the model adopted and the results obtained.
Section~\ref{sec:discuss} discuss the results of the MIR spectral analysis and
Sect.~\ref{sec:conc} summarizes our conclusions.

The \textit{Wilkinson Microwave Anisotropy Probe} (\textsl{WMAP}) concordance
cosmology has been adopted along the paper: $H_0=70$ km s$^{-1}$ Mpc$^{-1}$,
$\Omega_m=0.27, \Omega_{\Lambda}=0.73$ \citep{Komatsu09}.

\section{The sample}
\label{sec:sample}
This paper is based on the analysis of the F02 sample of HLIRG. Nine out of ten
sources in that sample\footnote{There were eleven sources in the original
sample, but we excluded one object, IRAS\,13279+3401, because the redshift in
the literature \citep[z = 0.36,][]{Rowan00} is incorrect, misclassifying the
source as a HLIRG.  We have shown, using recent optical and MIR spectra
\citep{Ruiz10}, that its redshift is $z \sim 0.02$.  Our estimated IR luminosity
for this source is hence $3\times 10^{10}~\Lsun$.} have been observed with the
\Spitzer-IRS in its low-resolution mode and the data are publicly available in
the \Spitzer Archive.

The F02 sample was originally selected in a manner independent of obscuration,
inclination or AGN content. Moreover, its parent sample is composed by
objects found from direct optical follow-up of 60~\micr or 850~\micr surveys
\citep{Rowan00} and hence it is statistically homogeneous and complete. 

Given these selection criteria, F02 claim that the sample should be
entirely free from AGN bias. However, since the parent sample is composed of
sources selected from surveys at different wavelenghts, this could produce
significant biases in favour or against 'hot' dust, which is more common in AGN
than in SB \citep{deGrijp85,Soifer87b}. Thus, objects from  60~\micr surveys are
more likely to contain hot dust heated by AGN, while those sources from
850~\micr surveys will be biased towards cold dust heated by star formation.
However, we must note that only a small fraction of the \citet{Rowan00} sample
is selected from 850~\micr data (less than $10\%$), and a similar fraction is
found in the F02 sample. Moreover, the effect of the AGN hot dust in composite
sources with equal AGN and SB bolometric output is severely diminished beyond
$\sim30$~\micr rest-frame \citep[cf. Fig.~1 of][]{Nardini09}. Only three HLIRG
in the F02 sample are at $z\sim1$, so the effect of hot dust is minimal. The
potential AGN bias is therefore negligible and we consider the F02 sample
suitable for drawing global conclusions about the HLIRG population.

Only five sources of this sample have been observed by \XMM, and just two of
these five HLIRG were detected \citep{Ruiz07}. In order to study the relation
between the X-ray and MIR emission for this kind of sources, we included four
additional HLIRG to our sample. These objects were selected from the
X-ray-selected HLIRG sample we analysed in \citet{Ruiz07}. They all were
detected by \XMM and observed with \Spitzer-IRS. Three of these additional HLIRG
were found via comparisons to known AGN (Tables 2-4 of \citealt{Rowan00}) and
therefore its inclusion leads to a non-negligible bias toward AGN content in our
final sample. Nevertheless, in some ocassions we can derive conclusions
about the global population of HLIRG restricting our results to the F02 sample
(nine out of ten F02 sources are included in our sample).

There are thirteen HLIRG in our final sample. Accordingly to their optical
spectra, six are type 1 AGN (Seyfert 1 or QSO), four are type 2 AGN (Seyfert 2
or QSO2) and three are SB galaxies (see Table~\ref{tab:sources}). The IR SED of
all sources have been observed and studied using radiative transfer models
(RTM) through several papers \citep[F02;][]{Rowan00,Verma02}. AGN and SB components
are both needed to reproduce the IR emission of these HLIRG, the AGN
output being dominant in most sources (eight out of thirteen).

Regarding only to the objects in the F02 sample, five are type 1 AGN and four
are optically classified as narrow-line (NL) objects. All type 1 sources and one
NL object show an AGN-dominated IR SED, and the remaining (three) NL sources
show an SB-dominated IR SED, accordingly to the F02 analysis.

The broad band SED (from radio to X-rays) of most sources (nine out of thirteen)
have also been studied using observational templates for the AGN and SB
components, offering consistent results with those obtained using RTM \citep{Ruiz10}.

\section{Data reduction}
\label{sec:reduction}
All sources from our HLIRG sample have been observed by the IRS onboard \Spitzer
in the low-resolution mode (see Table~\ref{tab:sources}) and the data were
publicly available. All observations were operated in the standard staring mode
using the two low-resolution modules (SL and LL). The Basic Calibrated Data
(BCD) files, bad pixels masks (BMASK) and flux uncertainty files were downloaded
from the \Spitzer Heritage Archive v1.5. All files are products of the official
Spitzer Science Center pipeline (version S18.18).

The BCD pipeline reduces the raw detector images from individual
exposures: removes the electronic and optical artifacts (e.g. dark current,
droop effect, non-linear effects, detection of cosmic-ray events), jail-bar
pattern and stray light, and perform a flat-field correction. BCD files, along with
the corresponding uncertainty images and BMASK files, are reliable inputs for
SMART, an IDL-based processing and analysis tool for IRS data \citep{smart2}.

We processed the BCD pipeline files using SMART 8.2.3. We cleaned the
BCD files of bad pixels using IRSCLEAN 2.1 and we combined the individual
exposures for a given ExpID (uncertainty files were used to perform a
weighted-average). In order to remove the background and low-level rogue pixels
we subtracted the two observations in the nodding cycle for each module and
order (SL1, SL2, LL1, LL2). We checked that all sources were point-like and well
centered in the slits and finally we ran an automatic, tappered column  (window
adjusted to the source extent and scaling with wavelength)
extraction\footnote{We also tested the advanced optimal (PSF weighted) method
\citep{smart1} to extract the spectra. Since all sources are bright and
point-like, the results from both extraction methods were similar.
However, we found that the spectra obtained through the advanced optimal method
show a strong sinusoidal artifact in most SL2 spectra. We compared our spectra
with the Cornell Atlas of Spitzer/IRS Sources \citep[CASIS,][]{CASIS} and we
found no similar features at those wavelengths. We chose then the tapered column
extraction to perform the subsequent analysis.} to obtain the flux and
wavelength calibrated spectra. We checked that individual orders did match in
flux after extraction, so no further scaling corrections were needed.

\begin{figure*}[ht]
   \begin{center} 
    \mbox{ 
     \subfigure{
      \includegraphics[angle=-90,width=.3\linewidth]{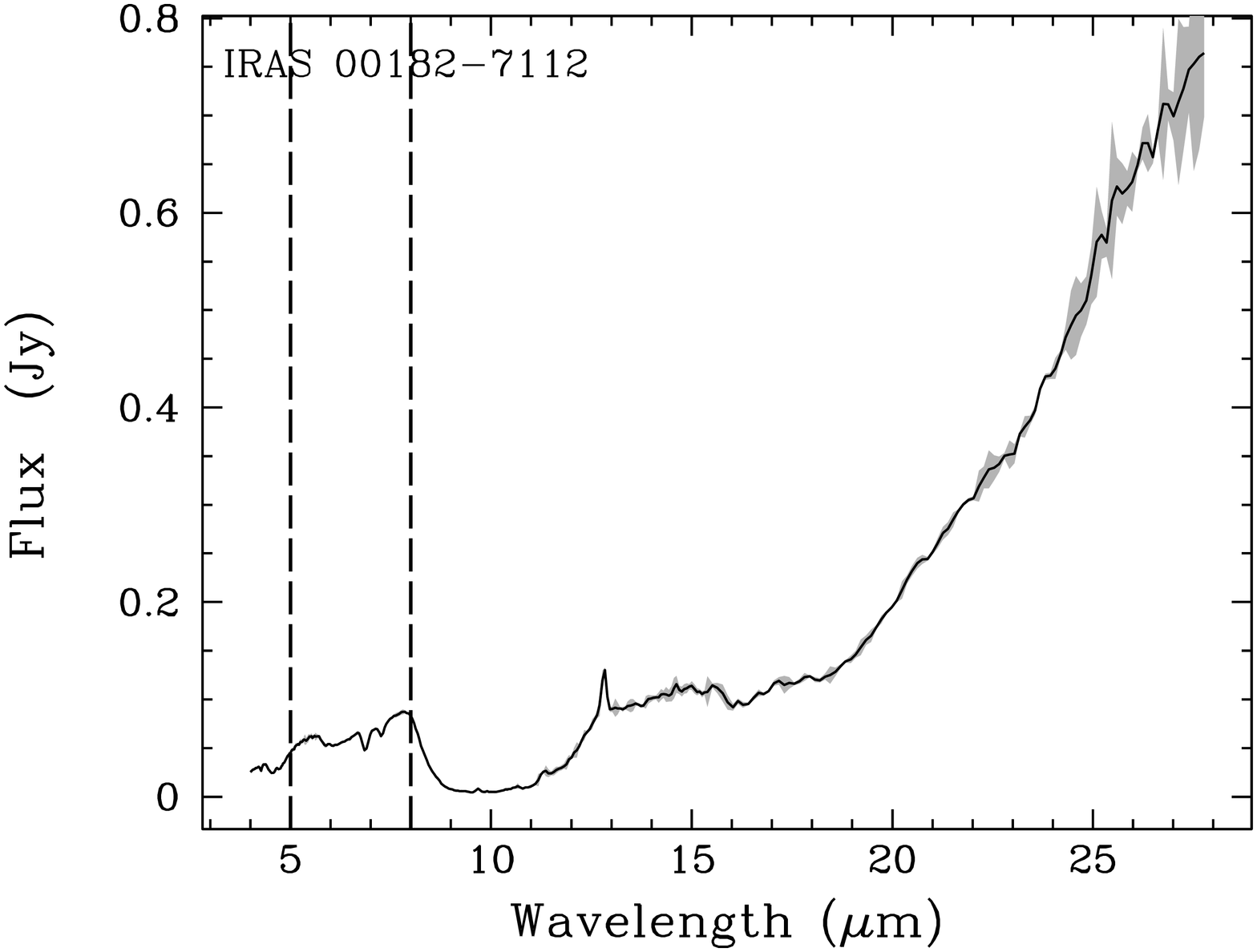}
      \label{fig:iras00182} 
     } 
     \subfigure{
      \includegraphics[angle=-90,width=.3\linewidth]{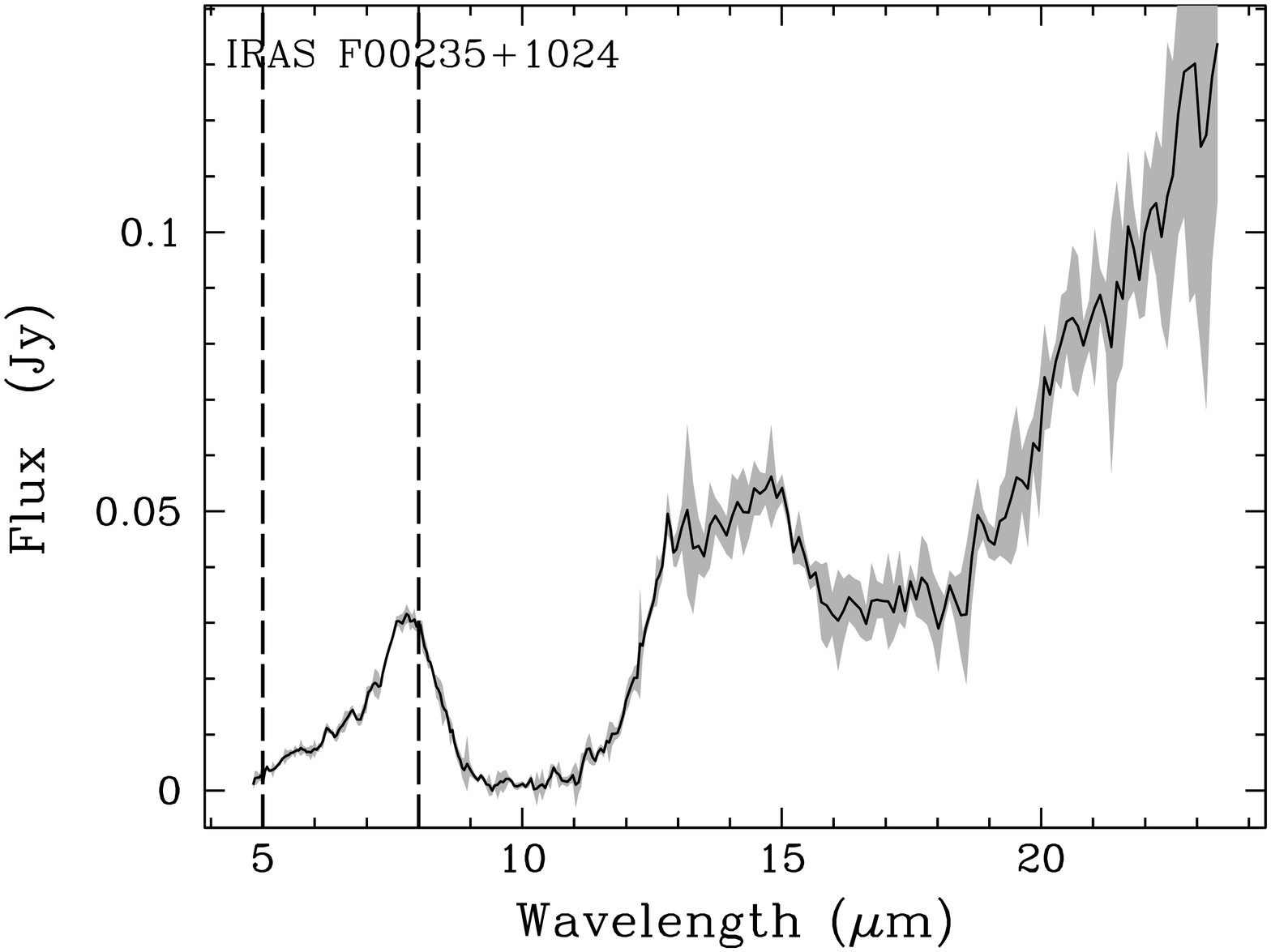}
      \label{fig:irasF00235} 
     }
     \subfigure{
      \includegraphics[angle=-90,width=.3\linewidth]{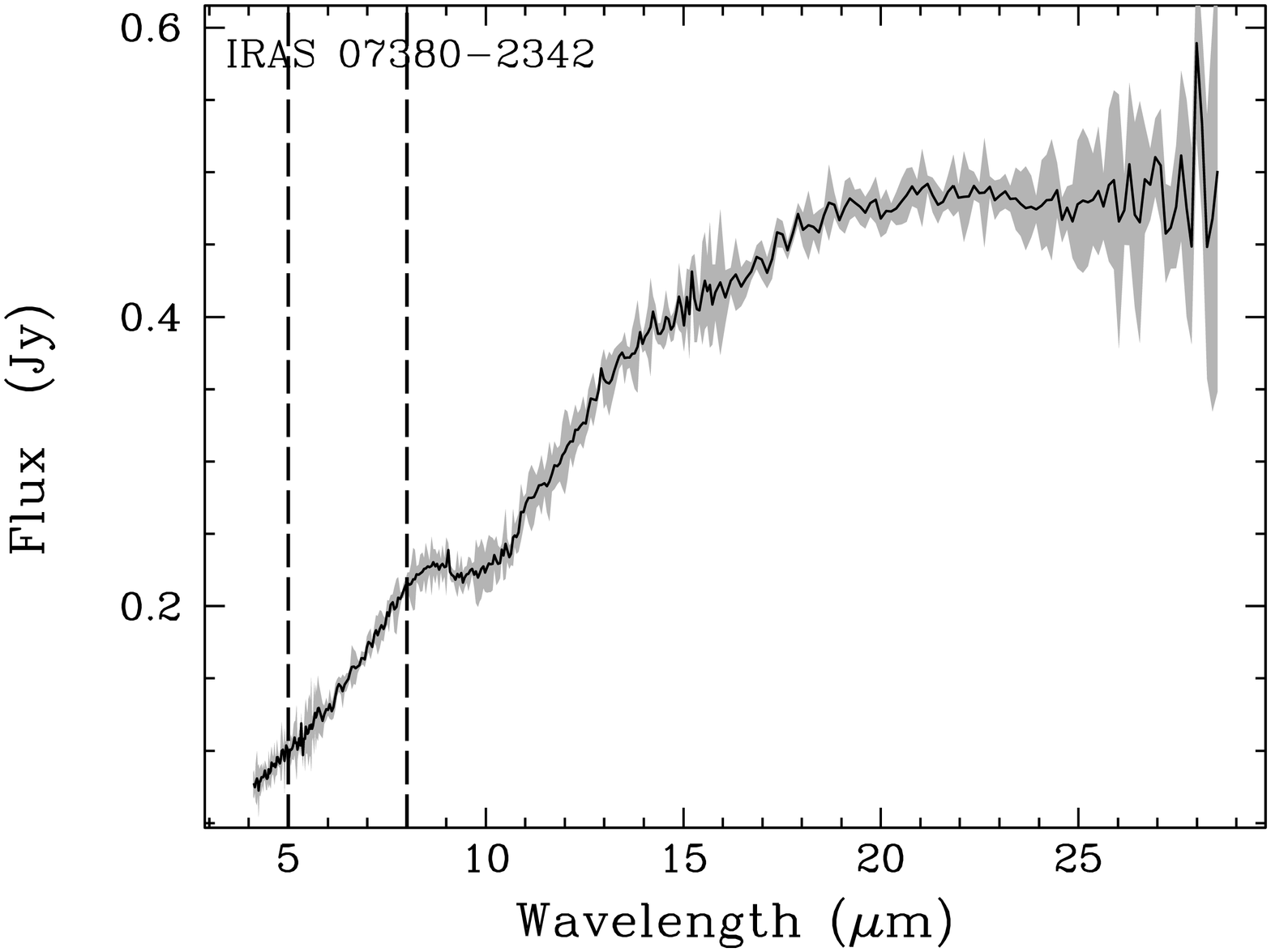}
      \label{fig:iras07380} 
     }     
    } 
    \mbox{ 
     \subfigure{
      \includegraphics[angle=-90,width=.3\linewidth]{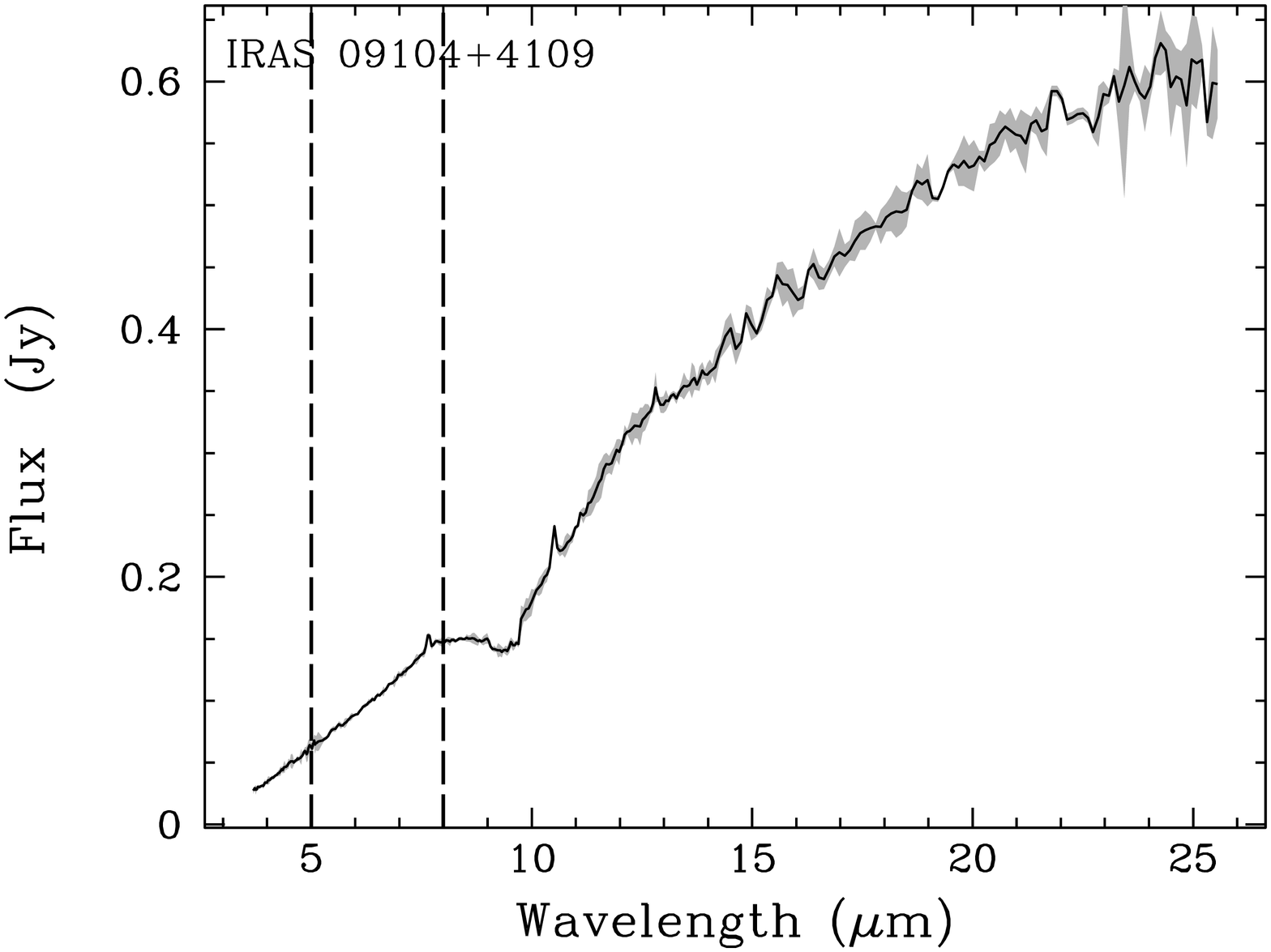}
      \label{fig:iras09104} 
     } 
     \subfigure{
      \includegraphics[angle=-90,width=.3\linewidth]{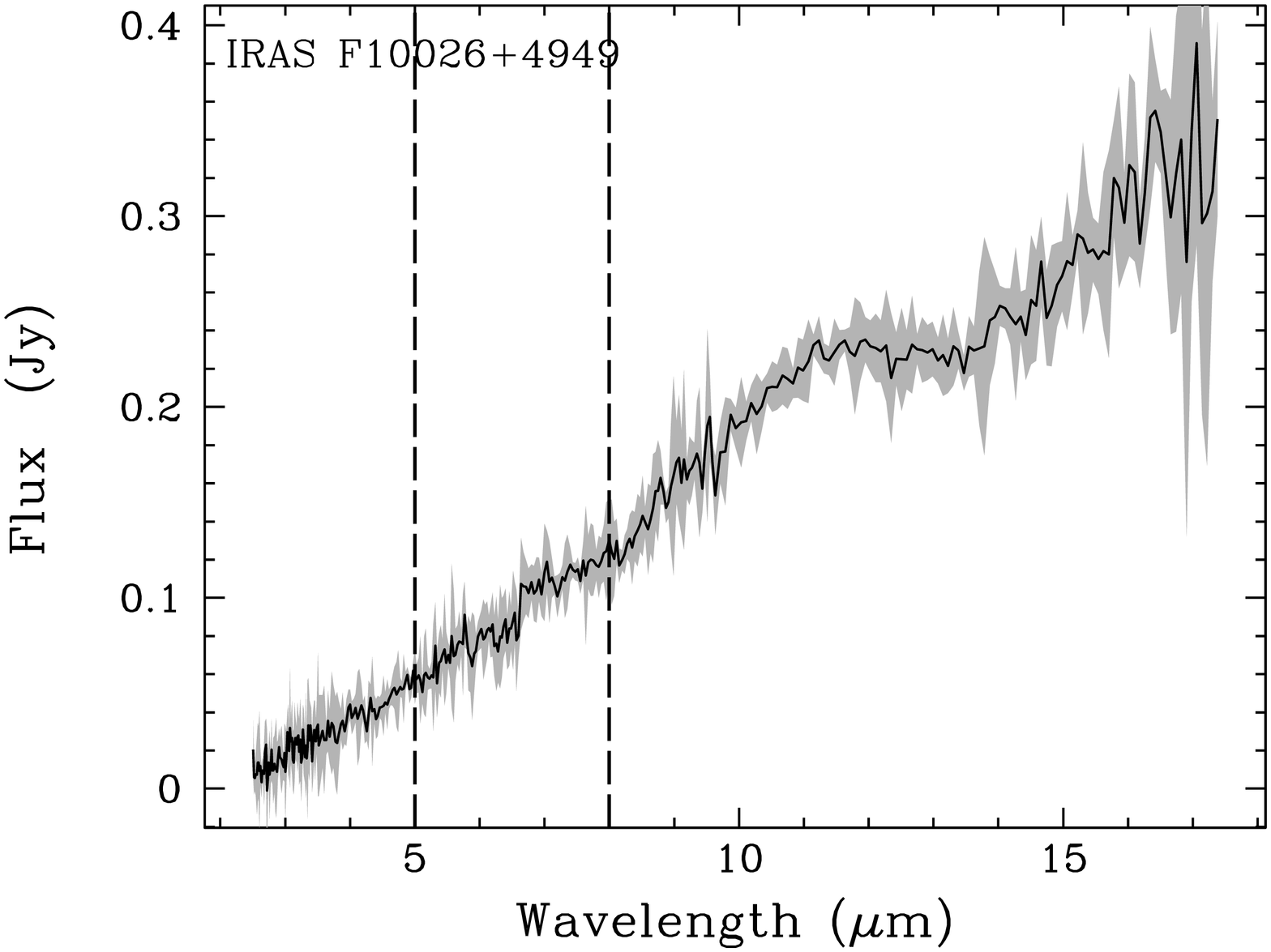}
      \label{fig:irasF10026} 
     } 
     \subfigure{
      \includegraphics[angle=-90,width=.3\linewidth]{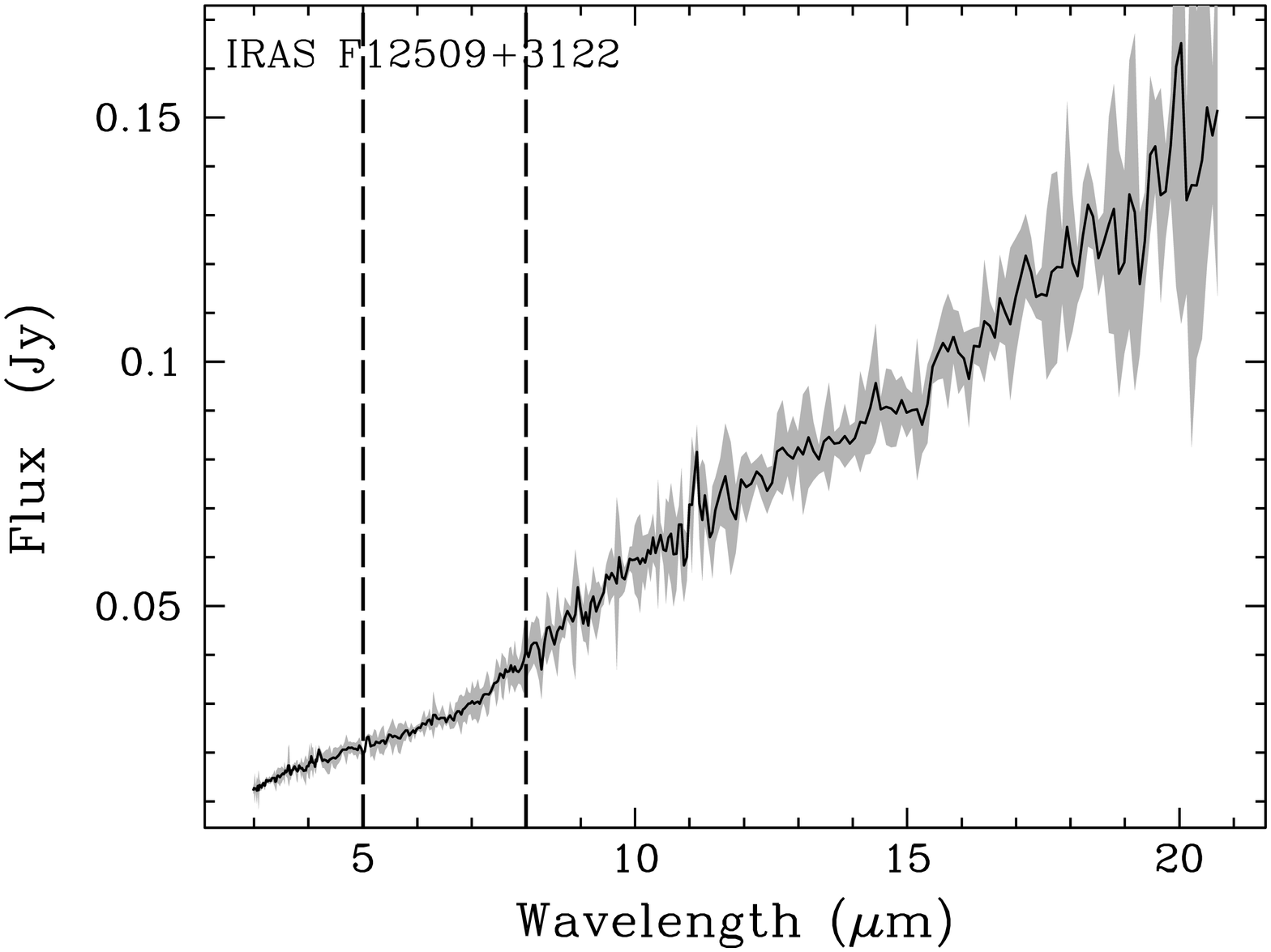}
      \label{fig:irasF12509} 
     } 
    } 
    \mbox{ 
     \subfigure{
      \includegraphics[angle=-90,width=.3\linewidth]{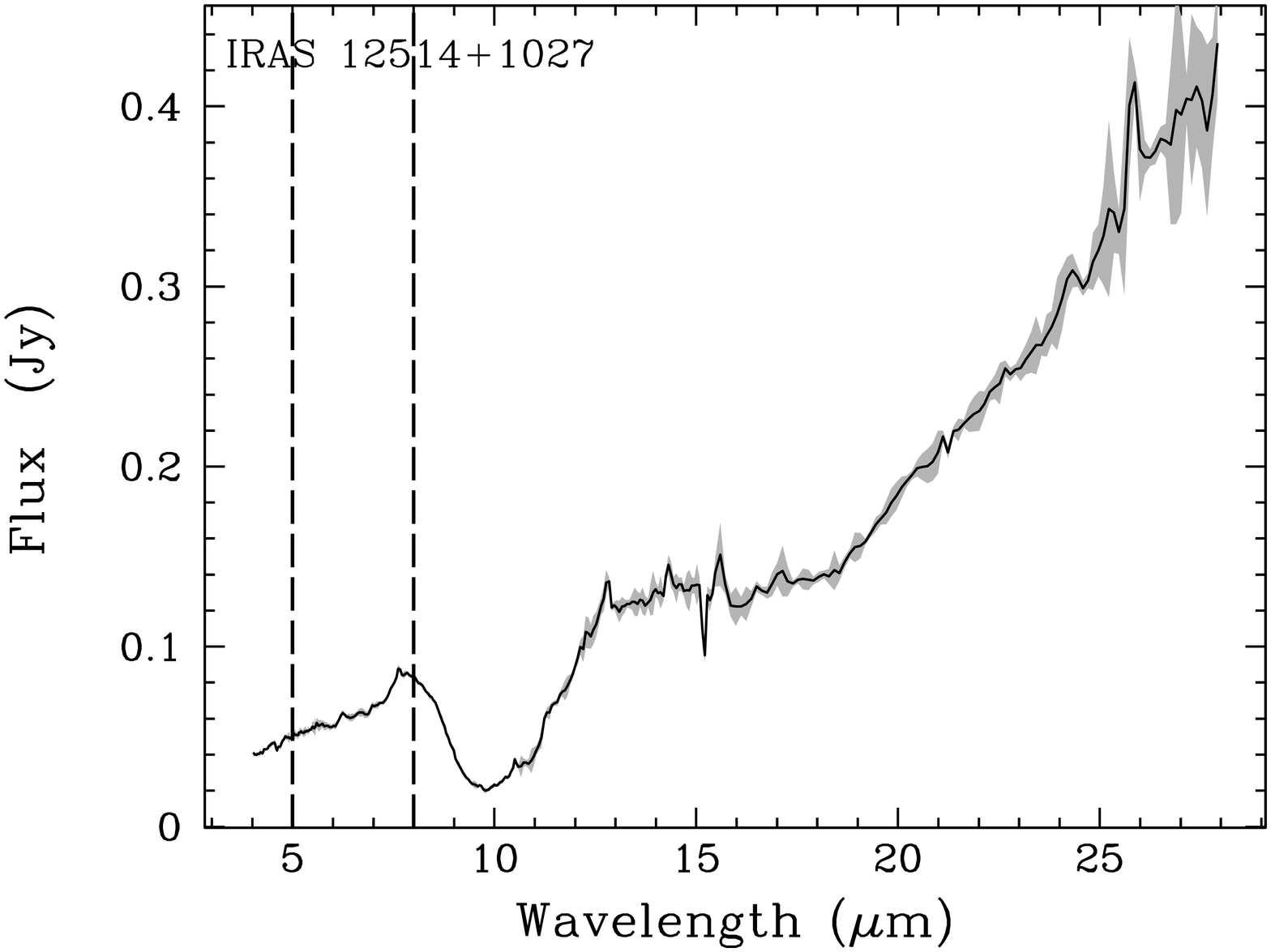}
      \label{fig:iras12514} 
     } 
     \subfigure{
      \includegraphics[angle=-90,width=.3\linewidth]{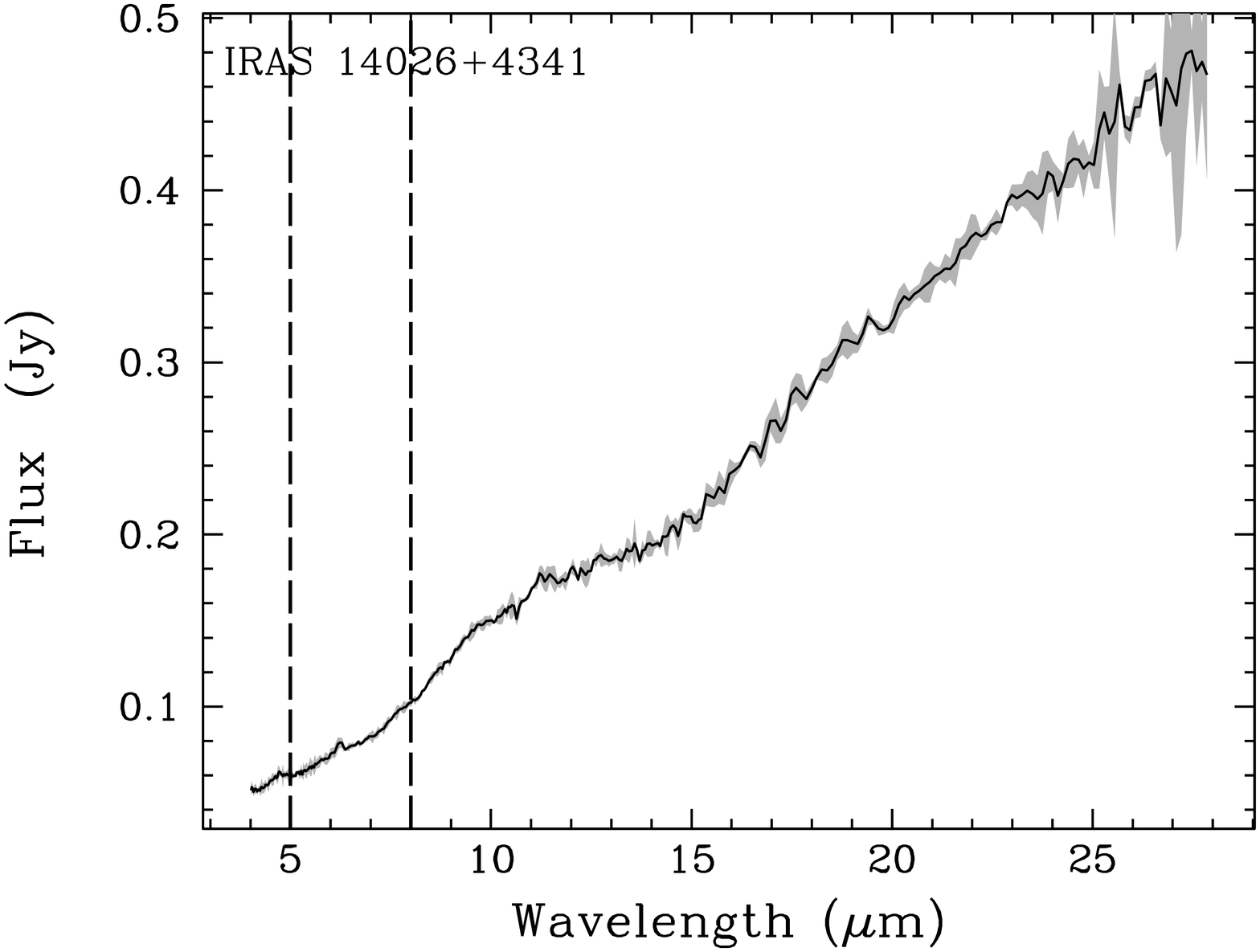}
      \label{fig:iras14026} 
     }
     \subfigure{
      \includegraphics[angle=-90,width=.3\linewidth]{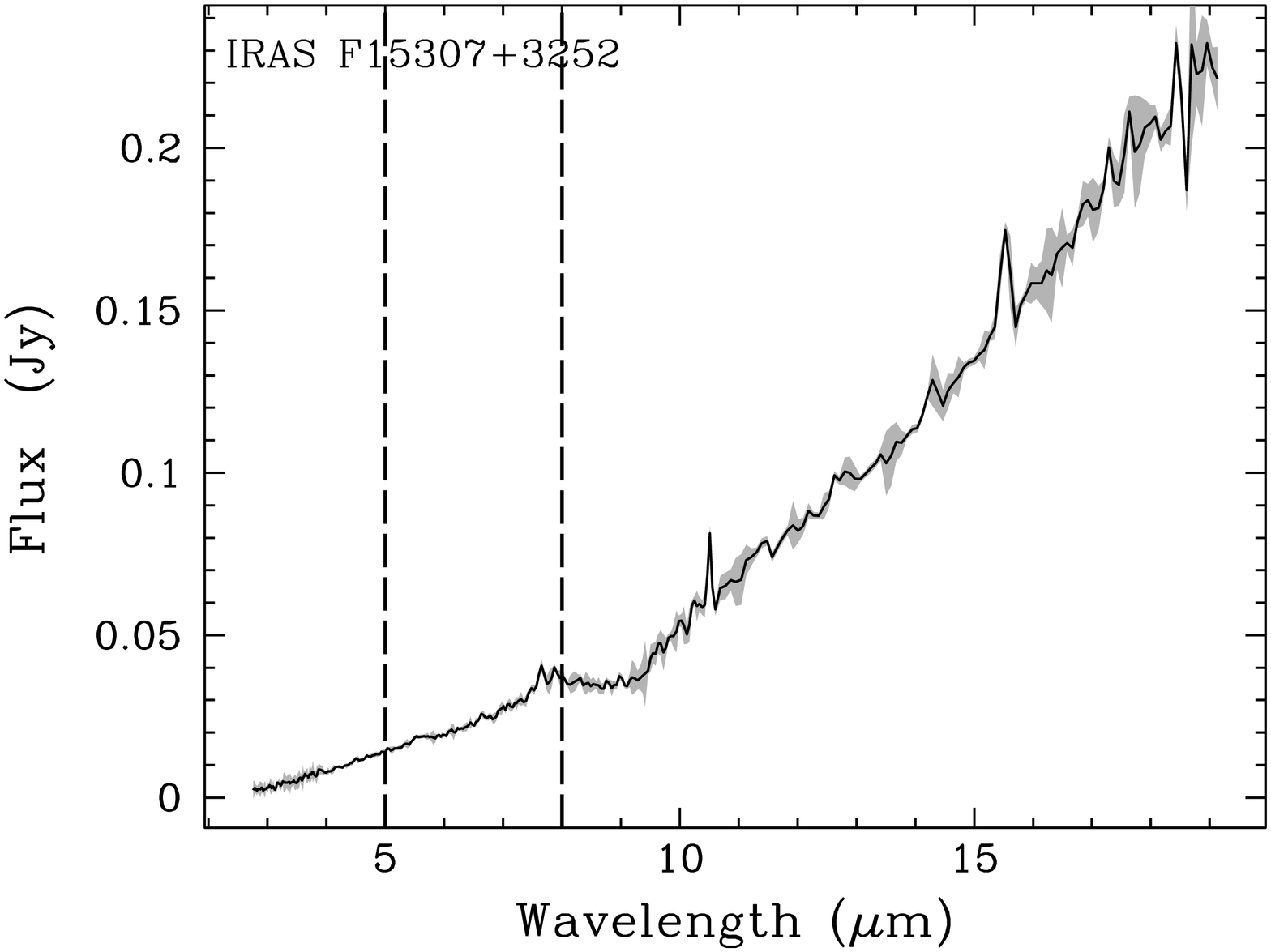}
      \label{fig:irasF15307}
     }
    }
    \mbox{ 
     \subfigure{
      \includegraphics[angle=-90,width=.3\linewidth]{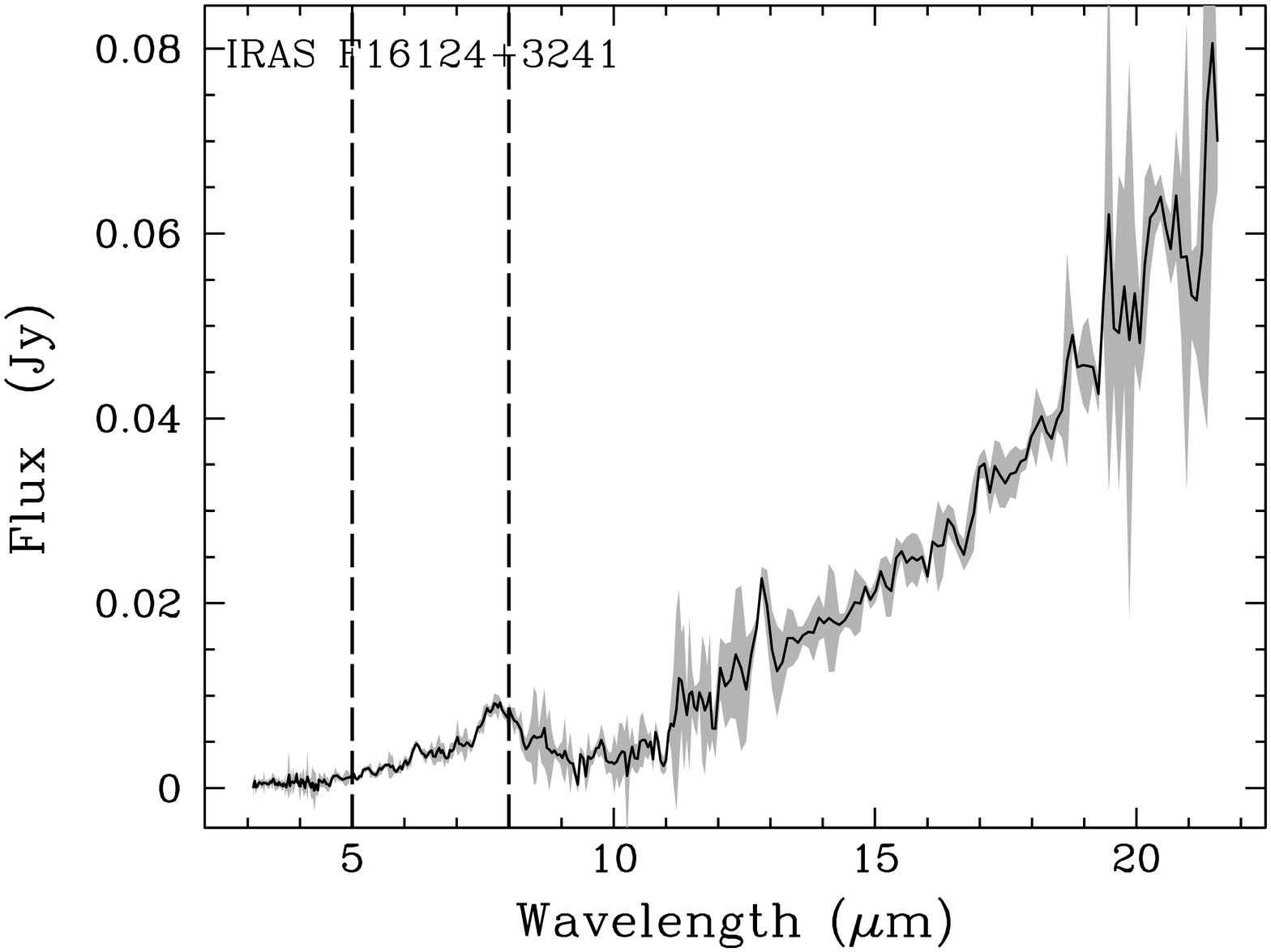}
      \label{fig:irasF16124} 
     } 
     \subfigure{
      \includegraphics[angle=-90,width=.3\linewidth]{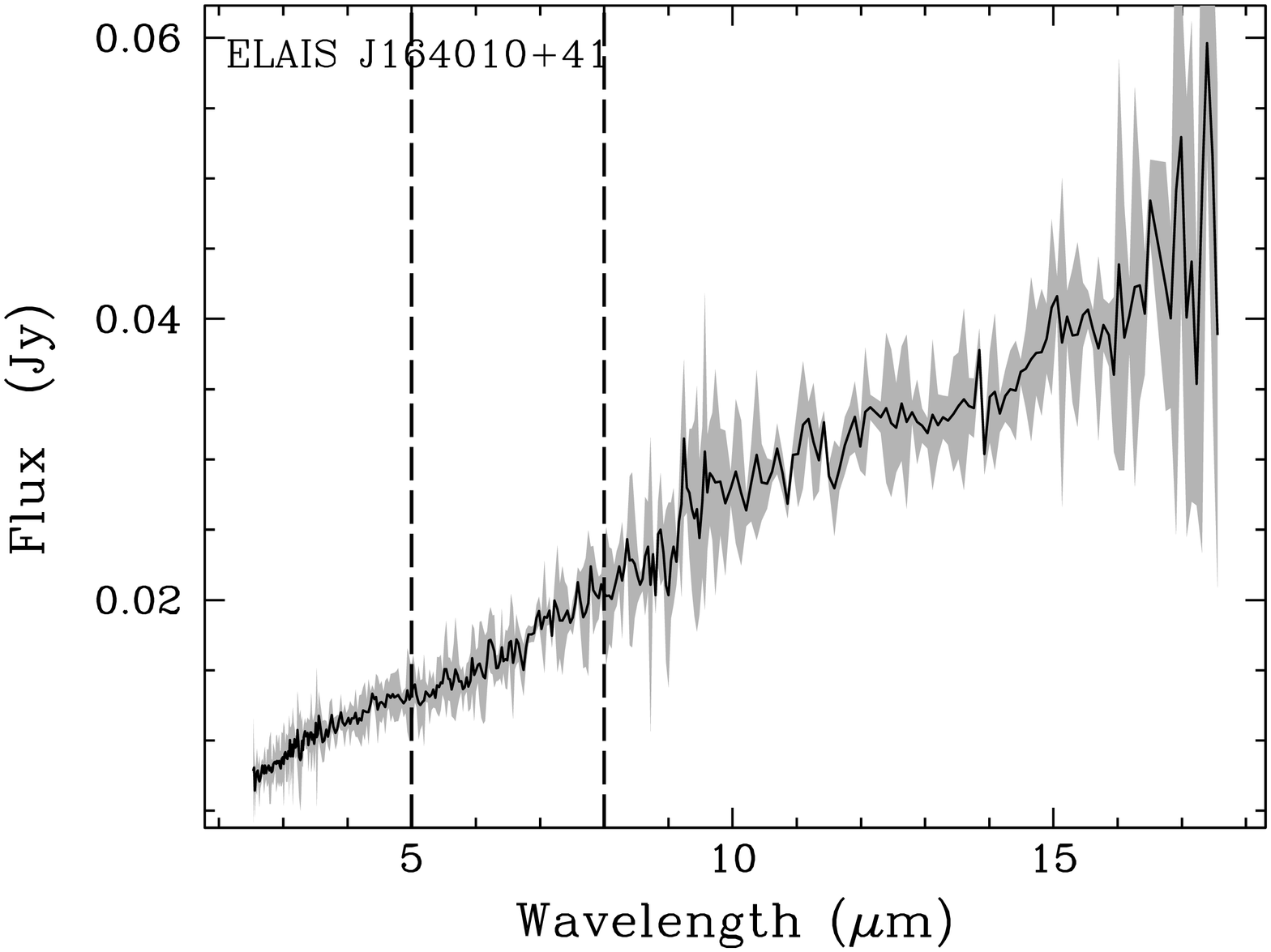}
      \label{fig:ej1640} 
     }
     \subfigure{
      \includegraphics[angle=-90,width=.3\linewidth]{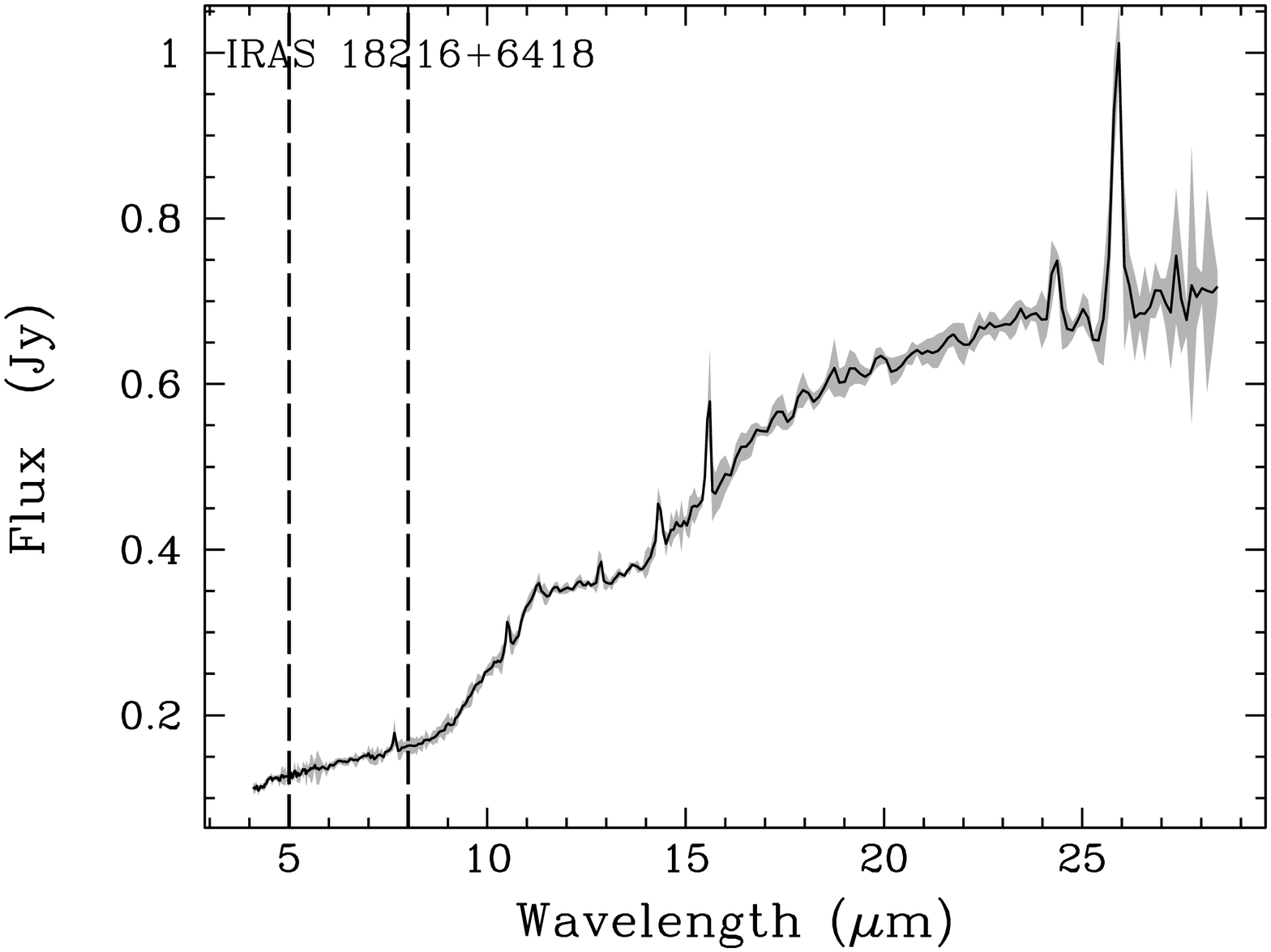}
      \label{fig:iras18216}
     }
    }
    \mbox{ 
     \subfigure{
      \includegraphics[angle=-90,width=.3\linewidth]{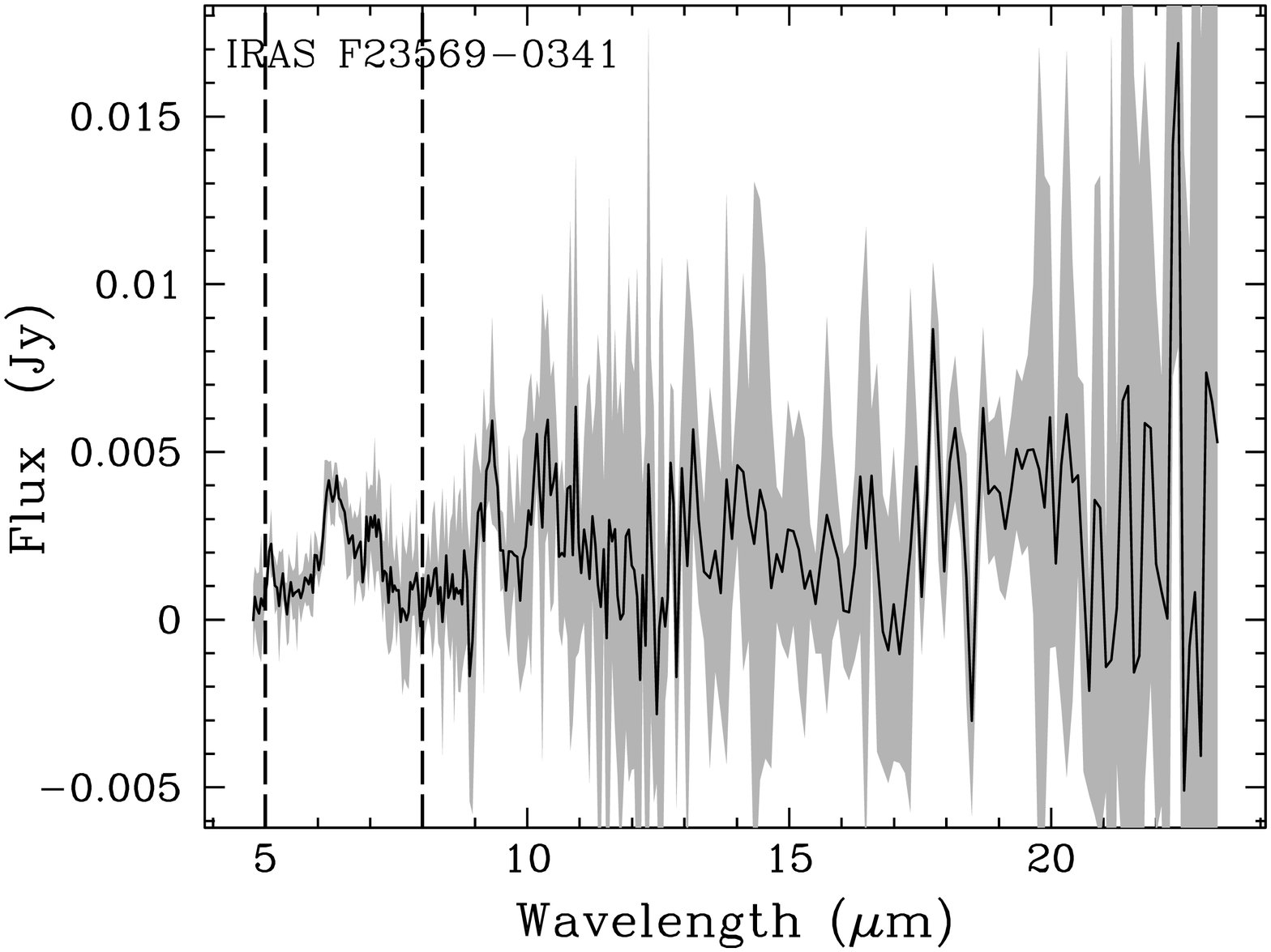}
      \label{fig:irasF23569} 
     } 
    }
   \end{center}
 \caption{MIR rest-frame spectra of HLIRG obtained with the Infrared
  Spectrograph onboard \Spitzer. The grey shaded area is the
  $1\sigma$ uncertainty region. The vertical dashed lines limit the spectral
  region where our decomposition technique was applied.}
 \label{fig:irsspec}
\end{figure*}

Two spectra were obtained for each source, one for each nod. We
calculated an error-weighted average of the two nod spectra\footnote{Averaging
the spectra introduces a systematic error related to the flux differences
between the two nod spectra. This difference is usually due to the presence of
other sources in the slit affecting the background subtraction, or to pointing
errors resulting in a slight shift in the dispersion direction \citep{CASIS}.}
and they were de-redshifted to rest frame. Figure~\ref{fig:irsspec} shows the 
final spectra.

\section{SB/AGN spectral decomposition}
\label{sec:decomp}
\subsection{Model for the MIR emission of HLIRG}
\label{sec:model}
Thanks to the unprecedent sensitiviy of \Spitzer-IRS, MIR studies of
large samples of luminous AGN and SB have found a significant spectral
homogeneity in the SED of the two separate classes
\citep{Brandl06,Netzer07,Mullaney11,HernanCaballero11}. For high luminosity
sources, this is particularly true within the 5-8~\micr range, showing nearly
constants shapes of both the PAH complex in SB \citep{Brandl06} and the
continuum in AGN \citep{Netzer07}.

This small spectral dispersion allows the use of universal AGN/SB templates to
parameterize the observed energy output of objects where both phenomena are
present, like ULIRG or HLIRG, in the wavelength range below 8 \micr. This
technique has been extensively tested in ULIRG, showing that the observed
differences of the PAH equivalent width and strength are caused, in most
sources, by the relative contribution of the AGN continuum and its obscuration
\citep{Nardini08, Nardini09, Nardini10}.

The bolometric luminosities of HLIRG and ULIRG differ no more than about an
order of magnitude (even less for the SB component). Moreover, a significant
fraction of HLIRG seems to share common properties with local ULIRG
\citep{Ruiz10}. Thus, we can rely with reasonable confidence on the model
proposed by \citet[][hereafter N08]{Nardini08} to reproduce the 5-8~\micr
emission in HLIRG. The observed flux at wavelength $\lambda$, $\Fobs$, is given
by:
\begin{equation}
\label{eq:model}
\Fobs = \Fint \left[\alfa~\tempAGN \mathrm{e}^{-\tau(\lambda)}  + (1 - \alfa)~\tempSB \right],
\end{equation}
where $\alfa$ is the AGN contribution to $\Fint$ (the intrinsic flux density at
6~\micr) while $\tempAGN$ and $\tempSB$ are the AGN and SB emission normalized
at 6 \micr. This model has only three free parameters: $\alfa$,
$\tau_6$ (the optical depth at 6~\micr) and $\Fint$ (the flux
normalization). Additional high-ionization emission lines and molecular
absorption features (due to ices and aliphatic hydrocarbons), whenever present,
were fitted using Gaussian profiles.

\subsubsection*{Starburst emission}
\begin{figure*}[ht]
 \includegraphics[width=0.9\linewidth]{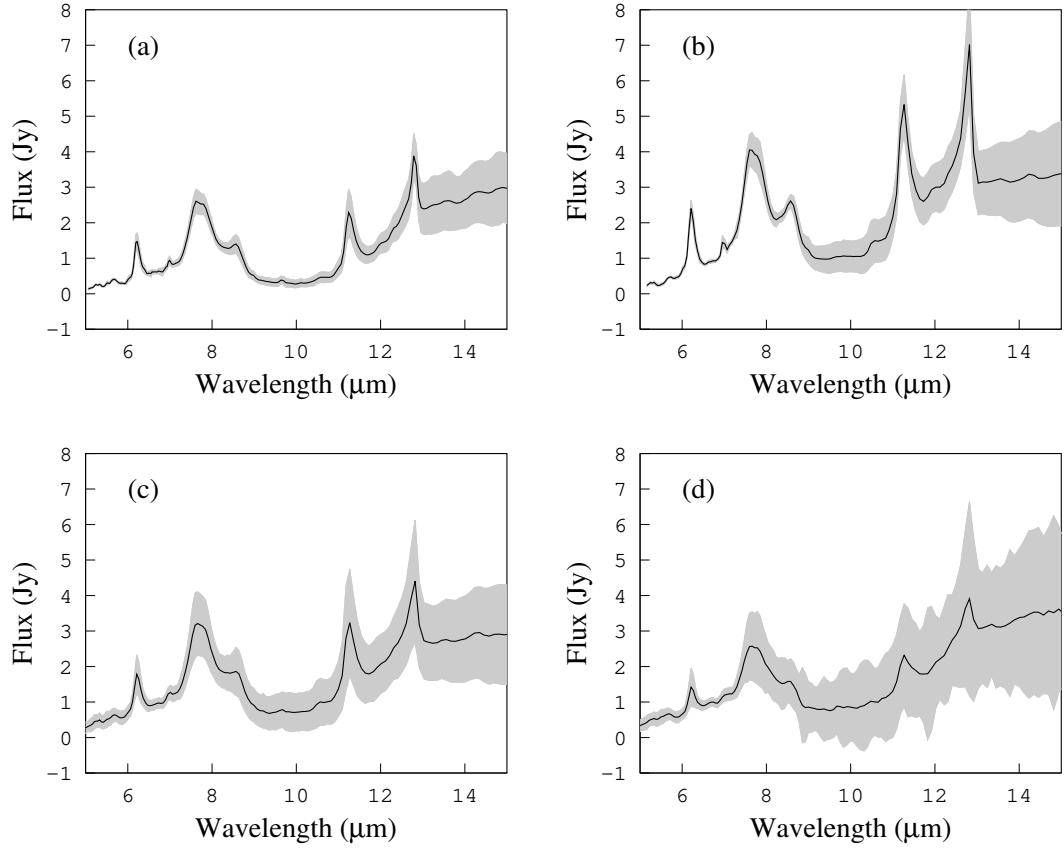}
 \caption{Average SB spectra used as templates in our model: Template (a) is from N08;
          templates (b), (c) and (d) from \citet{HernanCaballero11} (see Sect.~\ref{sec:decomp} 
          for details). The grey shaded area is the estimated $1\sigma$ dispersion.}
 \label{fig:sbtemplates}
\end{figure*}

As stated above, starbursts of different luminosities present a small spectral
dispersion in their MIR emission below 8~\micr. Given the large number of
physical variables (e.g. the initial mass function, the duration and evolution
of the burst, dust grain properties) involved in determining the observational
properties of an SB, it is in fact surprising finding such similarity.
\citet{Brandl06} suggested that this homogeneity can derive from the spatial
integration of unresolved star-forming spots. We can then assume that an
observational template will accurately reproduce the SB emission of our sources
in the 5-8~\micr range. This emission is characterized by two prominent PAH
features at 6.2 and 7.7~\micr (Fig.~\ref{fig:sbtemplates}).

To increase the robustness of our analysis we have employed four different SB
templates from N08 (a) and from \citealt{HernanCaballero11} (b, c, d): a) the
average MIR spectrum from the five brightest ULIRG among pure SB sources in the
N08 sample; b) the average spectrum of objects optically classified as SB or HII
in NED and with $\nu L_{\nu}$(7 \micr)$<10^{44}~\mathrm{erg s^{-1}}$ (16
sources); c) the average spectrum of SB-dominated objects in the MIR (257
sources) and d) the average spectrum of objects classified as ULIRG (184
sources). As expected, all SB templates show similar features\footnote{Slight
differences in the shape of the lines and their intensity ratio are probably
caused by variations of the obscuration in the original sources: higher
luminosity sources usually show higher obscuration \citep{Rigop99}.} and small
dispersion below 8~\micr (see Fig.~\ref{fig:sbtemplates}).

\subsubsection*{AGN emission}
In our range of interest the AGN spectrum is dominated by the cooling
of small dust grains that are transiently heated up to temperatures close to the
sublimation limit. This process is expected to produce a nearly featureless
power law continuum. However, PAH features are found in several quasars
\citep{Schweitzer06,Lutz07,Lutz08}. \citet{Netzer07} adscribe the FIR emission
of a set of luminous PG QSO to cold dust in extended regions, thus probably of
stellar origin. After removing this galactic contamination, the average spectrum
can be modeled with a single power law from $\sim3$~\micr to the 9.7~\micr
silicate bump. Since the spectral dispersion within this range is small for pure
luminous QSO, we can model the AGN emission in our sources with the same power
law with a fixed spectral index: $f_{\lambda}\propto \lambda^{0.8}$. Again, to
increase the robustness of our results, we alternatively fitted all sources
allowing the slope of the power law to vary in the range 0.2-1.8. This interval
covers the dispersion found in the spectral index for the AGN emission in ULIRG
\citep{Nardini10}.

Our model includes an exponential attenuation in the AGN emission to take into
account the reddening of the NIR radiation due to any compact absorber in the
line of sight. The optical depth follows the conventional law
$\tau(\lambda)\propto\lambda^{-1.75}$ \citep{Draine89}. The SB emission does not
need a similar correction because the possibles effects of obscuration are
already accounted in the observational templates.

\subsection{Results}
\label{sec:results}
The model presented in the previous section was implemented in Sherpa
\citep{sherpa}, the modelling and fitting tool included in the software package
CIAO 4.3 \citep{ciao}. The 5-8 \micr spectra were fitted using a
Levenberg-Marquardt $\chi^2$ minimization algorithm. The errors of the
parameters (within $1\sigma$ confidence limit) were calculated through
Montecarlo simulations.

\begin{figure*}[ht]
   \begin{center} 
    \mbox{ 
     \subfigure{
      \includegraphics[angle=0,width=.33\linewidth]{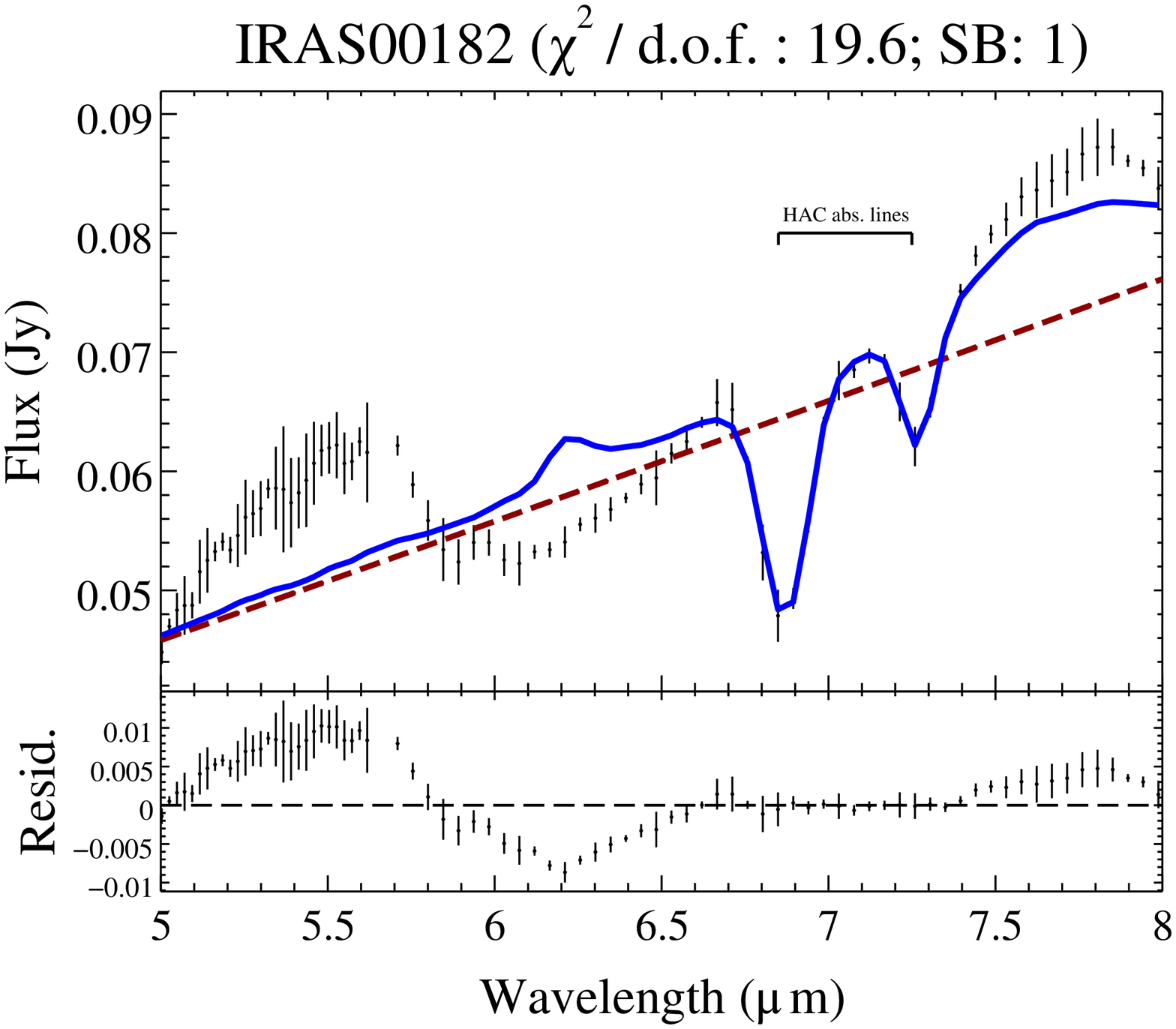}
      \label{fig:iras00182fit} 
     } 
     \subfigure{
      \includegraphics[angle=0,width=.33\linewidth]{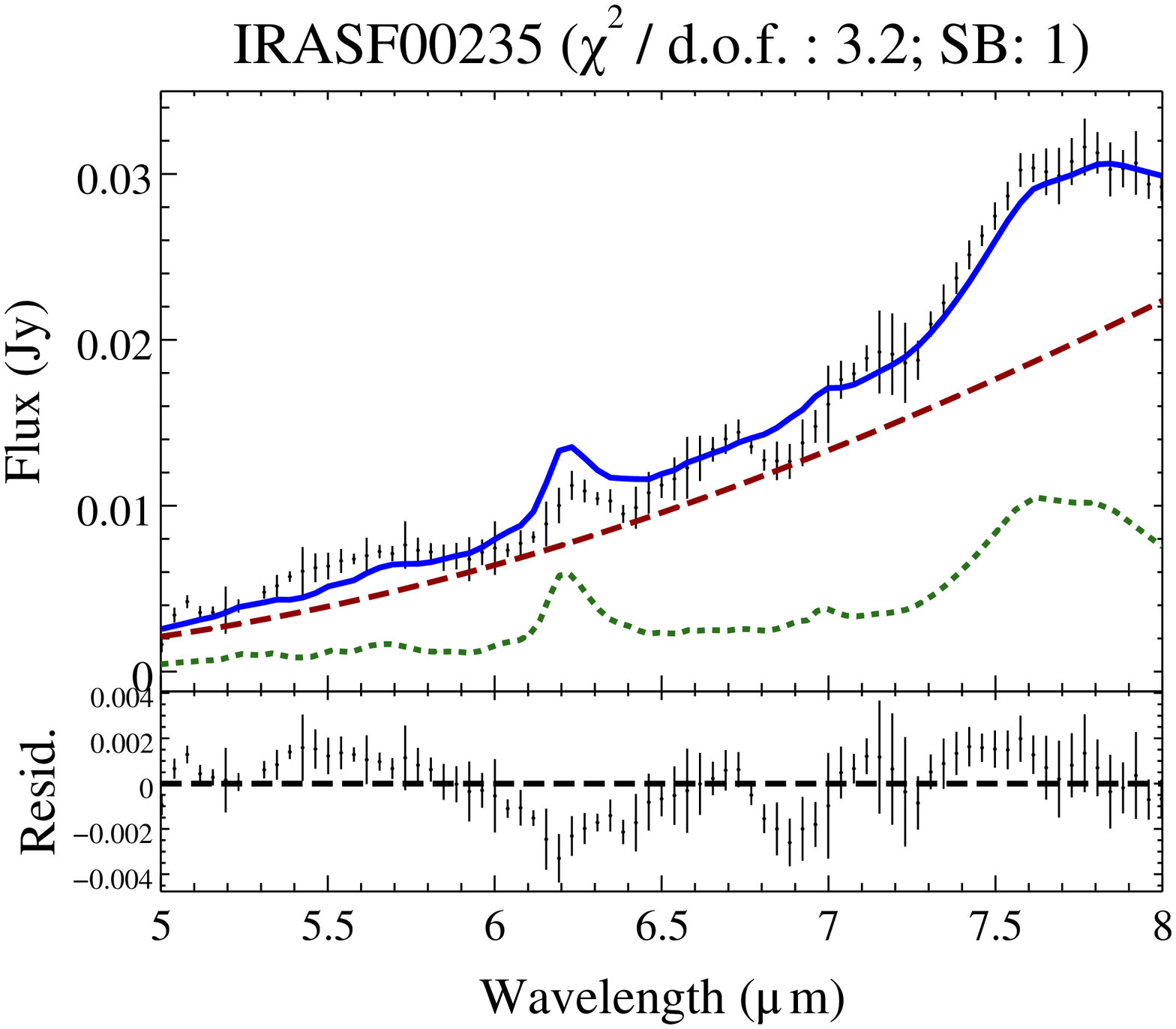}
      \label{fig:irasF00235fit} 
     } 
     \subfigure{
      \includegraphics[angle=0,width=.33\linewidth]{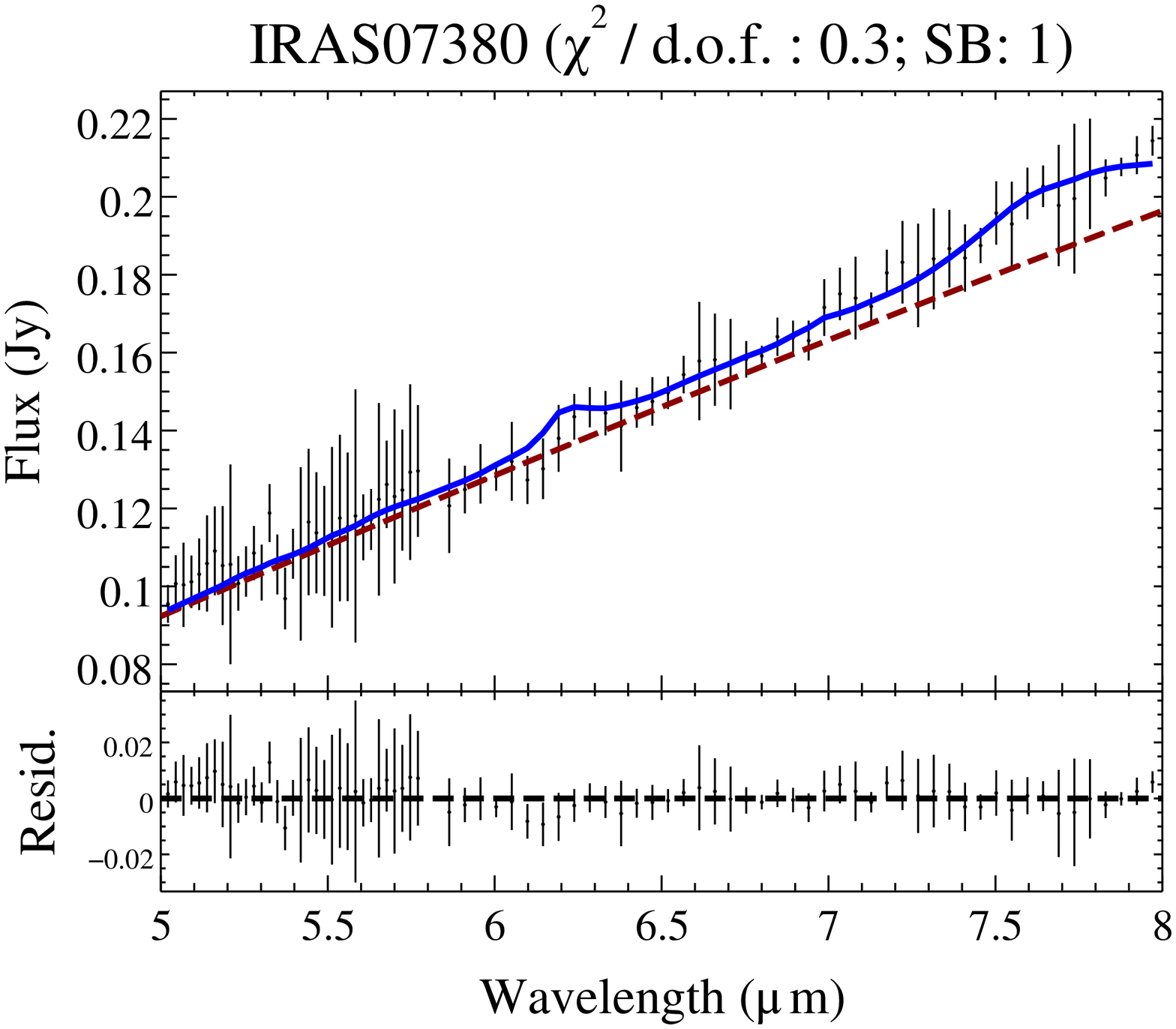}
      \label{fig:iras07380fit} 
     } 
    } 
    \mbox{ 
     \subfigure{
      \includegraphics[angle=0,width=.33\linewidth]{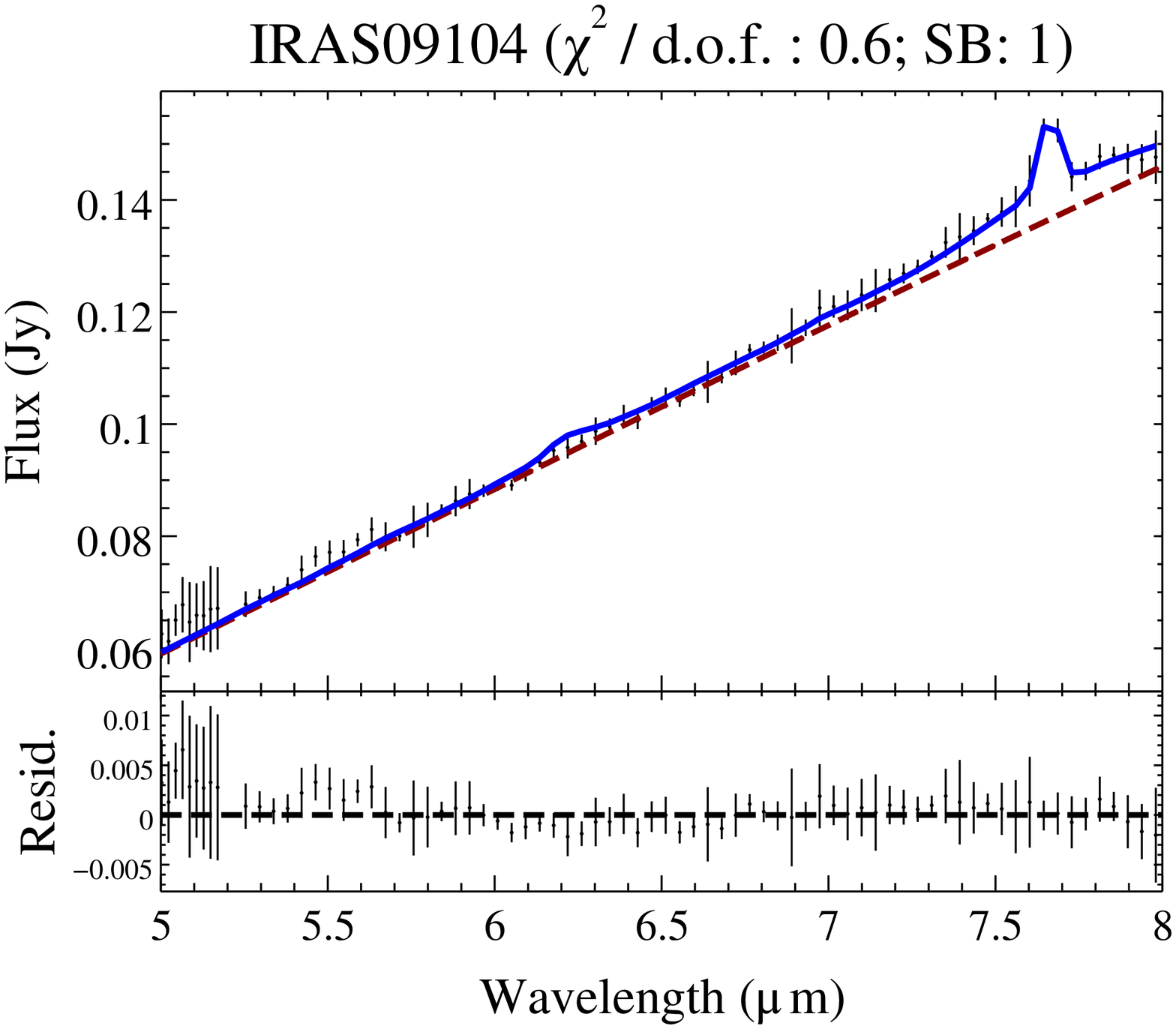}
      \label{fig:iras09104fit} 
     } 
     \subfigure{
      \includegraphics[angle=0,width=.33\linewidth]{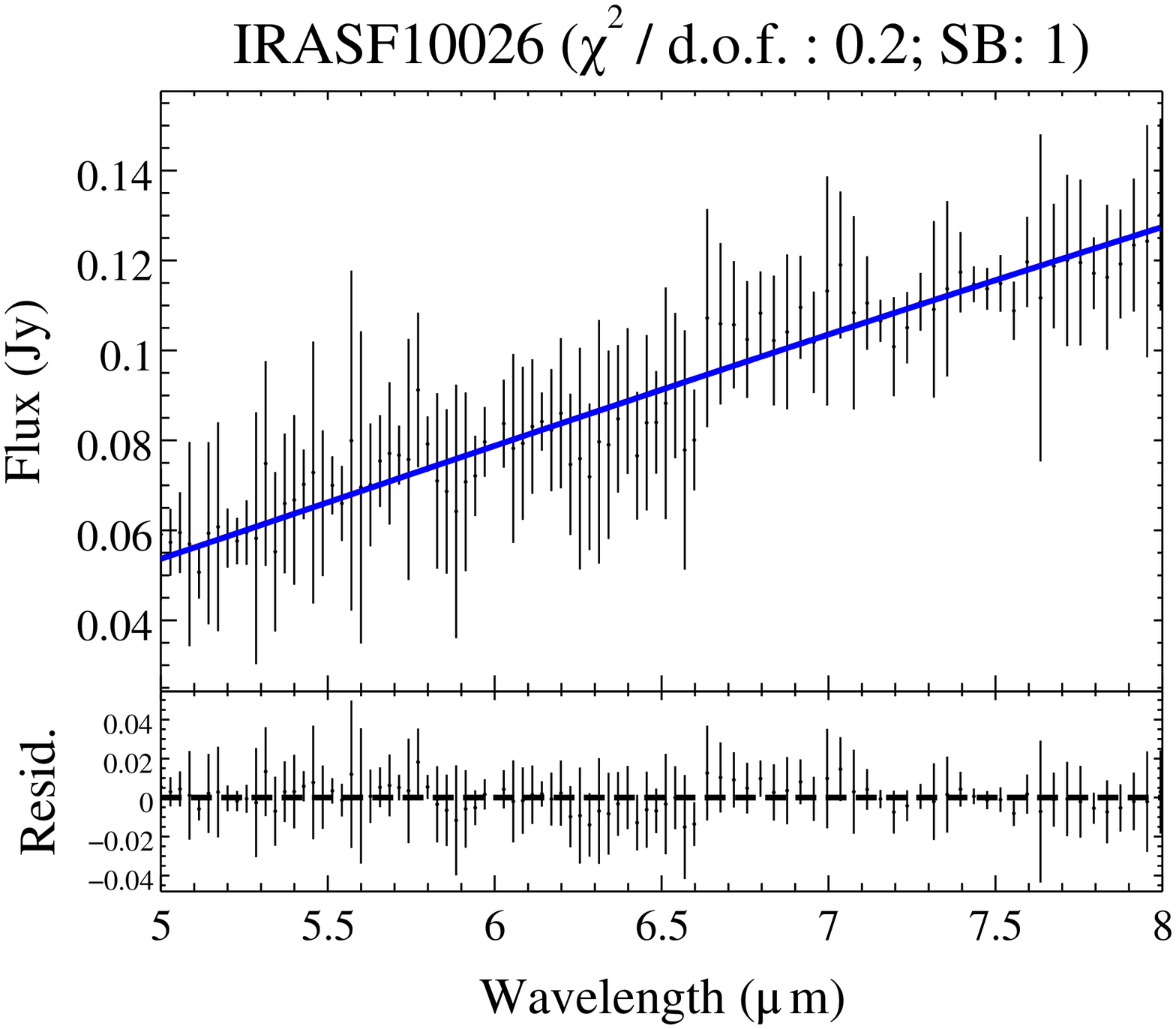}
      \label{fig:irasF10026fit} 
     } 
     \subfigure{
      \includegraphics[angle=0,width=.33\linewidth]{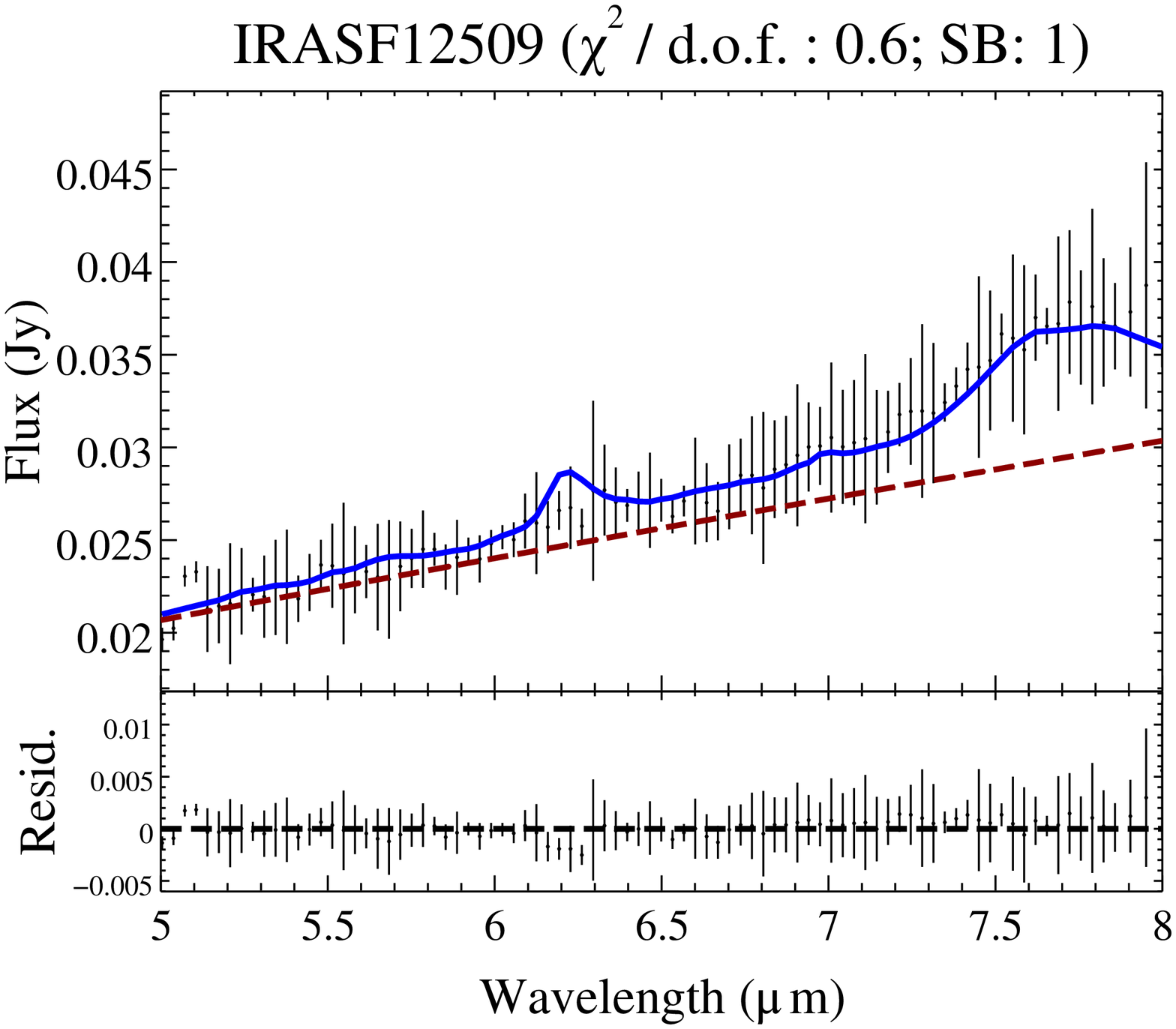}
      \label{fig:irasF12509fit} 
     } 
    } 
    \mbox{ 
     \subfigure{
      \includegraphics[angle=0,width=.33\linewidth]{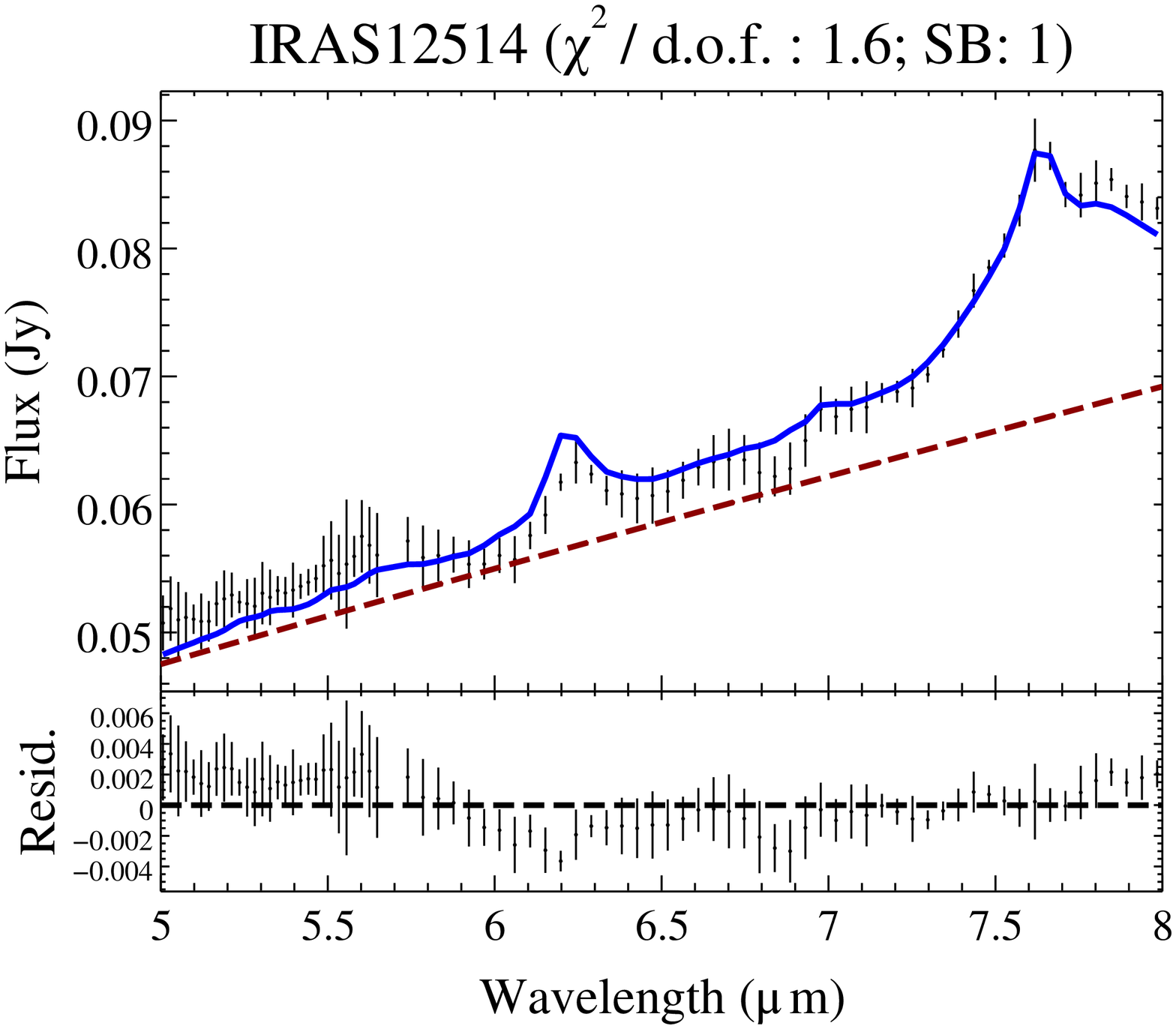}
      \label{fig:iras12514fit} 
    } 
     \subfigure{
      \includegraphics[angle=0,width=.33\linewidth]{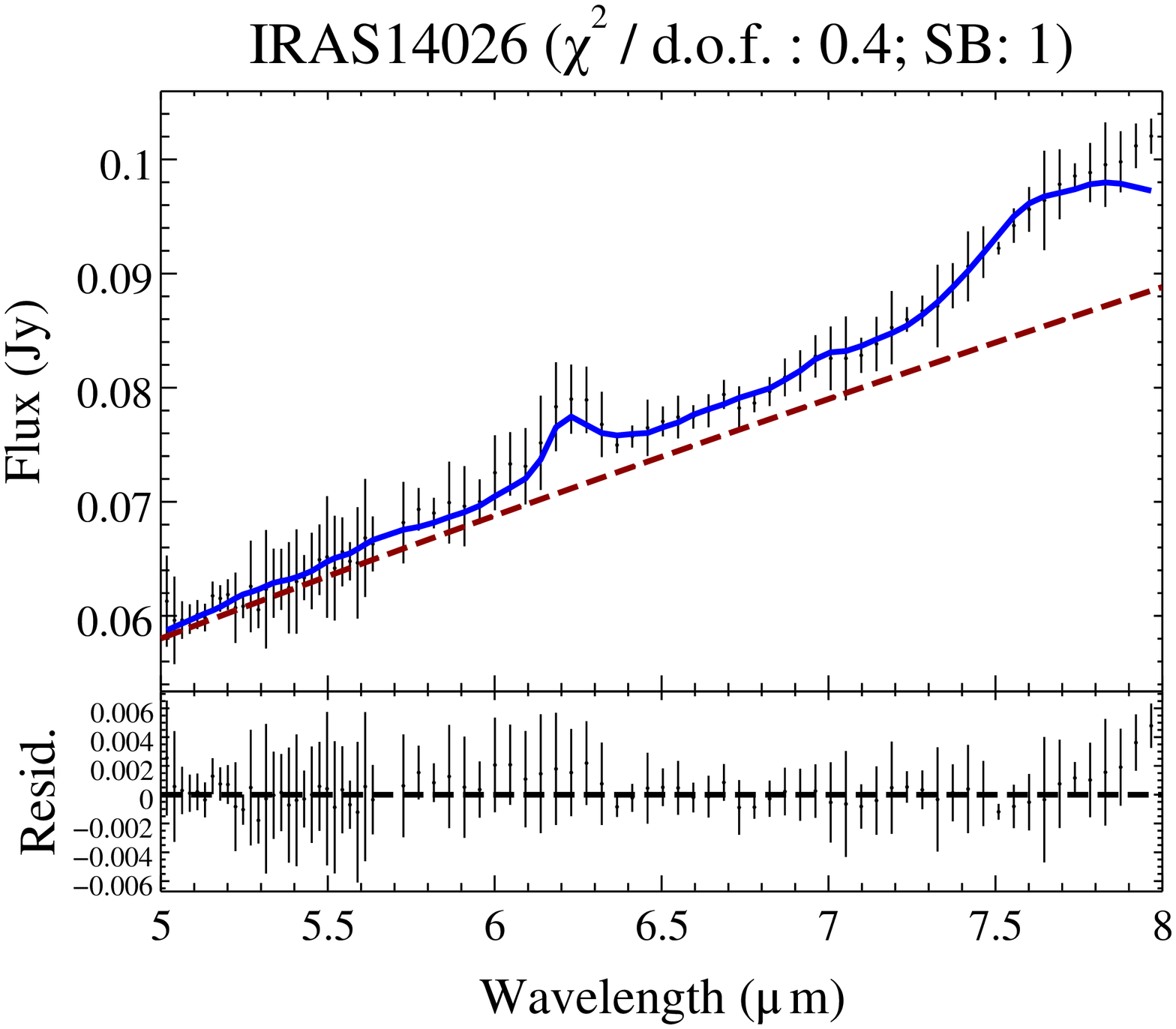}
      \label{fig:iras14026fit} 
     }
     \subfigure{        
      \includegraphics[angle=0,width=.33\linewidth]{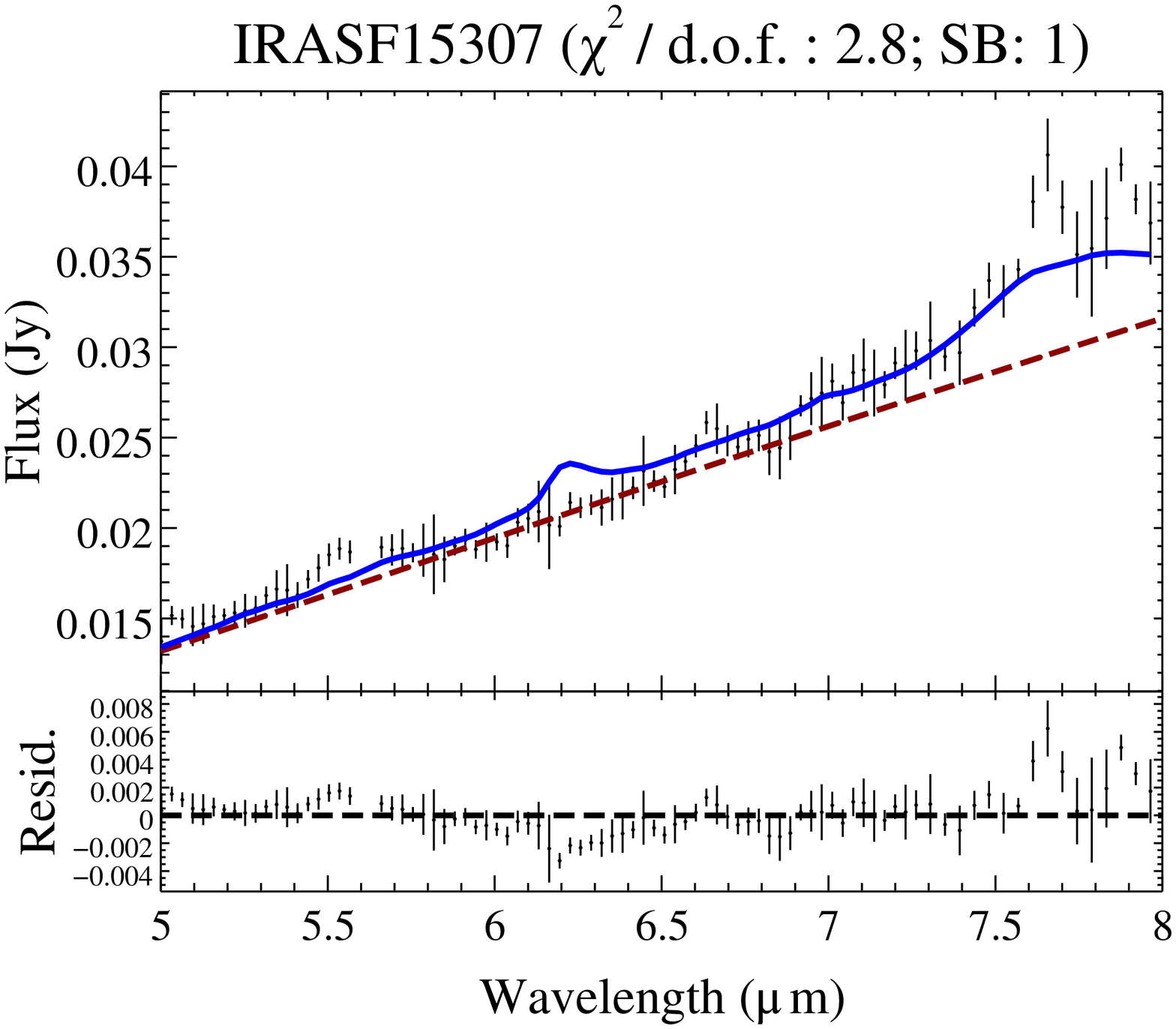}
      \label{fig:irasF15307fit}
     }
    } 
    \mbox{
     \subfigure{
      \includegraphics[angle=0,width=.33\linewidth]{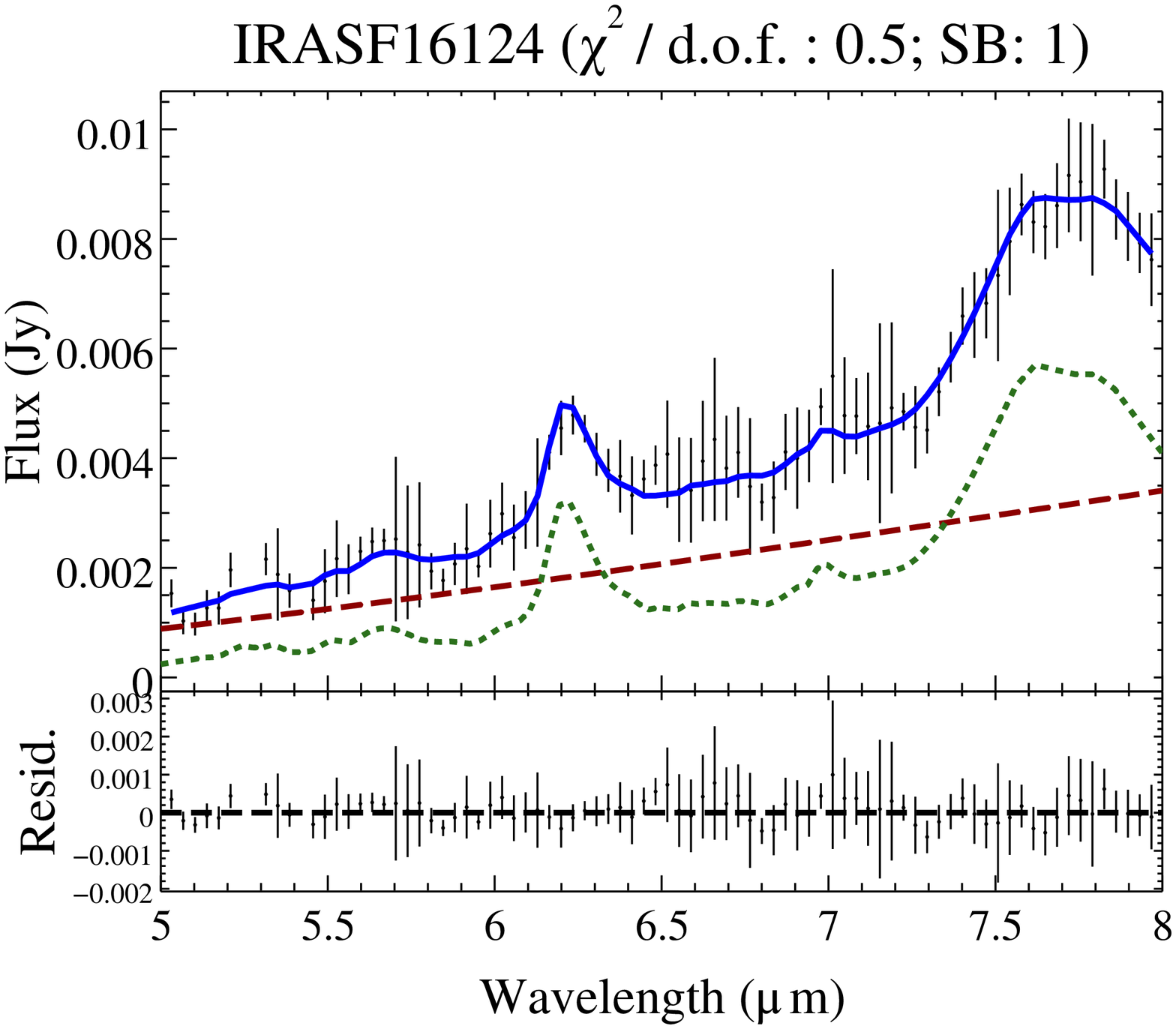}
      \label{fig:irasF16124fit}
     }
     \subfigure{
      \includegraphics[angle=0,width=.33\linewidth]{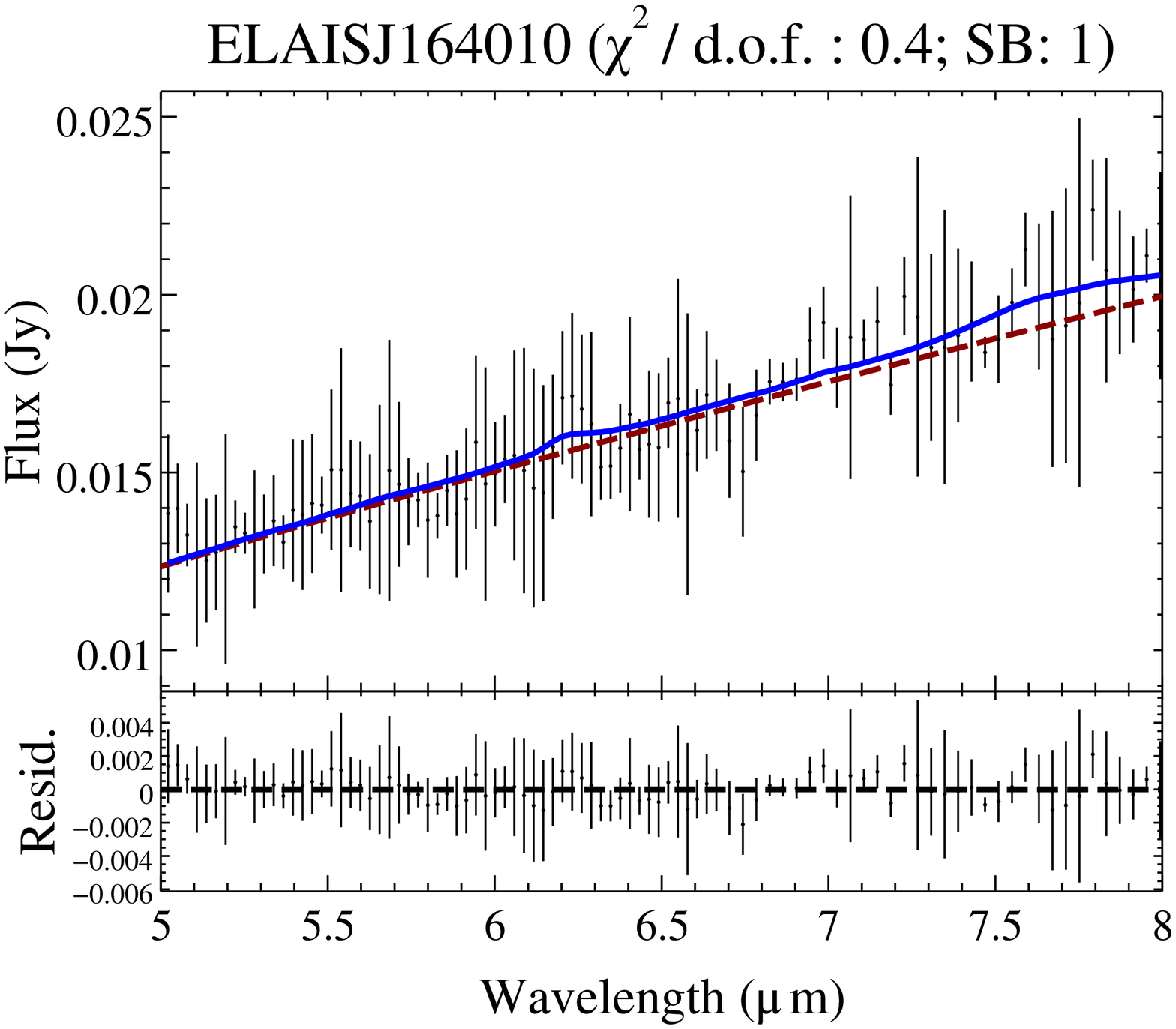}
      \label{fig:ej1640fit} 
     }      
     \subfigure{
      \includegraphics[angle=0,width=.33\linewidth]{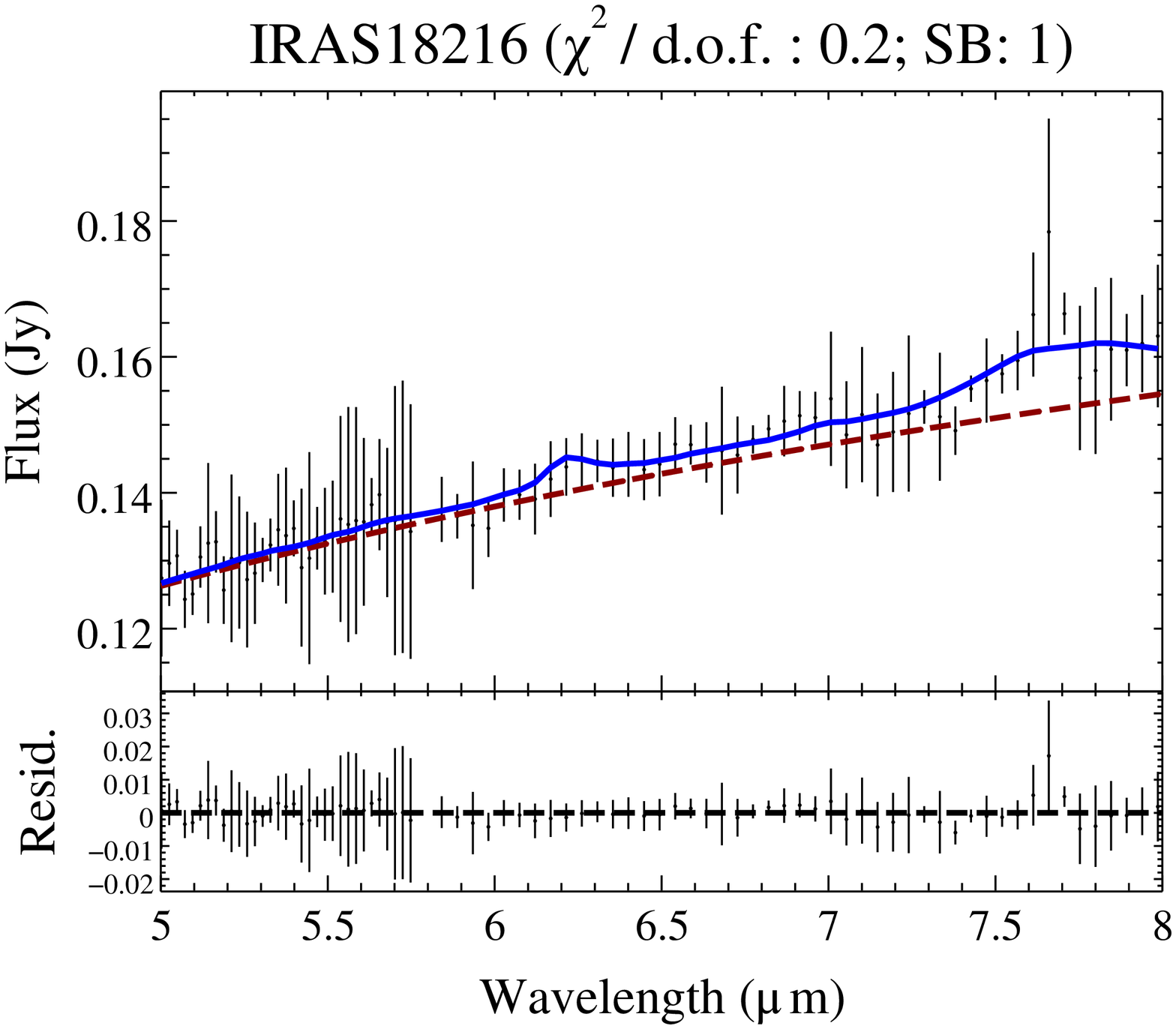}
      \label{fig:iras18216fit} 
     }
    }
   \end{center}
 \caption{5-8 \micr spectra and best fit models (blue solid line). The AGN component is ploted as 
          a red dashed line. For clarity, the SB component (green dotted line) is ploted only on 
          those sources where the AGN and SB contributions are comparable.}
 \label{fig:irsspecfit}
\end{figure*}

One object was barely detected by \Spitzer and in two additional sources our
model was not suitable (see Sect.~\ref{sec:sournotes} for a further discussion
on these objects), but the spectra of the remaining ten sources were well
reproduced by our model (see Fig.~\ref{fig:irsspecfit} and Table~\ref{tab:fit}).

We found no significant difference between the best-fit models using the four SB
templates. All templates offer a similar $\chi^2$ and consistent values for the
model parameters. We have therefore included in Table~\ref{tab:fit} only the
results obtained using the N08 SB template. However, we should note that the use
of the ULIRG template provides systematically lower values of $\alfa$ in all
sources. This result is probably caused by the presence of an AGN component in
the template, as expected in an average ULIRG spectrum including
high-luminosity objects \citep{Nardini10,HernanCaballero11}.

\begin{figure*}[ht]
\mbox{
\centering 
 \subfigure{
   \includegraphics[angle=0,width=0.5\linewidth]{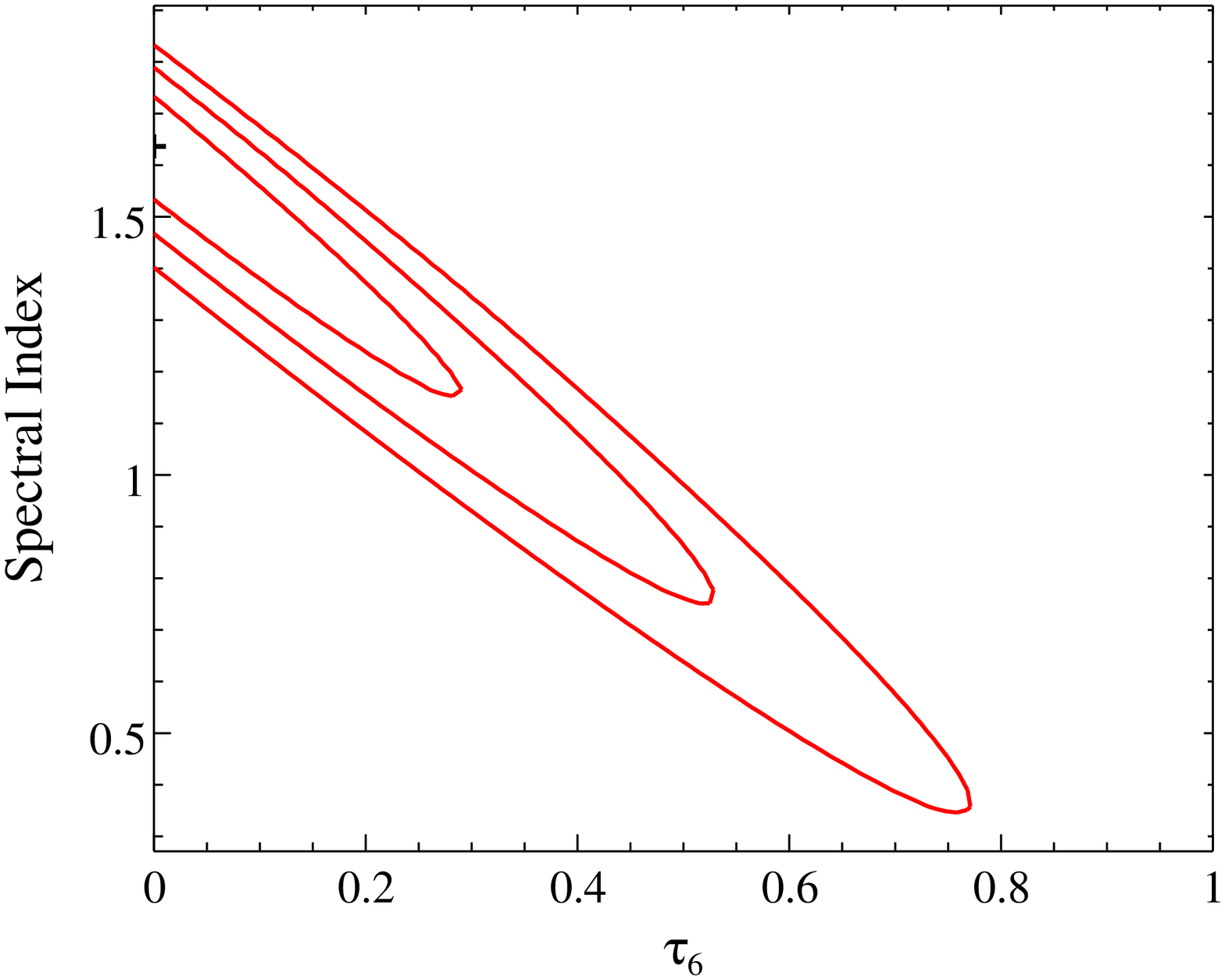}
   \label{fig:confregIRAS07380} 
 } 
 \subfigure{
   \includegraphics[angle=0,width=0.5\linewidth]{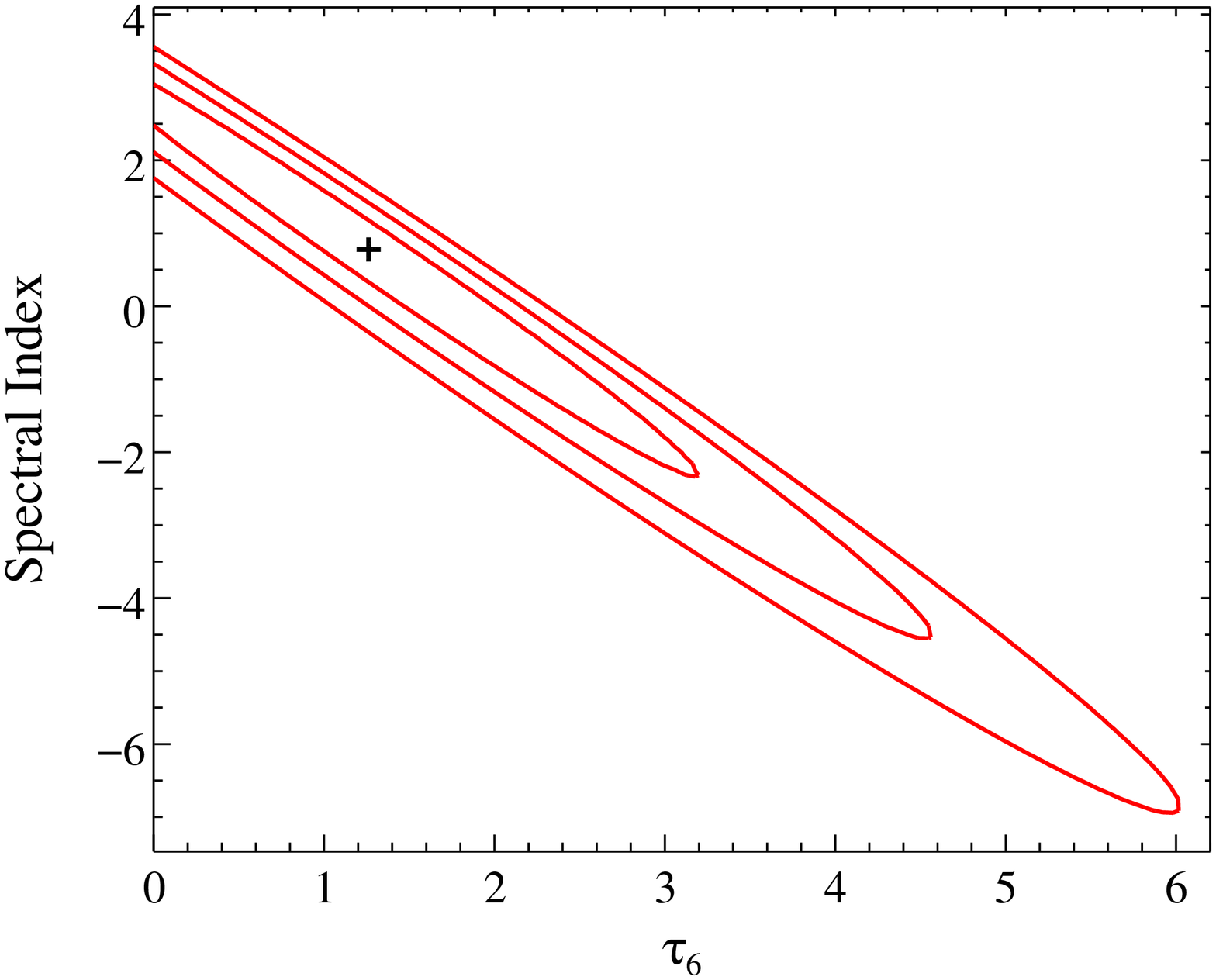}
   \label{fig:confregIRASF16124} 
  }
}
 \caption{Spectral Index-Optical depth confidence regions for IRAS~07380-2342 (left) and IRAS~F16124+3241 (right). 
          The contours represent $1\sigma$, $2\sigma$ and $3\sigma$ confidence levels.}
 \label{fig:confreg}
\end{figure*}

Allowing the spectral index to vary did not enhance the fit in most sources. A
statistically significant improvement (in terms of F-test) was found only in
three sources with high optical depth (beside IRAS\,18216+6418, which is
discussed below). However we must note that our model is severely degenerated in
the optical depth-spectral index plane (see Fig.~\ref{fig:confreg}). The shape
of the 5-8~\micr continuum cannot be well established in sources showing high
absorption. Hence, if the spectral index is a free parameter, the minimization
algorithm decreases the optical depth while increasing the spectral index. We
selected the model with fixed spectral index as our best fit in those sources
showing obscuration features outside the 5-8 \micr range (e.g. 9.7~\micr
silicate absorption).

The 6~\micr emission is largely dominated by the AGN component
($\alfa\gtrsim0.9$), which also dominates over the whole 5-8~\micr spectral
range. IRAS\,F00235+1024 and IRAS\,F16124+3241 are the only two sources where
the SB and AGN components are comparable, but the AGN emission is still above
the SB. The PAH emission at $\sim6.2$ and $\sim7.7$~\micr in these two sources
is well reproduced by our SB template, suggesting that our assumption, i.e. the
MIR SB emission of HLIRG can be modelled by the MIR emission of SB-dominated
ULIRG, is correct. 

All sources optically classified as type I AGN show minor absorption (except
IRAS\,F10026+4949), while all sources classified as NL or type II AGN show
high absorption, (except IRAS\,12514+1027, which shows a strong 9.7~\micr
silicate absorption feature).

In two sources (IRAS\,09104+4109 and IRAS\,12514+1027) we detected a spectral
feature at $\sim7.7$~\micr that can be modelled as a gaussian emission line. In
both cases the line is statistically significant (in terms of F-test), although
it is not needed to obtain a good fit. The line can be interpreted as the
unresolved [NeVI] fine-structure line at 7.65~\micr. Due to its high ionization
energy (126 eV), it seems to be connected to the hardest nuclear activity. This
emission feature has been detected in previous studies of local ULIRG
\citep{Nardini09,Veilleux09}.

\subsection{Notes on particular sources}
\label{sec:sournotes}
\subsubsection*{IRAS~00182-7112}
This Compton-Thick QSO2 shows a heavily absorbed MIR spectrum. At first sight,
our model seems to reproduce well the spectrum above 6.5~\micr (see
Fig.~\ref{fig:iras00182fit}) if two absorption lines at 6.8 and 7.3~\micr are
included (C-H stretching mode of hydrogenated amorphous carbon 
\citep[HAC;][]{Furton99}). However, the lack of the 6.2~\micr PAH emission feature and
the overall shape of the whole MIR spectrum (see Fig.~\ref{fig:iras00182})
strongly suggest that the 7.7~\micr feature is actually not associated with PAH
emission. \citet{Spoon04} suggest that the 5.7-7.7~\micr range is dominated by a
broad absorption complex due to water ice and hydrocarbons. Thus, our simple
model cannot reproduce the complex spectra of this object (as suggested by the
poor fit we obtain, with a reduced $\chi^2\sim20$).

The spectrum shows no signs of SB activity within the 5-8~\micr range, since the
PAH features are totally suppressed, and it is very similar to the MIR spectra
of deeply obscured AGN (e.g. NGC 4418, NGC 1068). This result, along additional
evidence from other wavelengths \citep[e.g. strong iron K$_\alpha$ emission line
with $\sim 1$~keV equivalent width in the X-ray spectrum,][]{Ruiz07}, suggests
that the power source hiding behind the optically thick material could be a
buried AGN.

\citet{Spoon04} present a detailed study of the whole IRS spectrum: the
strong absorption and weak emission features in the 4-27 \micr spectrum suggest
the existence of a dense warm gas cloud close to the nucleus of the source.
Based on the strengh of the 11.2~\micr PAH feature, there is also
evidence of star formation activity away from the absorbing region, that could be
responsible for up to 30\% of the IR luminosity of the system. For
further analysis and comparison we have employed this estimate of the SB
relative contribution.

\subsubsection*{IRAS~09104+4109}
This is a radio loud QSO2 and the MIR spectrum is probably dominated by
synchrotron radiation, modifying the shape of the AGN emission. Although our
model offers an excellent fit in a pure statistical sense ($\chi^2 /
\mathrm{d.o.f.} \simeq 0.6$), the selected AGN template just models the
reprocessed emission by dust. Hence, in this case, the physics predicted by our
model could be inaccurate.

\subsubsection*{IRAS~18216+6418}
The continuum of this QSO seems to be flatter than the adopted power
law. This is the only non-obscured source where a free spectral index model
offers a clear improvement in the best-fit. The flatter continuum could be just
the natural dispersion around the template: a flatter AGN slope has been
detected in $\sim10\%$ of ULIRG \citep{Nardini10}. However, we must note that
this object, although classified as a radio-quiet AGN, show many properties of
radio-loud sources (F02). Thus we cannot reject the posibility that the flatter
continuum is caused by synchrotron contamination.

\subsubsection*{IRAS~F23569-0341}
This source was barely detected by \Spitzer and the extracted spectrum had a
very poor S/N ratio, so it was rejected in the subsequent analysis. This object
has not been detected neither with \ISO or SCUBA (F02).

\section{Discussion}
\label{sec:discuss} 
The 5-8~\micr spectra of most ULIRG show signatures of AGN activity, but the
relative contribution of the AGN emission at 6~\micr spans a broad range,
from complete SB-dominated to complete AGN-dominated output
\citep{Nardini08,Nardini09,Nardini10}. Our spectral decomposition, on the other
hand, clearly states that 5-8~\micr spectra of HLIRG are strongly dominated by
the nuclear AGN emission reprocessed by the dusty torus. This is consistent with
previous studies showing that the fraction of ULIRG harbouring an AGN and the
relative contribution of this component increase with IR luminosity
\citep{Veilleux99,Veilleux02,Nardini10}.

This result is also in agreement with previous studies in other wavelengths
which found AGN activity in most HLIRG
\citep[F02;][]{Verma02,Rowan00,Rowan10,Ruiz07,Ruiz10}. Moreover, based on the
completeness and non AGN bias of the F02's sample, our analysis is a strong
direct evidence that all HLIRG harbour an AGN.

\subsection{Star formation rate from PAH emission}
\label{sec:sfr}
Many studies have shown the tight correlation between the presence of Polycyclic
Aromatic Hydrocarbon (PAH) features in the MIR spectrum and the presence of
starburst activity \citep{Genzel98,Rigop00,Brandl06}. The PAH emission arises in
the photo-dissociation region that lies between the HII region of an SB and the
surrounding molecular cloud where the stars are formed. The PAH spectral
features are, in addition, very uniform, especially below 8 \micr (cf. Fig.~1 of
N08 or Fig.~2 from \citealt{Sargsyan}).

The 5-8~\micr spectra of our HLIRG are strongly dominated by the AGN emission,
so we cannot obtain a reliable direct measure of the PAH emission. Instead, we
estimated the emission of this spectral feature through the SB component
obtained with our spectral decomposition. For each source, we calculated a rough
estimate of the PAH peak flux at 7.7~\micr using the best-fit
model:\footnote{$\tempSB (7.7~\upmu\mathrm{m})$ was estimated interpolating the
N08 SB template at 7.7~\micr.}
\begin{equation}
\label{eq:sfr}
 f_{\lambda}^\mathrm{SB}(7.7~\upmu\mathrm{m}) = \Fint~(1 - \alfa)~\tempSB(7.7~\upmu\mathrm{m}).
\end{equation}

After converting these fluxes into luminosities, we estimated the star
formation rate (SFR) using the relation obtained by \citet{Sargsyan}:
\begin{equation}
\label{eq:sfr2}
 \log \mathrm{SFR}_{\mathrm{PAH}} = \log \left[\lambda L_\lambda^\mathrm{SB}(7.7~\upmu\mathrm{m})\right] - 42.57 \pm 0.2,
\end{equation}
where $\mathrm{SFR}_{\mathrm{PAH}}$ is the star formation rate in solar masses
per year and $\lambda L_\lambda^\mathrm{SB}(7.7~\upmu\mathrm{m})$ is the PAH
luminosity at 7.7~\micr in ergs per s. 

Our estimate of the 7.7~\micr flux includes both the PAH emission and the
underlying SB continuum due to dust emission, but Eq.~\ref{eq:sfr2} is calibrated
respect to the total 7.7~\micr flux, including also both components. In any
case, the continuum contribution is typically just $\sim10\%$ of the total
\citep{Sargsyan}. Since $\alfa$ is close to unity in all our sources, the
relative error of the SB contribution at 6~\micr is high. Thus our
SFR estimates cannot be accurate beyond the order of magnitude.

\begin{figure*}[ht]
\centering
 \includegraphics[angle=-90,width=.9\linewidth]{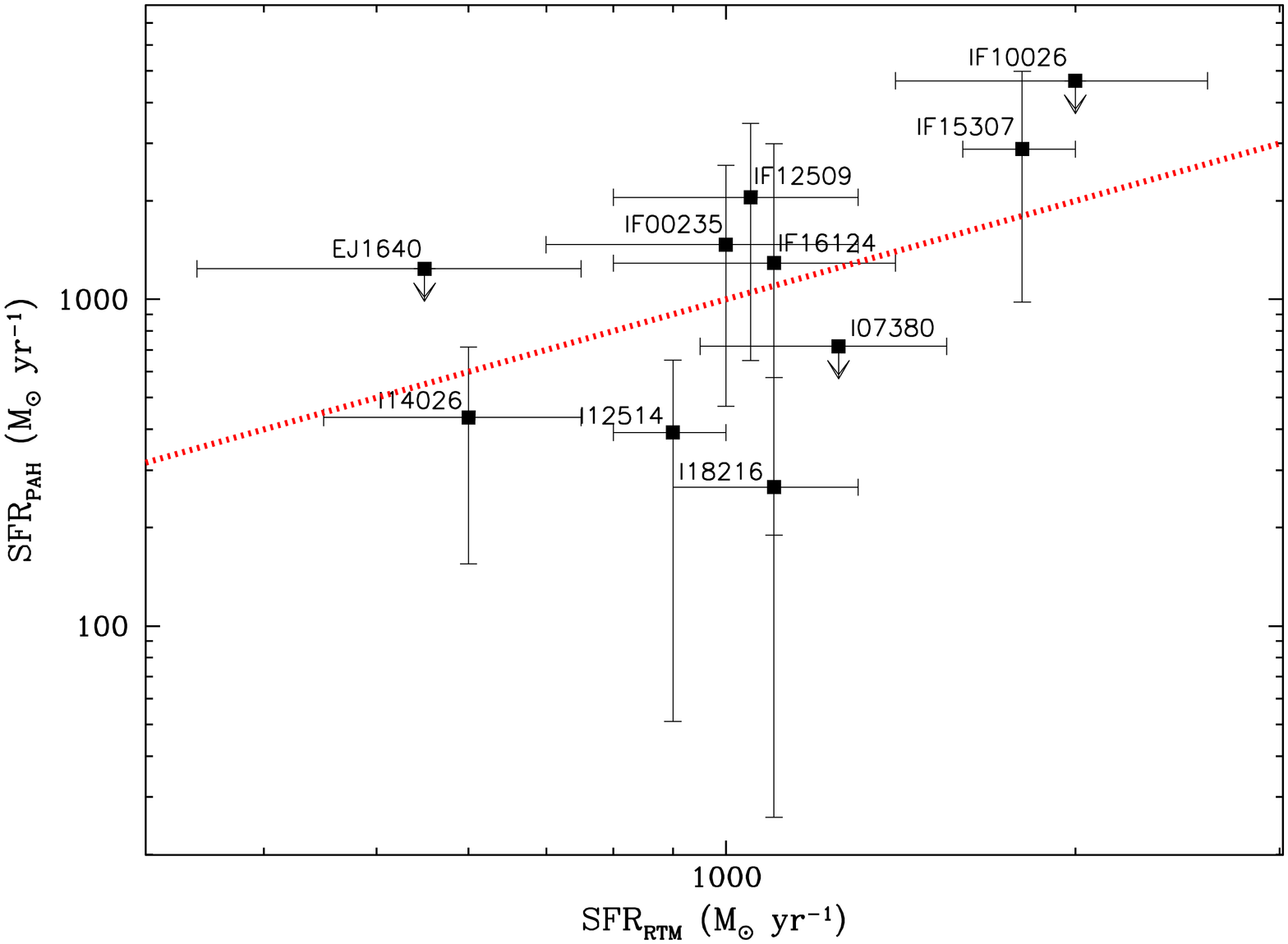}
 \caption{Comparison of the SFR estimated through PAH emission and through IR SED
	  modelling using RTM \citep{Farrah02submm,Rowan00,Verma02}.}
 \label{fig:SFRcomp}   
\end{figure*}

Figure~\ref{fig:SFRcomp} compares our SFR estimates with those obtained 
through modelling the IR SED using RTM \citep[F02]{Rowan00,Verma02}. Both
techniques seem to find consistent results. Despite most
being AGN-dominated objects (see Sect.~\ref{sec:alfabol} below), we still found
high star forming activity, with SFR $\sim300-3000~\Msun~\mathrm{yr}^{-1}$ (see
Table~\ref{tab:luminIR}).

\begin{figure*}[ht]
\centering
   \mbox{
     \subfigure[]{
      \includegraphics[angle=-90,width=0.5\linewidth]{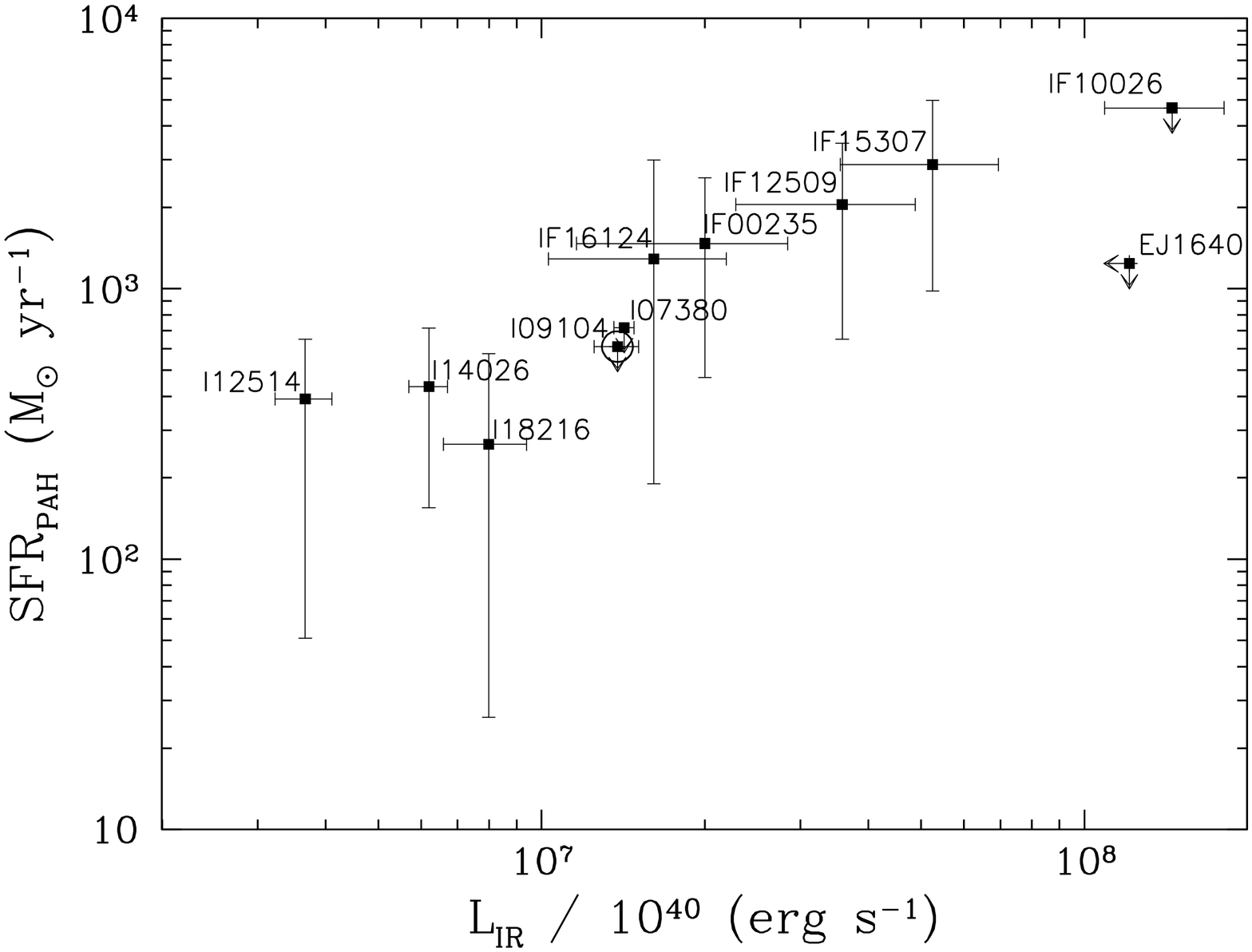}
      \label{fig:LIRtot}   
     }
     \subfigure[]{
      \includegraphics[angle=-90,width=0.5\linewidth]{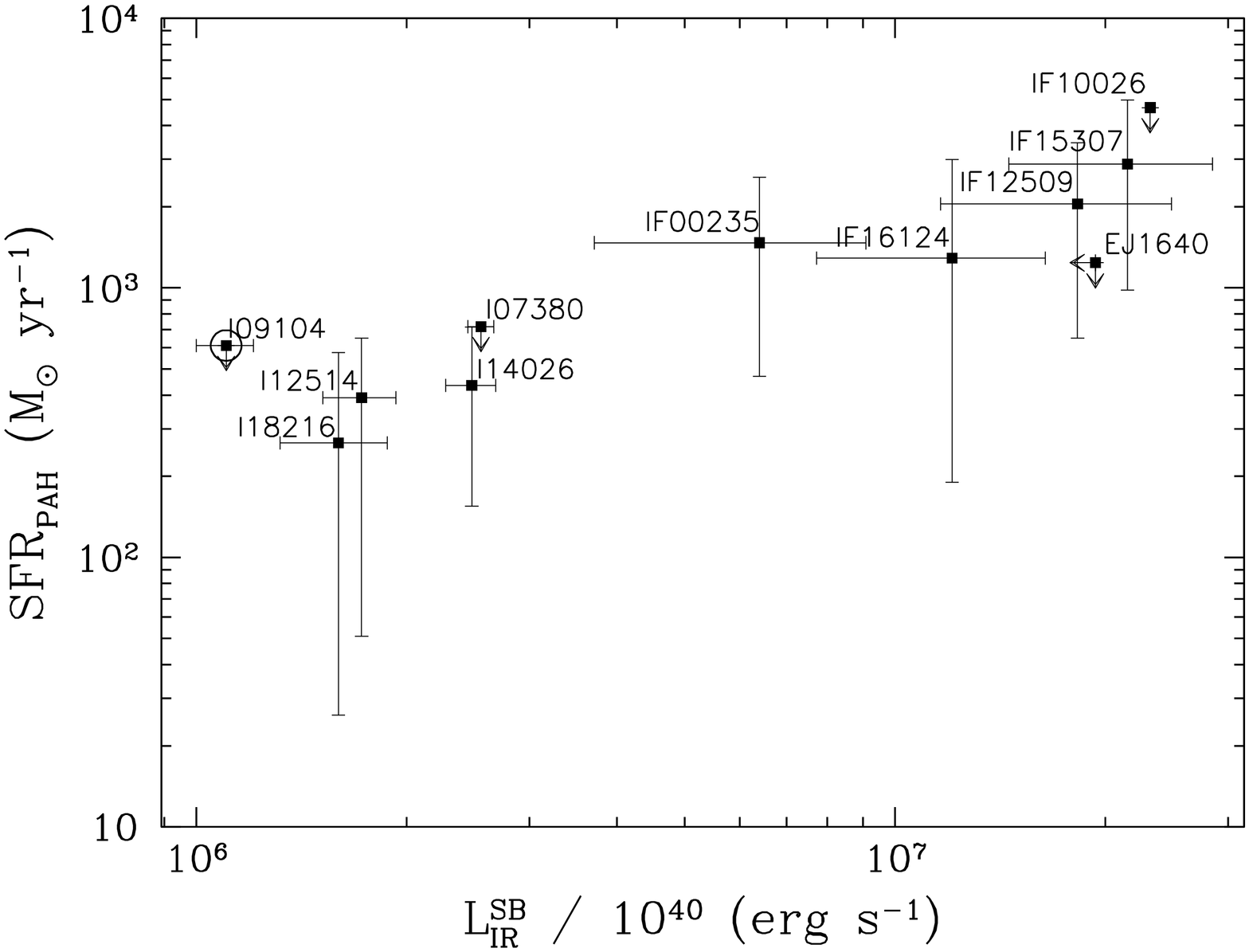}
      \label{fig:LIRsb}
     }
    }
\caption{SFR estimated using the PAH emission at 7.7~\micr versus (a) total and
         (b) starburst IR luminosities. The open circles mark those sources where our 
         model is inaccurate.}
\label{fig:SFRvsLIR}
\end{figure*}

Figure~\ref{fig:SFRvsLIR} shows the SFR we calculated versus (a) total
and (b) SB  IR luminosity. We estimated the SB luminosity as:
\begin{equation}
 L_\mathrm{IR}^\mathrm{SB} = (1 - \alpha_\mathrm{IR})~L_\mathrm{IR},
\end{equation}
where $\alpha_\mathrm{IR}$ is the fractional contribution of the AGN to the
total IR luminosity (see Sect.~\ref{sec:alfabol}) and $L_\mathrm{IR}$ is the
total IR emission (8-1000~\micr) calculated through the \IRAS fluxes
\citep{Sanders96}. When only an upper limit is available for the \IRAS fluxes at
12 or 25~\micr, we have simply assumed the one half of this limit. This
convention provides a fair aproximation of the actual flux densities even for
sources at moderate redshift \citep{Nardini09,Nardini10}.

We performed several statistical tests to check any possible correlation 
between the SFR and the IR luminosities. Since our estimates of
$L_\mathrm{IR}^\mathrm{SB}$ and SFR are both dependent of the parameter $\alfa$
(see Eqs.~\ref{eq:sfr} and \ref{eq:alfaIR}), we used a generalized Kendall's Tau
partial correlation test\footnote{Kendall's Tau test is a non-parametric
correlation test based on the Kendall rank coefficient \citep{Kendall90}. The
generalized version is an extension to include censored data (e.g. upper
limits). We used the code developed by the Center for Astrostatistics to perform
this test (\url{http://astrostatistics.psu.edu/statcodes/cens_tau}).}
\citep{Akritas96}. The result shows that both quantities are correlated with a
probability $\sim99.99\% > 3\sigma$. This tight correlation between the star
formation and SB IR luminosity is expected from several theoretical results
\citep[and references therein]{Kenn98,Draine01,Draine07}.

In the case of total IR luminosity versus SFR we can apply a generalized
Kendall's Tau test\footnote{We used the ASURV software for this test
\citep{asurv2}.}, since the estimate of the IR luminosity is completely
independent of our estimate of the SFR. They show a weaker, but slightly
significant, correlation (the probability is $98\%>2\sigma$). This slight
correlation and the high SFR suggest that SB emission could be a significant
contributor to the IR output (see Sect.~\ref{sec:alfabol}).

\subsection{AGN contribution to the IR luminosity}
\label{sec:alfabol}
We define the 6~\micr-to-IR bolometric ratio as:
\begin{equation}
\label{eq:Rdef}
R=\dfrac{\nu_6\Fint}{F_\mathrm{IR}},
\end{equation}
where $F_\mathrm{IR}$ is the total IR flux (8-1000~\micr) estimated as in
\citet{Sanders96}. The hypothesis that the IR and total bolometric fluxes are
almost coincident is typically a fair assumption in ULIRG, but in principle we
cannot adopt this hypothesis in our sources. The study of the HLIRG's broad band
SED has revealed that, for an important number of sources classified as HLIRG, a
non-negligible fraction of the bolometric luminosity is emitted outside the IR
range \citep{Ruiz10}. Therefore we limited our analysis to the IR luminosity.

We can derive a connection between $R$ and the parameter $\alfa$:
\begin{equation}
\label{eq:Ralpha6}
R=\dfrac{\RAGN~\RSB}{\alfa~\RSB + (1-\alfa)~\RAGN},
\end{equation}
where $\RAGN$ and $\RSB$ are the equivalents of $R$ for pure (unobscured) AGN
and pure SB, as defined in Eq.~\ref{eq:Rdef}. 

\begin{figure*}[ht]
\centering
 \includegraphics[angle=-90,width=\linewidth]{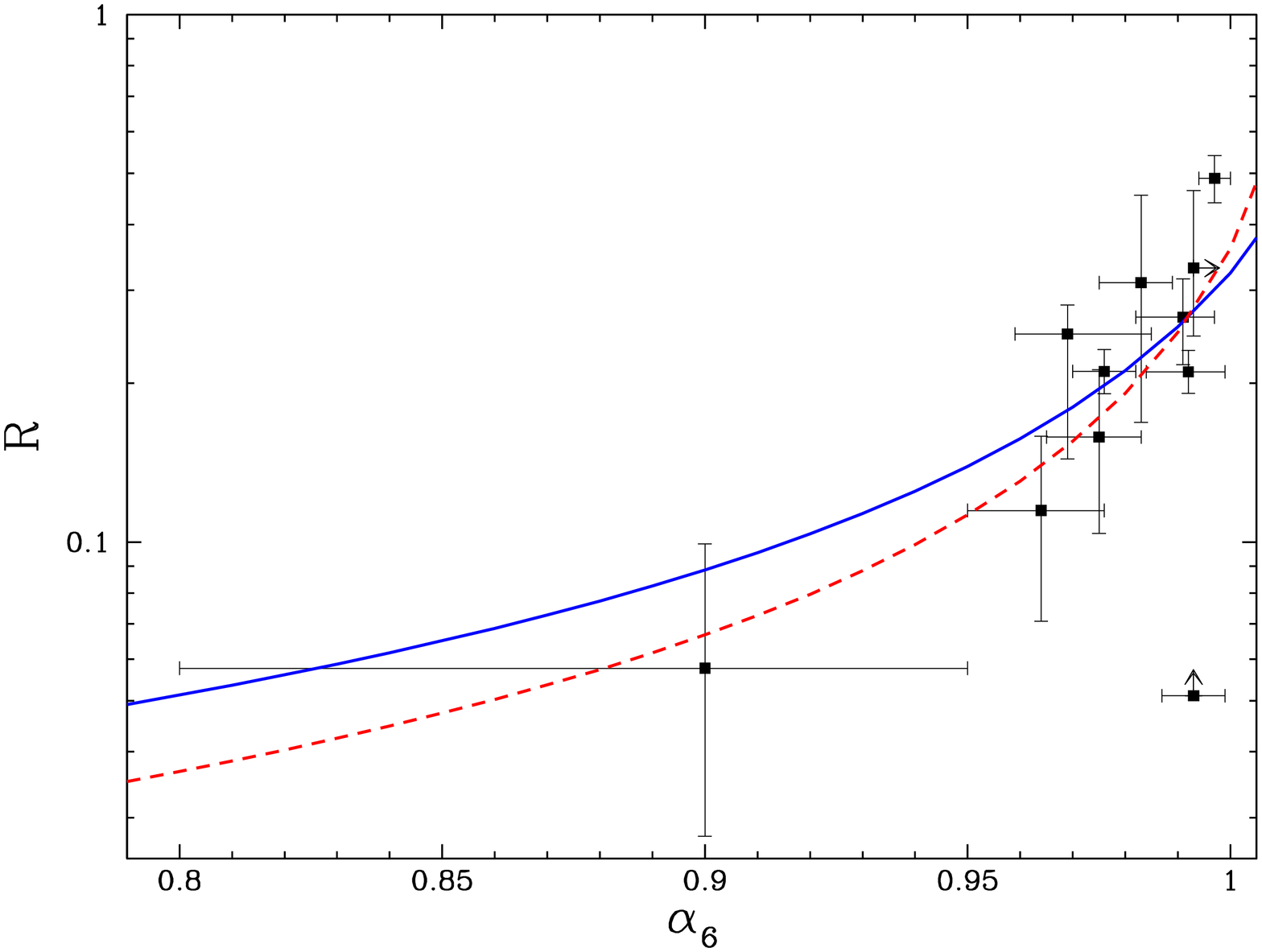}
 \caption{6~\micr-to-IR ratio (R) versus $\alfa$. The red-dashed line is our best fit 
          for the relation between R and $\alfa$ given by Eq.~\ref{eq:Ralpha6}. The 
          blue-solid line is the best fit obtained by N08 for a sample of ULIRG.}
 \label{fig:Rplot}
\end{figure*}

$R$ and $\alfa$ are both known quantities. Assuming $\alfa$ as the independent
variable and $R$ as the dependent variable, we can fit Eq.~\ref{eq:Ralpha6} to
the values observed in our sources, considering $\RAGN$ and $\RSB$ as free
parameters. We found $\RAGN=0.36\pm0.08, \RSB=\left(8\pm3\right) \times
10^{-3}$. In spite of the limited size of our sample and the narrow range for
$\alfa$, these results are in good agreement with those obtained by N08 for a
larger sample of ULIRG: $\RAGN=0.32^{+0.11}_{-0.08},
\RSB=\left(11.7^{+0.9}_{-0.7}\right) \times 10^{-3}$. Fig.~\ref{fig:Rplot} shows
$R$ versus $\alfa$ for our sample of HLIRG, including our best-fit and the
result from N08.

Using $\alfa$, $\RAGN$ and $\RSB$ (using the N08 values for the last
two quantities), we could estimate the fractional AGN contribution
($\alpha_\mathrm{IR}$) to the IR output of each
source:
\begin{equation}
\label{eq:alfaIR}
\alpha_\mathrm{IR} = \dfrac{F_\mathrm{IR}^\mathrm{AGN}}{F_\mathrm{IR}} =
\dfrac{\alfa}{\alfa + (\RAGN/\RSB)(1-\alfa)}.
\end{equation}

\begin{figure*}[ht]
 \centering
 \includegraphics[angle=-90,width=0.75\linewidth]{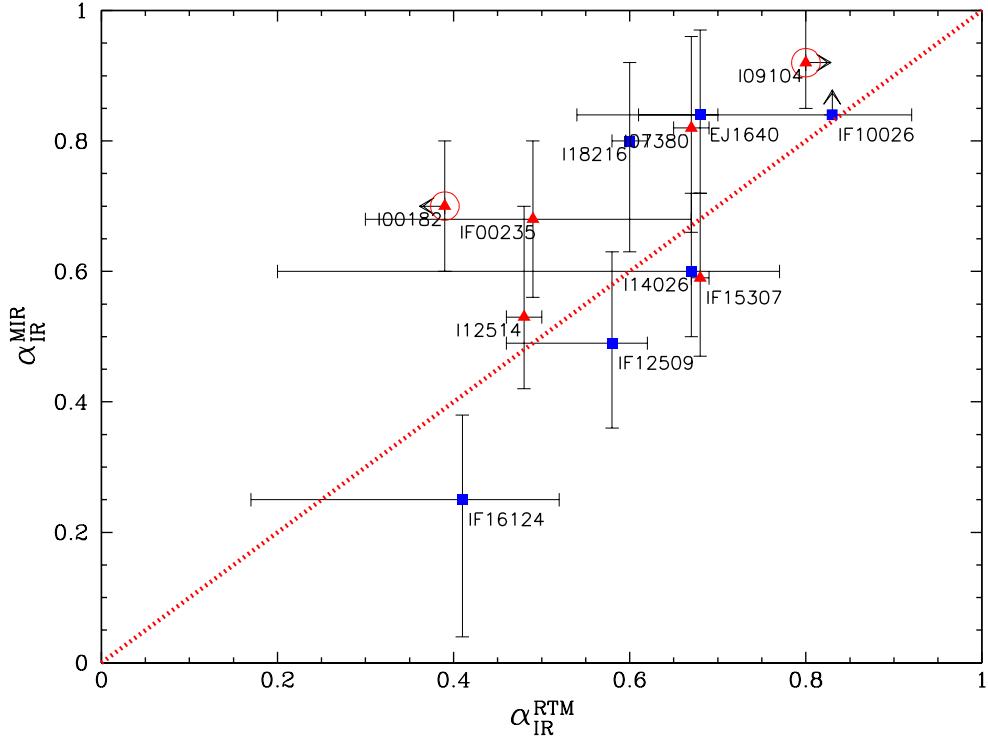}
 \caption{Comparison of the AGN contribution to the IR luminosity estimated through
          MIR spectral decomposition and through RTM. The blue squares are type I AGN and 
          the red triangles are type II AGN and SB. The open circles mark those sources 
          where our model is inaccurate. MIR contribution for IRAS~00182-7112 from 
          \citet{Spoon04}.}
 \label{fig:RTMvsMIR}
\end{figure*}

Figure \ref{fig:RTMvsMIR} compares $\alpha_{\mathrm{IR}}$ estimated through our
MIR spectral analysis and through the analysis of the IR SED using RTM
\citep[F02;]{Rowan00,Verma02}. Both estimates seem to be consistent in most
sources. Letting aside those sources where our model is not suitable (IRAS\,09104+4109
and IRAS\,00182-7112), we found that the IR luminosity of seven out of ten HLIRG
is AGN dominated. However, the SB contribution is also significant for all
sources, spanning from $\sim20\% - 70\%$ (see Table~\ref{tab:luminIR}). Using
only the sources from the F02 sample, the mean SB contribution is $\sim30\%$,
close to $35\%$ obtained by F02. This significant SB contribution is consistent
with the correlation we found between SFR and IR luminosity and the high SFR
estimated (see Sect.~\ref{sec:sfr}). Our analysis confirms the idea that star
formation and accretion into SMBH are both crucial phenomena to explain the
properties of these extreme objects.

\subsection{Covering Factor}
\label{sec:cf}
An important physical parameter to unveil the distribution of dust in the
nuclear environment of an AGN is the dust covering factor (CF), i.e. the
fraction of sky covered by dust viewed from the central engine of the AGN. The
CF is also critical to understand the fraction of direct nuclear emission that
is absorbed by dust and re-emitted in the IR range and hence has a significant
effect in the bolometric luminosity corrections of AGN.

The ratio between the thermal AGN luminosity (i.e., the IR reprocessed emission
due to heated dust), $L^\mathrm{AGN}_\mathrm{TH}$, and the primary AGN
luminosity (above $\sim1$~\micr, i.e., the accretion disk bolometric emission),
$L^\mathrm{AGN}_\mathrm{BOL}$, is commonly interpreted as an estimate of the
dust CF \citep{Maiolino07,Rowan09,Roseboom12}:
\begin{equation}
  \mathrm{CF(dust)} \approx \dfrac{L^\mathrm{AGN}_\mathrm{TH}}{L^\mathrm{AGN}_\mathrm{BOL}}.
\end{equation}

The thermal AGN luminosity can be estimated through the MIR continuum emission.
Assuming that the IR AGN luminosity is dominated by the dust emission and using
the best-fit parameters obtained in our MIR spectral decomposition, the thermal
emission is computed as:
\begin{equation}
 L^\mathrm{AGN}_\mathrm{TH} \sim \dfrac{\alpha_6~\lambda~L_\lambda(6~\mu\mathrm{m})}{\RAGN}.
\end{equation}

X-ray emission is a primary product of the central engine of AGN, so X-ray
luminosity is usually a good proxy to estimate the bolometric AGN luminosity
\citep{Maiolino07,Rowan09}. We applied the bolometric correction estimated 
by \citet{Marconi04}:
\begin{equation}
\label{eq:bolcor}
 \log \dfrac{L^\mathrm{AGN}_\mathrm{BOL}}{L_\mathrm{X}} = 1.54 + 0.24\mathcal{L} + 0.012\mathcal{L}^2 - 0.0015\mathcal{L}^3,
\end{equation}
where $\mathcal{L}=\log \left(
L^\mathrm{AGN}_\mathrm{BOL}/\Lsun \right) - 12$ and $L_\mathrm{X}$ is the
intrinsic (i.e., absorption-corrected) 2-10~keV luminosity.

Therefore we can estimate the CF of those sources for which MIR and X-ray data
are both available. Eight out of thirteen HLIRG in our sample fulfill this
condition. The X-ray luminosities were obtained from \citealt{Ruiz07}.
IRAS\,F00235+1024 and IRAS\,07380-2342 are probably Compton-Thick (CT) sources
\citep{Ruiz07,Ruiz10} so the upper limits on their X-ray luminosities have been
increased by a factor of 60, the average ratio between intrinsic and observed
X-ray luminosities in CT sources \citep{Panessa06}. IRAS\,14026+4341 was not
detected in our X-ray analysis, but it has a counterpart in the 2XMMi catalogue
\citep{2xmmi}. We estimated its 2-10~keV luminosity using the 2XMMi X-ray
fluxes. Since this object seems to be an X-ray absorbed QSO \citep{Ruiz10}, we
applied the correction for CT objects described above to calculate its intrinsic
X-ray luminosity.

\begin{figure*}[ht]
 \centering
 \includegraphics[angle=-90,width=0.75\linewidth]{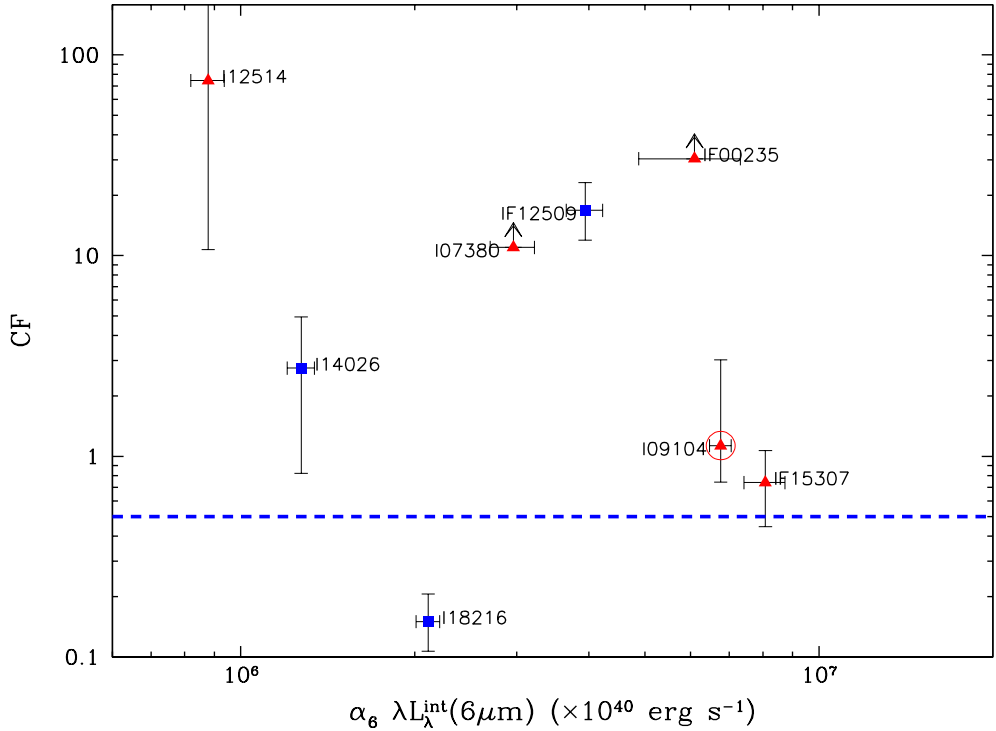}
 \caption{CF versus the AGN luminosity at 6~\micr. The blue dashed line is
          the estimated average CF for local QSO (see Sect.~\ref{sec:cf}).
          Symbols as in Fig.~\ref{fig:RTMvsMIR}.}
 \label{fig:CFvsLMIR}   
\end{figure*}

Table~\ref{tab:covfactor} shows all the derived AGN luminosities and final
estimates of the dust CF. The covering factors versus the AGN luminosities at
6~\micr are plotted in Fig.~\ref{fig:CFvsLMIR}. For comparison, the plot
includes the average CF estimated for local QSO.\footnote{The average CF was
estimated through a direct integration of the \citet{Richards06} and
\citet{Hopkins07} average AGN SED. The direct AGN emission was calculated
integrating the SED at wavelengths below 1~\micr, and the thermal emission
integrating the SED at wavelengths above 2~\micr.}

Four out of eight objects show a dust CF consistent with $\sim1$ (at 2$\sigma$
level) and sistematically above the average CF of local QSO ($\sim0.5$). This
result and the heavy X-ray absorption shown by most of these sources (all but
one source, IRAS\,14026+4341, show signatures of CT absorption in their X-ray
emission, see \citealt{Wilman03,Iwasawa05,Ruiz07,Ruiz10,Vignali11}) point
towards large amounts of gas and dust enshrouding their nuclear environment as
it has been found in local ULIRG \citep{Spoon04arp220,Verma05,Yan10}.

We found one source with $\mathrm{CF}\sim0.2$. This source, IRAS\,18216+6418, is
a QSO with no sign of X-ray or MIR obscuration. The low CF is consistent with
previous studies finding a decrease in the CF of QSO with increasing luminosity
\citep{Maiolino07,Treister08}. The decrease could be explained by ``receding torus''
models \citep{Lawrence91,Maiolino07,Hasinger08}: low-luminosity AGN
are surrounded by a dust torus of obscuring material covering a large fraction
of the central source. High-luminosity AGN would be able to clean out the
environment ionizing the surrounding medium or blowing it away through
outflowing winds. Hence the opening angle of the torus would be larger and the
covered solid angle would be lower (assuming the height of the torus is not
luminosity-dependent).

The remaining three sources show $\mathrm{CF}>>1$. A dust CF above unity is
usually interpreted as evidence of CT obscuration \citep{Rowan09}. As stated
above, two objects (IRAS\,F00235+1024 and IRAS\,07380-2342) were not detected in
X-rays \citep{Ruiz07} and their broadband SED suggest that their low X-ray
emission is due to high obscuration \citep{Ruiz10}. Hence the bolometric AGN
emission is probably underestimated in these sources. Observational and
theoretical studies of CT AGN predict even larger values (i.e., $>>60$) for the
intrinsic-to-observed X-ray luminosity ratio, depending on the amount of
absorption and on the viewing angle with respect to the obscuring torus
\citep{Haardt94,Iwasawa97}. The dust CF of these sources would be consistent
with unity if the largest correction factor ($\sim1000$) were applied.

The third remaining source, IRAS\,F12509+3122, is a QSO showing an X-ray emission
significantly lower than the one predicted for a high-luminosity QSO, given its
IR luminosity \citep{Ruiz10}. This suggests that its bolometric AGN emission is
also underestimated. However, the low X-ray luminosity of this object cannot be
related to X-ray absorption \citep{Ruiz07}. Alternatively, an increase with Eddington ratio has been
observed in the X-ray bolometric correction of AGN 
($\lambda_{Edd}=L_{BOL}/L_{Edd}$), from $\kappa_X\sim15-30$ for
$\lambda_{Edd}\lesssim0.1$ to $\kappa_X\sim70-120$ for $\lambda_{Edd}\gtrsim0.2$
\citep{Vasudevan09}. Using the SDSS optical spectrum of IRAS\,F12509+3122 and the
\citet{McLure02MgII} relation between black hole masses and MgII emission line
widths, we estimated $\lambda_{Edd}\sim0.5$ for this object. This high Eddington
ratio suggests that the X-ray bolometric correction for this source could be
higher than that obtained in Eq.~\ref{eq:bolcor}. Further studies are needed to
investigate how the Eddington ratio influences the X-ray luminosity of HLIRG.

\section{Conclusions}
\label{sec:conc}
We studied low resolution MIR spectra of thirteen HLIRG observed with \Spitzer,
nine of them also observed with \XMM. We modelled their 5-8~\micr spectra, using
the AGN/SB spectral decomposition technique developed by N08, and we estimated
the contribution of each component to the total IR luminosity.

We found that all HLIRG in the sample harbour an AGN that clearly dominates the
5-8~\micr spectrum. Given the completeness of the F02 sample (nine out of ten F02
objects are included in our sample), this is strong evidence suggesting that all
HLIRG harbour an AGN. We have also found that the AGN component seems to be the
dominant power source of the total IR output in most sources. However, all
sources, even the AGN-dominated HLIRG, show a significant SB activity, with a
mean SB contribution of $\sim30\%$. The SFR, estimated through the PAH emission,
is also very high in all sources ($\sim300-3000~\Msun\mathrm{yr}^{-1}$).

These results are in agreement with previous studies of HLIRG using theoretical
models to reproduce their IR SED \citep[F02;][]{Rowan00,Verma02,Rowan10},
providing further support to the assumptions of these detailed models. The AGN
relative contribution to the total IR output is also consistent with those
estimated through SED modeling. The large mean AGN contribution we found
($\sim70\%$) is consistent with previous studies of ULIRG, pointing towards an
increase of the AGN emission with increasing luminosity
\citep{Veilleux02,Nardini10}. Our work confirms the crucial role of both AGN and
SB phenomena to explain the properties of these extreme sources.

Using X-ray and MIR data we were able to estimate the covering factor (CF) of
these HLIRG, finding that a significant fraction (seven out of eight) 
have CF$\gtrsim1$. Most of these sources with large CF also show heavy X-ray absorption
and high optical depth or absorption features in their MIR spectrum. This
strongly suggests that the nuclear environment of these sources is heavily
enshrouded by large amounts of gas and dust, as it has been observed in local
ULIRG.

F02 proposed that HLIRG could be divided in two populations: (1) mergers between
gas-rich galaxies, as found in the ULIRG population, and (2) young active
galaxies going through their maximal star formation periods whilst harbouring an
AGN. In \citet{Ruiz10} we found strong evidence supporting the two-population
hypothesis. \citet{Han12} studied the same HLIRG sample using AGN and SB
theoretical models and bayesian techniques to fit their IR SED. They found
further evidence that the physical properties of AGN and SB harboured by HLIRG
could be divided into two populations: SB-dominated HLIRG (corresponding to our
``ULIRG-like'' objects) show a higher fraction of OB stars and the starburst
region is more compact, while AGN-dominated objects show a dustier thorus.
Additional evidence favouring this two-populations idea is presented by
\citet{Draper12}. They suggest, based on recent Herschel observations, that
major mergers cannot explain the total population of luminous high-$z$ ULIRG 
harbouring AGN.

All sources showing $\mathrm{CF}\sim1$ (except IRAS\,14026+4341) were classified
as ``ULIRG-like'' objects in \citet{Ruiz10} while IRAS 18216+6418, with a
significantly lower CF, was classified as a ``non-ULIRG'' source. Furthermore,
IRAS\,F00235+1024 and IRAS\,F15307+3252, both sources showing a high CF, were
observed by the Hubble Space Telescope (HST) and both present signs of recent
mergers. On the other hand, IRAS 18216+6418, also observed by HST, shows no
signs of interactions or mergers \citep{Farrah02hst}. The study of the dust CF
adds another piece of evidence favouring that HLIRG can be separated into two
differentiated populations, although further studies with larger samples of
HLIRG are needed to obtain stronger conclusions.

\begin{acknowledgements}
We are grateful to the anonymous referee for the constructive comments and
suggestions that improved this paper. A.R. acknowledges support from an IUCAA
postdoctoral fellowship and from ASI grant n. ASI I/088/06/0. Financial support
for A.R. and F.J.C. was provided by the Spanish Ministry of Education and
Science, under project ESP2003-00812 and ESP2006-13608-C02-01. F.J.C
acknowledges financial support under the project AYA2010-21490-C02-01. F.P.
acknowledges financial support under the project ASI INAF I/08/07/0.

This work is based on observations made with the Spitzer Space Telescope, which
is operated by the Jet Propulsion Laboratory, California Institute of Technology
under a contract with NASA, and with \XMM, an ESA science mission with instruments
and contributions directly funded by ESA Member States and NASA. The IRS was a
collaborative venture between Cornell University and Ball Aerospace Corporation
funded by NASA through the Jet Propulsion Laboratory and Ames Research Center.
SMART was developed by the IRS Team at Cornell University.
\end{acknowledgements}

\bibliographystyle{aa}

\begin{thebibliography}{109}
\expandafter\ifx\csname natexlab\endcsname\relax\def\natexlab#1{#1}\fi

\bibitem[{{Akritas} \& {Siebert}(1996)}]{Akritas96}
{Akritas}, M.~G. \& {Siebert}, J. 1996, \mnras, 278, 919

\bibitem[{{Armus} {et~al.}(2007){Armus}, {Charmandaris}, {Bernard-Salas},
  {Spoon}, {Marshall}, {Higdon}, {Desai}, {Teplitz}, {Hao}, {Devost}, {Brandl},
  {Wu}, {Sloan}, {Soifer}, {Houck}, \& {Herter}}]{Armus07}
{Armus}, L., {Charmandaris}, V., {Bernard-Salas}, J., {et~al.} 2007, \apj, 656,
  148

\bibitem[{{Brandl} {et~al.}(2006){Brandl}, {Bernard-Salas}, {Spoon}, {Devost},
  {Sloan}, {Guilles}, {Wu}, {Houck}, {Weedman}, {Armus}, {Appleton}, {Soifer},
  {Charmandaris}, {Hao}, {Higdon}, \& {Herter}}]{Brandl06}
{Brandl}, B.~R., {Bernard-Salas}, J., {Spoon}, H.~W.~W., {et~al.} 2006, \apj,
  653, 1129

\bibitem[{{Caputi} {et~al.}(2007){Caputi}, {Lagache}, {Yan}, {Dole},
  {Bavouzet}, {Le Floc'h}, {Choi}, {Helou}, \& {Reddy}}]{Caputi07}
{Caputi}, K.~I., {Lagache}, G., {Yan}, L., {et~al.} 2007, \apj, 660, 97

\bibitem[{{Carico} {et~al.}(1990){Carico}, {Sanders}, {Soifer}, {Matthews}, \&
  {Neugebauer}}]{Carico90}
{Carico}, D.~P., {Sanders}, D.~B., {Soifer}, B.~T., {Matthews}, K., \&
  {Neugebauer}, G. 1990, \aj, 100, 70

\bibitem[{{Condon} {et~al.}(1991){Condon}, {Huang}, {Yin}, \&
  {Thuan}}]{Condon91}
{Condon}, J.~J., {Huang}, Z., {Yin}, Q.~F., \& {Thuan}, T.~X. 1991, \apj, 378,
  65

\bibitem[{{Daddi} {et~al.}(2005){Daddi}, {Dickinson}, {Chary}, {Pope},
  {Morrison}, {Alexander}, {Bauer}, {Brandt}, {Giavalisco}, {Ferguson}, {Lee},
  {Lehmer}, {Papovich}, \& {Renzini}}]{Daddi05}
{Daddi}, E., {Dickinson}, M., {Chary}, R., {et~al.} 2005, \apjl, 631, L13

\bibitem[{{Dasyra} {et~al.}(2006){Dasyra}, {Tacconi}, {Davies}, {Genzel},
  {Lutz}, {Naab}, {Burkert}, {Veilleux}, \& {Sanders}}]{Dasyra06}
{Dasyra}, K.~M., {Tacconi}, L.~J., {Davies}, R.~I., {et~al.} 2006, \apj, 638,
  745

\bibitem[{{de Grijp} {et~al.}(1985){de Grijp}, {Miley}, {Lub}, \& {de
  Jong}}]{deGrijp85}
{de Grijp}, M.~H.~K., {Miley}, G.~K., {Lub}, J., \& {de Jong}, T. 1985, \nat,
  314, 240

\bibitem[{{Draine}(1989)}]{Draine89}
{Draine}, B.~T. 1989, in ESA Special Publication, Vol. 290, Infrared
  Spectroscopy in Astronomy, ed. E.~{B{\"o}hm-Vitense}, 93--98

\bibitem[{{Draine} \& {Li}(2001)}]{Draine01}
{Draine}, B.~T. \& {Li}, A. 2001, \apj, 551, 807

\bibitem[{{Draine} \& {Li}(2007)}]{Draine07}
{Draine}, B.~T. \& {Li}, A. 2007, \apj, 657, 810

\bibitem[{{Draper} \& {Ballantyne}(2012)}]{Draper12}
{Draper}, A.~R. \& {Ballantyne}, D.~R. 2012, ArXiv e-prints

\bibitem[{{Elbaz} {et~al.}(2002){Elbaz}, {Cesarsky}, {Chanial}, {Aussel},
  {Franceschini}, {Fadda}, \& {Chary}}]{Elbaz02}
{Elbaz}, D., {Cesarsky}, C.~J., {Chanial}, P., {et~al.} 2002, \aap, 384, 848

\bibitem[{{Evans} {et~al.}(1998){Evans}, {Sanders}, {Cutri}, {Radford},
  {Surace}, {Solomon}, {Downes}, \& {Kramer}}]{Evans98}
{Evans}, A.~S., {Sanders}, D.~B., {Cutri}, R.~M., {et~al.} 1998, \apj, 506, 205

\bibitem[{{Farrah} {et~al.}(2007){Farrah}, {Bernard-Salas}, {Spoon}, {Soifer},
  {Armus}, {Brandl}, {Charmandaris}, {Desai}, {Higdon}, {Devost}, \&
  {Houck}}]{Farrah07}
{Farrah}, D., {Bernard-Salas}, J., {Spoon}, H.~W.~W., {et~al.} 2007, \apj, 667,
  149

\bibitem[{{Farrah} {et~al.}(2006){Farrah}, {Lonsdale}, {Borys}, {Fang},
  {Waddington}, {Oliver}, {Rowan-Robinson}, {Babbedge}, {Shupe}, {Polletta},
  {Smith}, \& {Surace}}]{Farrah06}
{Farrah}, D., {Lonsdale}, C.~J., {Borys}, C., {et~al.} 2006, \apjl, 641, L17

\bibitem[{{Farrah} {et~al.}(2002{\natexlab{a}}){Farrah}, {Serjeant},
  {Efstathiou}, {Rowan-Robinson}, \& {Verma}}]{Farrah02submm}
{Farrah}, D., {Serjeant}, S., {Efstathiou}, A., {Rowan-Robinson}, M., \&
  {Verma}, A. 2002{\natexlab{a}}, \mnras, 335, 1163

\bibitem[{{Farrah} {et~al.}(2002{\natexlab{b}}){Farrah}, {Verma}, {Oliver},
  {Rowan-Robinson}, \& {McMahon}}]{Farrah02hst}
{Farrah}, D., {Verma}, A., {Oliver}, S., {Rowan-Robinson}, M., \& {McMahon}, R.
  2002{\natexlab{b}}, \mnras, 329, 605

\bibitem[{{Franceschini} {et~al.}(2001){Franceschini}, {Aussel}, {Cesarsky},
  {Elbaz}, \& {Fadda}}]{Fran01}
{Franceschini}, A., {Aussel}, H., {Cesarsky}, C.~J., {Elbaz}, D., \& {Fadda},
  D. 2001, \aap, 378, 1

\bibitem[{{Franceschini} {et~al.}(1994){Franceschini}, {Mazzei}, {de Zotti}, \&
  {Danese}}]{Fran94}
{Franceschini}, A., {Mazzei}, P., {de Zotti}, G., \& {Danese}, L. 1994, \apj,
  427, 140

\bibitem[{{Frayer} {et~al.}(1999){Frayer}, {Ivison}, {Scoville}, {Evans},
  {Yun}, {Smail}, {Barger}, {Blain}, \& {Kneib}}]{Frayer99}
{Frayer}, D.~T., {Ivison}, R.~J., {Scoville}, N.~Z., {et~al.} 1999, \apjl, 514,
  L13

\bibitem[{{Frayer} {et~al.}(1998){Frayer}, {Ivison}, {Scoville}, {Yun},
  {Evans}, {Smail}, {Blain}, \& {Kneib}}]{Frayer98}
{Frayer}, D.~T., {Ivison}, R.~J., {Scoville}, N.~Z., {et~al.} 1998, \apjl, 506,
  L7

\bibitem[{{Freeman} {et~al.}(2001){Freeman}, {Doe}, \&
  {Siemiginowska}}]{sherpa}
{Freeman}, P., {Doe}, S., \& {Siemiginowska}, A. 2001, in \procspie, ed.
  {J.~L.~Starck \& F.~D.~Murtagh}, Vol. 4477, 76--87

\bibitem[{{Fruscione} {et~al.}(2006){Fruscione}, {McDowell}, {Allen},
  {Brickhouse}, {Burke}, {Davis}, {Durham}, {Elvis}, {Galle}, {Harris},
  {Huenemoerder}, {Houck}, {Ishibashi}, {Karovska}, {Nicastro}, {Noble},
  {Nowak}, {Primini}, {Siemiginowska}, {Smith}, \& {Wise}}]{ciao}
{Fruscione}, A., {McDowell}, J.~C., {Allen}, G.~E., {et~al.} 2006, in
  \procspie, Vol. 6270

\bibitem[{{Furton} {et~al.}(1999){Furton}, {Laiho}, \& {Witt}}]{Furton99}
{Furton}, D.~G., {Laiho}, J.~W., \& {Witt}, A.~N. 1999, \apj, 526, 752

\bibitem[{{Genzel} \& {Cesarsky}(2000)}]{Genzel00}
{Genzel}, R. \& {Cesarsky}, C.~J. 2000, \araa, 38, 761

\bibitem[{{Genzel} {et~al.}(1998){Genzel}, {Lutz}, {Sturm}, {Egami}, {Kunze},
  {Moorwood}, {Rigopoulou}, {Spoon}, {Sternberg}, {Tacconi-Garman}, {Tacconi},
  \& {Thatte}}]{Genzel98}
{Genzel}, R., {Lutz}, D., {Sturm}, E., {et~al.} 1998, \apj, 498, 579

\bibitem[{{Genzel} {et~al.}(2001){Genzel}, {Tacconi}, {Rigopoulou}, {Lutz}, \&
  {Tecza}}]{Genzel01}
{Genzel}, R., {Tacconi}, L.~J., {Rigopoulou}, D., {Lutz}, D., \& {Tecza}, M.
  2001, \apj, 563, 527

\bibitem[{{Granato} {et~al.}(1996){Granato}, {Danese}, \&
  {Franceschini}}]{Granato96}
{Granato}, G.~L., {Danese}, L., \& {Franceschini}, A. 1996, \apjl, 460, L11

\bibitem[{{Haardt} {et~al.}(1994){Haardt}, {Maraschi}, \&
  {Ghisellini}}]{Haardt94}
{Haardt}, F., {Maraschi}, L., \& {Ghisellini}, G. 1994, \apjl, 432, L95

\bibitem[{{Han} \& {Han}(2012)}]{Han12}
{Han}, Y. \& {Han}, Z. 2012, \apj, 749, 123

\bibitem[{{Hasinger}(2008)}]{Hasinger08}
{Hasinger}, G. 2008, \aap, 490, 905

\bibitem[{{Hern{\'a}n-Caballero} \& {Hatziminaoglou}(2011)}]{HernanCaballero11}
{Hern{\'a}n-Caballero}, A. \& {Hatziminaoglou}, E. 2011, \mnras, 414, 500

\bibitem[{{Higdon} {et~al.}(2004){Higdon}, {Devost}, {Higdon}, {Brandl},
  {Houck}, {Hall}, {Barry}, {Charmandaris}, {Smith}, {Sloan}, \&
  {Green}}]{smart2}
{Higdon}, S.~J.~U., {Devost}, D., {Higdon}, J.~L., {et~al.} 2004, \pasp, 116,
  975

\bibitem[{{Holland} {et~al.}(1999){Holland}, {Robson}, {Gear}, {Cunningham},
  {Lightfoot}, {Jenness}, {Ivison}, {Stevens}, {Ade}, {Griffin}, {Duncan},
  {Murphy}, \& {Naylor}}]{Scuba}
{Holland}, W.~S., {Robson}, E.~I., {Gear}, W.~K., {et~al.} 1999, \mnras, 303,
  659

\bibitem[{{Hopkins} {et~al.}(2007){Hopkins}, {Richards}, \&
  {Hernquist}}]{Hopkins07}
{Hopkins}, P.~F., {Richards}, G.~T., \& {Hernquist}, L. 2007, \apj, 654, 731

\bibitem[{{Houck} {et~al.}(2004){Houck}, {Roellig}, {van Cleve}, {Forrest},
  {Herter}, {Lawrence}, {Matthews}, {Reitsema}, {Soifer}, {Watson}, {Weedman},
  {Huisjen}, {Troeltzsch}, {Barry}, {Bernard-Salas}, {Blacken}, {Brandl},
  {Charmandaris}, {Devost}, {Gull}, {Hall}, {Henderson}, {Higdon}, {Pirger},
  {Schoenwald}, {Sloan}, {Uchida}, {Appleton}, {Armus}, {Burgdorf},
  {Fajardo-Acosta}, {Grillmair}, {Ingalls}, {Morris}, \&
  {Teplitz}}]{Spitzer-IRS}
{Houck}, J.~R., {Roellig}, T.~L., {van Cleve}, J., {et~al.} 2004, \apjs, 154,
  18

\bibitem[{{Hurley} {et~al.}(2012){Hurley}, {Oliver}, {Farrah}, {Wang}, \&
  {Efstathiou}}]{Hurley12}
{Hurley}, P.~D., {Oliver}, S., {Farrah}, D., {Wang}, L., \& {Efstathiou}, A.
  2012, ArXiv e-prints

\bibitem[{{Isobe} {et~al.}(1986){Isobe}, {Feigelson}, \& {Nelson}}]{asurv2}
{Isobe}, T., {Feigelson}, E.~D., \& {Nelson}, P.~I. 1986, \apj, 306, 490

\bibitem[{{Iwasawa} {et~al.}(2005){Iwasawa}, {Crawford}, {Fabian}, \&
  {Wilman}}]{Iwasawa05}
{Iwasawa}, K., {Crawford}, C.~S., {Fabian}, A.~C., \& {Wilman}, R.~J. 2005,
  \mnras, 362, L20

\bibitem[{{Iwasawa} {et~al.}(1997){Iwasawa}, {Fabian}, \& {Matt}}]{Iwasawa97}
{Iwasawa}, K., {Fabian}, A.~C., \& {Matt}, G. 1997, \mnras, 289, 443

\bibitem[{Kendall \& Gibbons(1990)}]{Kendall90}
Kendall, M. \& Gibbons, J. 1990, Rank correlation methods, A Charles Griffin
  Book (E. Arnold)

\bibitem[{{Kennicutt}(1998)}]{Kenn98}
{Kennicutt}, R.~C. 1998, \apj, 498, 541

\bibitem[{{Komatsu} {et~al.}(2009){Komatsu}, {Dunkley}, {Nolta}, {Bennett},
  {Gold}, {Hinshaw}, {Jarosik}, {Larson}, {Limon}, {Page}, {Spergel},
  {Halpern}, {Hill}, {Kogut}, {Meyer}, {Tucker}, {Weiland}, {Wollack}, \&
  {Wright}}]{Komatsu09}
{Komatsu}, E., {Dunkley}, J., {Nolta}, M.~R., {et~al.} 2009, \apjs, 180, 330

\bibitem[{{Laurent} {et~al.}(2000){Laurent}, {Mirabel}, {Charmandaris},
  {Gallais}, {Madden}, {Sauvage}, {Vigroux}, \& {Cesarsky}}]{Laurent00}
{Laurent}, O., {Mirabel}, I.~F., {Charmandaris}, V., {et~al.} 2000, \aap, 359,
  887

\bibitem[{{Lawrence}(1991)}]{Lawrence91}
{Lawrence}, A. 1991, \mnras, 252, 586

\bibitem[{{Lebouteiller} {et~al.}(2011){Lebouteiller}, {Barry}, {Spoon},
  {Bernard-Salas}, {Sloan}, {Houck}, \& {Weedman}}]{CASIS}
{Lebouteiller}, V., {Barry}, D.~J., {Spoon}, H.~W.~W., {et~al.} 2011, \apjs,
  196, 8

\bibitem[{{Lebouteiller} {et~al.}(2010){Lebouteiller}, {Bernard-Salas},
  {Sloan}, \& {Barry}}]{smart1}
{Lebouteiller}, V., {Bernard-Salas}, J., {Sloan}, G.~C., \& {Barry}, D.~J.
  2010, \pasp, 122, 231

\bibitem[{{Lilly} {et~al.}(1999){Lilly}, {Eales}, {Gear}, {Hammer}, {Le
  F{\`e}vre}, {Crampton}, {Bond}, \& {Dunne}}]{Lilly99}
{Lilly}, S.~J., {Eales}, S.~A., {Gear}, W.~K.~P., {et~al.} 1999, \apj, 518, 641

\bibitem[{{Lonsdale} {et~al.}(2004){Lonsdale}, {Polletta}, {Surace}, {Shupe},
  {Fang}, {Xu}, {Smith}, {Siana}, {Rowan-Robinson}, {Babbedge}, {Oliver},
  {Pozzi}, {Davoodi}, {Owen}, {Padgett}, {Frayer}, {Jarrett}, {Masci},
  {O'Linger}, {Conrow}, {Farrah}, {Morrison}, {Gautier}, {Franceschini},
  {Berta}, {Perez-Fournon}, {Hatziminaoglou}, {Afonso-Luis}, {Dole}, {Stacey},
  {Serjeant}, {Pierre}, {Griffin}, \& {Puetter}}]{Lonsdale04}
{Lonsdale}, C., {Polletta}, M.~d.~C., {Surace}, J., {et~al.} 2004, \apjs, 154,
  54

\bibitem[{{Lonsdale} {et~al.}(2006){Lonsdale}, {Farrah}, \&
  {Smith}}]{Lonsdale06}
{Lonsdale}, C.~J., {Farrah}, D., \& {Smith}, H.~E. 2006, {Ultraluminous
  Infrared Galaxies} (Springer Verlag), 285

\bibitem[{{Lutz} {et~al.}(2007){Lutz}, {Sturm}, {Tacconi}, {Valiante},
  {Schweitzer}, {Netzer}, {Maiolino}, {Andreani}, {Shemmer}, \&
  {Veilleux}}]{Lutz07}
{Lutz}, D., {Sturm}, E., {Tacconi}, L.~J., {et~al.} 2007, \apjl, 661, L25

\bibitem[{{Lutz} {et~al.}(2008){Lutz}, {Sturm}, {Tacconi}, {Valiante},
  {Schweitzer}, {Netzer}, {Maiolino}, {Andreani}, {Shemmer}, \&
  {Veilleux}}]{Lutz08}
{Lutz}, D., {Sturm}, E., {Tacconi}, L.~J., {et~al.} 2008, \apj, 684, 853

\bibitem[{{Magdis} {et~al.}(2010){Magdis}, {Elbaz}, {Hwang}, {Amblard},
  {Arumugam}, {Aussel}, {Blain}, {Bock}, {Boselli}, {Buat},
  {Castro-Rodr{\'{\i}}guez}, {Cava}, {Chanial}, {Clements}, {Conley},
  {Conversi}, {Cooray}, {Dowell}, {Dwek}, {Eales}, {Farrah}, {Franceschini},
  {Glenn}, {Griffin}, {Halpern}, {Hatziminaoglou}, {Huang}, {Ibar}, {Isaak},
  {Le Floc'h}, {Lagache}, {Levenson}, {Lonsdale}, {Lu}, {Madden}, {Maffei},
  {Mainetti}, {Marchetti}, {Morrison}, {Nguyen}, {O'Halloran}, {Oliver},
  {Omont}, {Owen}, {Page}, {Pannella}, {Panuzzo}, {Papageorgiou}, {Pearson},
  {P{\'e}rez-Fournon}, {Pohlen}, {Rigopoulou}, {Rizzo}, {Roseboom},
  {Rowan-Robinson}, {Schulz}, {Scott}, {Seymour}, {Shupe}, {Smith}, {Stevens},
  {Strazzullo}, {Symeonidis}, {Trichas}, {Tugwell}, {Vaccari}, {Valtchanov},
  {Vigroux}, {Wang}, {Wright}, {Xu}, \& {Zemcov}}]{Magdis10}
{Magdis}, G.~E., {Elbaz}, D., {Hwang}, H.~S., {et~al.} 2010, \mnras, 409, 22

\bibitem[{{Magdis} {et~al.}(2011){Magdis}, {Elbaz}, {Hwang}, {Pep Team}, \&
  {Hermes Team}}]{Magdis11}
{Magdis}, G.~E., {Elbaz}, D., {Hwang}, H.~S., {Pep Team}, \& {Hermes Team}.
  2011, in Astronomical Society of the Pacific Conference Series, Vol. 446,
  Galaxy Evolution: Infrared to Millimeter Wavelength Perspective, ed.
  W.~{Wang}, J.~{Lu}, Z.~{Luo}, Z.~{Yang}, H.~{Hua}, \& Z.~{Chen}, 221

\bibitem[{{Maiolino} {et~al.}(2007){Maiolino}, {Shemmer}, {Imanishi}, {Netzer},
  {Oliva}, {Lutz}, \& {Sturm}}]{Maiolino07}
{Maiolino}, R., {Shemmer}, O., {Imanishi}, M., {et~al.} 2007, \aap, 468, 979

\bibitem[{{Marconi} {et~al.}(2004){Marconi}, {Risaliti}, {Gilli}, {Hunt},
  {Maiolino}, \& {Salvati}}]{Marconi04}
{Marconi}, A., {Risaliti}, G., {Gilli}, R., {et~al.} 2004, \mnras, 351, 169

\bibitem[{{McLure} \& {Jarvis}(2002)}]{McLure02MgII}
{McLure}, R.~J. \& {Jarvis}, M.~J. 2002, \mnras, 337, 109

\bibitem[{{Mihos} \& {Hernquist}(1994)}]{Mihos94}
{Mihos}, J.~C. \& {Hernquist}, L. 1994, \apj, 437, 611

\bibitem[{{Mullaney} {et~al.}(2011){Mullaney}, {Alexander}, {Goulding}, \&
  {Hickox}}]{Mullaney11}
{Mullaney}, J.~R., {Alexander}, D.~M., {Goulding}, A.~D., \& {Hickox}, R.~C.
  2011, \mnras, 414, 1082

\bibitem[{{Mushotzky}(2004)}]{Mushotzky04b}
{Mushotzky}, R. 2004, in Astrophysics and Space Science Library, Vol. 308,
  Supermassive Black Holes in the Distant Universe, ed. A.~J. {Barger}, 53

\bibitem[{{Nardini} {et~al.}(2008){Nardini}, {Risaliti}, {Salvati}, {Sani},
  {Imanishi}, {Marconi}, \& {Maiolino}}]{Nardini08}
{Nardini}, E., {Risaliti}, G., {Salvati}, M., {et~al.} 2008, \mnras, 385, L130

\bibitem[{{Nardini} {et~al.}(2009){Nardini}, {Risaliti}, {Salvati}, {Sani},
  {Watabe}, {Marconi}, \& {Maiolino}}]{Nardini09}
{Nardini}, E., {Risaliti}, G., {Salvati}, M., {et~al.} 2009, \mnras, 399, 1373

\bibitem[{{Nardini} {et~al.}(2010){Nardini}, {Risaliti}, {Watabe}, {Salvati},
  \& {Sani}}]{Nardini10}
{Nardini}, E., {Risaliti}, G., {Watabe}, Y., {Salvati}, M., \& {Sani}, E. 2010,
  \mnras, 584

\bibitem[{{Netzer} {et~al.}(2007){Netzer}, {Lutz}, {Schweitzer}, {Contursi},
  {Sturm}, {Tacconi}, {Veilleux}, {Kim}, {Rupke}, {Baker}, {Dasyra},
  {Mazzarella}, \& {Lord}}]{Netzer07}
{Netzer}, H., {Lutz}, D., {Schweitzer}, M., {et~al.} 2007, \apj, 666, 806

\bibitem[{{Niemi} {et~al.}(2012){Niemi}, {Somerville}, {Ferguson}, {Huang},
  {Lotz}, \& {Koekemoer}}]{Niemi12}
{Niemi}, S.-M., {Somerville}, R.~S., {Ferguson}, H.~C., {et~al.} 2012, \mnras,
  421, 1539

\bibitem[{{Panessa} {et~al.}(2006){Panessa}, {Bassani}, {Cappi}, {Dadina},
  {Barcons}, {Carrera}, {Ho}, \& {Iwasawa}}]{Panessa06}
{Panessa}, F., {Bassani}, L., {Cappi}, M., {et~al.} 2006, \aap, 455, 173

\bibitem[{{Richards} {et~al.}(2006){Richards}, {Lacy}, {Storrie-Lombardi},
  {Hall}, {Gallagher}, {Hines}, {Fan}, {Papovich}, {Vanden Berk}, {Trammell},
  {Schneider}, {Vestergaard}, {York}, {Jester}, {Anderson}, {Budav{\'a}ri}, \&
  {Szalay}}]{Richards06}
{Richards}, G.~T., {Lacy}, M., {Storrie-Lombardi}, L.~J., {et~al.} 2006, \apjs,
  166, 470

\bibitem[{{Rieke} {et~al.}(1980){Rieke}, {Lebofsky}, {Thompson}, {Low}, \&
  {Tokunaga}}]{Rieke80}
{Rieke}, G.~H., {Lebofsky}, M.~J., {Thompson}, R.~I., {Low}, F.~J., \&
  {Tokunaga}, A.~T. 1980, \apj, 238, 24

\bibitem[{{Rigopoulou} {et~al.}(2000){Rigopoulou}, {Franceschini}, {Aussel},
  {Genzel}, {van der Werf}, {Cesarsky}, {Dennefeld}, {Oliver},
  {Rowan-Robinson}, {Mann}, {Perez-Fournon}, \& {Rocca-Volmerange}}]{Rigop00}
{Rigopoulou}, D., {Franceschini}, A., {Aussel}, H., {et~al.} 2000, \apjl, 537,
  L85

\bibitem[{{Rigopoulou} {et~al.}(1999){Rigopoulou}, {Spoon}, {Genzel}, {Lutz},
  {Moorwood}, \& {Tran}}]{Rigop99}
{Rigopoulou}, D., {Spoon}, H.~W.~W., {Genzel}, R., {et~al.} 1999, \aj, 118,
  2625

\bibitem[{{Risaliti} \& {Elvis}(2004)}]{Risaliti04}
{Risaliti}, G. \& {Elvis}, M. 2004, {A Panchromatic View of AGN} (ASSL
  Vol.~308: Supermassive Black Holes in the Distant Universe), 187

\bibitem[{{Risaliti} {et~al.}(2010){Risaliti}, {Imanishi}, \&
  {Sani}}]{Risaliti10}
{Risaliti}, G., {Imanishi}, M., \& {Sani}, E. 2010, \mnras, 401, 197

\bibitem[{{Risaliti} {et~al.}(2006){Risaliti}, {Maiolino}, {Marconi}, {Sani},
  {Berta}, {Braito}, {Ceca}, {Franceschini}, \& {Salvati}}]{Risaliti06}
{Risaliti}, G., {Maiolino}, R., {Marconi}, A., {et~al.} 2006, \mnras, 365, 303

\bibitem[{{Rodighiero} {et~al.}(2011){Rodighiero}, {Daddi}, {Baronchelli},
  {Cimatti}, {Renzini}, {Aussel}, {Popesso}, {Lutz}, {Andreani}, {Berta},
  {Cava}, {Elbaz}, {Feltre}, {Fontana}, {F{\"o}rster Schreiber},
  {Franceschini}, {Genzel}, {Grazian}, {Gruppioni}, {Ilbert}, {Le Floch},
  {Magdis}, {Magliocchetti}, {Magnelli}, {Maiolino}, {McCracken}, {Nordon},
  {Poglitsch}, {Santini}, {Pozzi}, {Riguccini}, {Tacconi}, {Wuyts}, \&
  {Zamorani}}]{Rodighiero11}
{Rodighiero}, G., {Daddi}, E., {Baronchelli}, I., {et~al.} 2011, \apjl, 739,
  L40

\bibitem[{{Roseboom} {et~al.}(2012){Roseboom}, {Lawrence}, {Elvis}, {Petty},
  {Shen}, \& {Hao}}]{Roseboom12}
{Roseboom}, I.~G., {Lawrence}, A., {Elvis}, M., {et~al.} 2012, ArXiv e-prints

\bibitem[{{Rowan-Robinson}(2000)}]{Rowan00}
{Rowan-Robinson}, M. 2000, \mnras, 316, 885

\bibitem[{{Rowan-Robinson} {et~al.}(2009){Rowan-Robinson}, {Valtchanov}, \&
  {Nandra}}]{Rowan09}
{Rowan-Robinson}, M., {Valtchanov}, I., \& {Nandra}, K. 2009, \mnras, 397, 1326

\bibitem[{{Rowan-Robinson} \& {Wang}(2010)}]{Rowan10}
{Rowan-Robinson}, M. \& {Wang}, L. 2010, \mnras, 406, 720

\bibitem[{{Ruiz} {et~al.}(2007){Ruiz}, {Carrera}, \& {Panessa}}]{Ruiz07}
{Ruiz}, A., {Carrera}, F.~J., \& {Panessa}, F. 2007, \aap, 471, 775

\bibitem[{{Ruiz} {et~al.}(2010){Ruiz}, {Miniutti}, {Panessa}, \&
  {Carrera}}]{Ruiz10}
{Ruiz}, A., {Miniutti}, G., {Panessa}, F., \& {Carrera}, F.~J. 2010, \aap, 515,
  A99

\bibitem[{{Sanders} \& {Mirabel}(1996)}]{Sanders96}
{Sanders}, D.~B. \& {Mirabel}, I.~F. 1996, \araa, 34, 749

\bibitem[{{Sargsyan} \& {Weedman}(2009)}]{Sargsyan}
{Sargsyan}, L.~A. \& {Weedman}, D.~W. 2009, \apj, 701, 1398

\bibitem[{{Schweitzer} {et~al.}(2006){Schweitzer}, {Lutz}, {Sturm}, {Contursi},
  {Tacconi}, {Lehnert}, {Dasyra}, {Genzel}, {Veilleux}, {Rupke}, {Kim},
  {Baker}, {Netzer}, {Sternberg}, {Mazzarella}, \& {Lord}}]{Schweitzer06}
{Schweitzer}, M., {Lutz}, D., {Sturm}, E., {et~al.} 2006, \apj, 649, 79

\bibitem[{{Soifer} {et~al.}(1984{\natexlab{a}}){Soifer}, {Neugebauer}, {Helou},
  {Lonsdale}, {Hacking}, {Rice}, {Houck}, {Low}, \&
  {Rowan-Robinson}}]{Soifer84b}
{Soifer}, B.~T., {Neugebauer}, G., {Helou}, G., {et~al.} 1984{\natexlab{a}},
  \apjl, 283, L1

\bibitem[{{Soifer} {et~al.}(1987{\natexlab{a}}){Soifer}, {Neugebauer}, \&
  {Houck}}]{Soifer87b}
{Soifer}, B.~T., {Neugebauer}, G., \& {Houck}, J.~R. 1987{\natexlab{a}}, \araa,
  25, 187

\bibitem[{{Soifer} {et~al.}(1984{\natexlab{b}}){Soifer}, {Rowan-Robinson},
  {Houck}, {de Jong}, {Neugebauer}, {Aumann}, {Beichman}, {Boggess}, {Clegg},
  {Emerson}, {Gillett}, {Habing}, {Hauser}, {Low}, {Miley}, \&
  {Young}}]{Soifer84}
{Soifer}, B.~T., {Rowan-Robinson}, M., {Houck}, J.~R., {et~al.}
  1984{\natexlab{b}}, \apjl, 278, L71

\bibitem[{{Soifer} {et~al.}(1987{\natexlab{b}}){Soifer}, {Sanders}, {Madore},
  {Neugebauer}, {Danielson}, {Elias}, {Lonsdale}, \& {Rice}}]{Soifer87}
{Soifer}, B.~T., {Sanders}, D.~B., {Madore}, B.~F., {et~al.}
  1987{\natexlab{b}}, \apj, 320, 238

\bibitem[{{Spoon} {et~al.}(2004{\natexlab{a}}){Spoon}, {Armus}, {Cami},
  {Tielens}, {Chiar}, {Peeters}, {Keane}, {Charmandaris}, {Appleton},
  {Teplitz}, \& {Burgdorf}}]{Spoon04}
{Spoon}, H.~W.~W., {Armus}, L., {Cami}, J., {et~al.} 2004{\natexlab{a}}, \apjs,
  154, 184

\bibitem[{{Spoon} {et~al.}(2007){Spoon}, {Marshall}, {Houck}, {Elitzur}, {Hao},
  {Armus}, {Brandl}, \& {Charmandaris}}]{Spoon07}
{Spoon}, H.~W.~W., {Marshall}, J.~A., {Houck}, J.~R., {et~al.} 2007, \apjl,
  654, L49

\bibitem[{{Spoon} {et~al.}(2004{\natexlab{b}}){Spoon}, {Moorwood}, {Lutz},
  {Tielens}, {Siebenmorgen}, \& {Keane}}]{Spoon04arp220}
{Spoon}, H.~W.~W., {Moorwood}, A.~F.~M., {Lutz}, D., {et~al.}
  2004{\natexlab{b}}, \aap, 414, 873

\bibitem[{{Sturm} {et~al.}(2010){Sturm}, {Verma}, {Graci{\'a}-Carpio},
  {Hailey-Dunsheath}, {Contursi}, {Fischer}, {Gonz{\'a}lez-Alfonso},
  {Poglitsch}, {Sternberg}, {Genzel}, {Lutz}, {Tacconi}, {Christopher}, \& {de
  Jong}}]{Sturm10}
{Sturm}, E., {Verma}, A., {Graci{\'a}-Carpio}, J., {et~al.} 2010, \aap, 518,
  L36

\bibitem[{{Tran} {et~al.}(2001){Tran}, {Lutz}, {Genzel}, {Rigopoulou}, {Spoon},
  {Sturm}, {Gerin}, {Hines}, {Moorwood}, {Sanders}, {Scoville}, {Taniguchi}, \&
  {Ward}}]{Tran01}
{Tran}, Q.~D., {Lutz}, D., {Genzel}, R., {et~al.} 2001, \apj, 552, 527

\bibitem[{{Treister} {et~al.}(2008){Treister}, {Krolik}, \&
  {Dullemond}}]{Treister08}
{Treister}, E., {Krolik}, J.~H., \& {Dullemond}, C. 2008, \apj, 679, 140

\bibitem[{{Vasudevan} \& {Fabian}(2009)}]{Vasudevan09}
{Vasudevan}, R.~V. \& {Fabian}, A.~C. 2009, \mnras, 392, 1124

\bibitem[{{Veilleux} {et~al.}(1999){Veilleux}, {Kim}, \&
  {Sanders}}]{Veilleux99}
{Veilleux}, S., {Kim}, D.-C., \& {Sanders}, D.~B. 1999, \apj, 522, 113

\bibitem[{{Veilleux} {et~al.}(2002){Veilleux}, {Kim}, \&
  {Sanders}}]{Veilleux02}
{Veilleux}, S., {Kim}, D.-C., \& {Sanders}, D.~B. 2002, \apjs, 143, 315

\bibitem[{{Veilleux} {et~al.}(2009){Veilleux}, {Rupke}, {Kim}, {Genzel},
  {Sturm}, {Lutz}, {Contursi}, {Schweitzer}, {Tacconi}, {Netzer}, {Sternberg},
  {Mihos}, {Baker}, {Mazzarella}, {Lord}, {Sanders}, {Stockton}, {Joseph}, \&
  {Barnes}}]{Veilleux09}
{Veilleux}, S., {Rupke}, D.~S.~N., {Kim}, D.-C., {et~al.} 2009, \apjs, 182, 628

\bibitem[{{Verma} {et~al.}(2005){Verma}, {Charmandaris}, {Klaas}, {Lutz}, \&
  {Haas}}]{Verma05}
{Verma}, A., {Charmandaris}, V., {Klaas}, U., {Lutz}, D., \& {Haas}, M. 2005,
  Space Science Reviews, 119, 355

\bibitem[{{Verma} {et~al.}(2002){Verma}, {Rowan-Robinson}, {McMahon}, \&
  {Andreas Efstathiou}}]{Verma02}
{Verma}, A., {Rowan-Robinson}, M., {McMahon}, R., \& {Andreas Efstathiou},
  A.~E. 2002, \mnras, 335, 574

\bibitem[{{Vignali} {et~al.}(2011){Vignali}, {Piconcelli}, {Lanzuisi},
  {Feltre}, {Feruglio}, {Maiolino}, {Fiore}, {Fritz}, {La Parola}, {Mignoli},
  \& {Pozzi}}]{Vignali11}
{Vignali}, C., {Piconcelli}, E., {Lanzuisi}, G., {et~al.} 2011, \mnras, 416,
  2068

\bibitem[{{Wang} {et~al.}(2011){Wang}, {Farrah}, {Connolly}, {Connolly},
  {Lebouteiller}, {Oliver}, \& {Spoon}}]{Wang11}
{Wang}, L., {Farrah}, D., {Connolly}, B., {et~al.} 2011, \mnras, 411, 1809

\bibitem[{{Watson} {et~al.}(2009){Watson}, {Schr{\"o}der}, {Fyfe}, {Page},
  {Lamer}, {Mateos}, {Pye}, {Sakano}, {Rosen}, {Ballet}, {Barcons}, {Barret},
  {Boller}, {Brunner}, {Brusa}, {Caccianiga}, {Carrera}, {Ceballos}, {Della
  Ceca}, {Denby}, {Denkinson}, {Dupuy}, {Farrell}, {Fraschetti}, {Freyberg},
  {Guillout}, {Hambaryan}, {Maccacaro}, {Mathiesen}, {McMahon}, {Michel},
  {Motch}, {Osborne}, {Page}, {Pakull}, {Pietsch}, {Saxton}, {Schwope},
  {Severgnini}, {Simpson}, {Sironi}, {Stewart}, {Stewart}, {Stobbart}, {Tedds},
  {Warwick}, {Webb}, {West}, {Worrall}, \& {Yuan}}]{2xmmi}
{Watson}, M.~G., {Schr{\"o}der}, A.~C., {Fyfe}, D., {et~al.} 2009, \aap, 493,
  339

\bibitem[{{Werner} {et~al.}(2004){Werner}, {Roellig}, {Low}, {Rieke}, {Rieke},
  {Hoffmann}, {Young}, {Houck}, {Brandl}, {Fazio}, {Hora}, {Gehrz}, {Helou},
  {Soifer}, {Stauffer}, {Keene}, {Eisenhardt}, {Gallagher}, {Gautier}, {Irace},
  {Lawrence}, {Simmons}, {Van Cleve}, {Jura}, {Wright}, \&
  {Cruikshank}}]{Spitzer}
{Werner}, M.~W., {Roellig}, T.~L., {Low}, F.~J., {et~al.} 2004, \apjs, 154, 1

\bibitem[{{Wilman} {et~al.}(2003){Wilman}, {Fabian}, {Crawford}, \&
  {Cutri}}]{Wilman03}
{Wilman}, R.~J., {Fabian}, A.~C., {Crawford}, C.~S., \& {Cutri}, R.~M. 2003,
  \mnras, 338, L19

\bibitem[{{Wright} {et~al.}(1984){Wright}, {Joseph}, \& {Meikle}}]{Wright84}
{Wright}, G.~S., {Joseph}, R.~D., \& {Meikle}, W.~P.~S. 1984, \nat, 309, 430

\bibitem[{{Yan} {et~al.}(2010){Yan}, {Tacconi}, {Fiolet}, {Sajina}, {Omont},
  {Lutz}, {Zamojski}, {Neri}, {Cox}, \& {Dasyra}}]{Yan10}
{Yan}, L., {Tacconi}, L.~J., {Fiolet}, N., {et~al.} 2010, \apj, 714, 100

\bibitem[{{Yun} \& {Scoville}(1998)}]{Yun98}
{Yun}, M.~S. \& {Scoville}, N.~Z. 1998, \apj, 507, 774

\end{thebibliography}

\begin{table*}
 \begin{minipage}[!ht]{0.85\textwidth}
 \caption{\Spitzer observations of HLIRG.}
 \label{tab:sources}
 \renewcommand{\footnoterule}{}  
  \begin{tabular}{@{}lllcclrrc@{}}
  \hline \hline
  Sources \tablefootmark{a} & Sample \tablefootmark{b} & Type \tablefootmark{c} & R.A. & DEC. & $z$ & AOR \tablefootmark{d} & Exposure (s) & Date \tablefootmark{e} \\
  \hline
  \object{IRAS 00182-7112}      & X     & QSO2  & 00 20 34.7 & -70 55 27 & 0.327 &  7556352 &  94.4 & 2003-11-14 \\
  \object{IRAS F00235+1024}\dag & F, X  & NL-SB & 00 26 06.5 & +10 41 32 & 0.575 & 12237056 &  94.4 & 2005-08-11 \\
  \object{IRAS 07380-2342}\dag  & F, X  & NL    & 07 40 09.8 & -23 49 58 & 0.292 & 12236032 &  12.6 & 2005-03-23 \\
  \object{IRAS 09104+4109}      & X     & QSO2  & 09 13 45.4 & +40 56 28 & 0.442 &  6619136 &  73.4 & 2003-11-29 \\
  \object{IRAS F10026+4949}     & F     & Sy1   & 10 05 52.9 & +49 34 42 & 1.120 &  4738560 &  14.7 & 2004-04-17 \\
  \object{IRAS F12509+3122}     & F, X  & QSO   & 12 53 17.6 & +31 05 50 & 0.780 & 12236800 &  62.9 & 2005-06-06 \\
  \object{IRAS 12514+1027}      & X     & Sy2   & 12 54 00.8 & +10 11 12 & 0.32  &  4978432 &  94.4 & 2005-02-06 \\
  \object{IRAS 14026+4341}\dag  & F, X  & QSO   & 14 04 38.8 & +43 27 07 & 0.323 &  4374016 & 102.8 & 2005-05-22 \\
  \object{IRAS F15307+3252}     & X     & QSO2  & 15 32 44.0 & +32 42 47 & 0.926 &  4983552 & 243.8 & 2004-03-04 \\
  \object{IRAS F16124+3241}     & F     & NL    & 16 14 22.1 & +32 34 04 & 0.71  &  4984832 & 243.8 & 2005-03-19 \\
  \object{ELAIS J1640+41}       & F     & QSO   & 16 40 10.1 & +41 05 22 & 1.099 & 11345664 & 243.8 & 2005-08-13 \\
  \object{IRAS 18216+6418}      & F, X  & QSO   & 18 21 57.3 & +64 20 36 & 0.297 &  4676096 &  31.5 & 2004-04-17 \\
  \object{IRAS F23569-0341}     & F     & NL    & 23 59 33.6 & -03 25 13 & 0.59  & 12235776 & 243.8 & 2004-12-13 \\
  \hline
  \end{tabular} \\ \\
 \textbf{Notes.} 
 \tablefoottext{a}{Sources marked with \dag~have been observed with \XMM, but not detected.}
 \tablefoottext{b}{F: sources from the F02 sample; X: sources from the \citet{Ruiz07} X-ray-selected sample.}
 \tablefoottext{c}{Optical spectrum classification.}
 \tablefoottext{d}{Astronomical Observation Request (AOR) of the \Spitzer observation.}
 \tablefoottext{e}{Date of the \Spitzer observation.}
 \end{minipage}
\end{table*}

\begin{table*}
 \begin{minipage}[!ht]{0.9\textwidth}
 \caption{Best-fit model parameters (c.f. Sect.~\ref{sec:decomp}).}
 \label{tab:fit}
 \renewcommand{\footnoterule}{}  
  \begin{tabular}{@{}lccrrrrc@{}}
  \hline \hline
  Sources \tablefootmark{a} & Model \tablefootmark{b} & $\chi^2/$d.o.f. & $\Fint$ (mJy) \tablefootmark{c} & $\alfa$ \tablefootmark{d}        & Spectral index \tablefootmark{e}         & $\tau_6$ \tablefootmark{f}       & Additional features \\
  \hline
  IRAS F00235+1024 & \textbf{A} & \textbf{238 / 75} & $\mathbf{ 86\pm 5}$   & $\mathbf{0.982\pm0.002}$  &  $\mathbf{0.8}$         & $\mathbf{2.57\pm0.06}$ & \textbf{...} \\
                   &         B  &         218 / 74  & $46^{+32}_{-2}$       & $0.969^{+0.012}_{-0.003}$ &  $>1.0$                 & $1.93^{+0.53}_{-0.05}$ &         ...  \\
  IRAS 07380-2342  & \textbf{A} & \textbf{ 22 / 76} & $\mathbf{213\pm12}$   & $\mathbf{0.988\pm0.006}$  &  $\mathbf{0.8}$         & $\mathbf{0.49\pm0.05}$ & \textbf{...} \\
                   &         B  &          16 / 75  & $130^{+43}_{-1}$      & $        0.990\pm0.009$   &  $1.64^{+0.03}_{-0.48}$ & $<0.3$                 &         ...  \\
  IRAS 09104+4109* & \textbf{A} & \textbf{ 44 / 69} & $\mathbf{176\pm 3}$   & $\mathbf{0.995\pm0.002}$  &  $\mathbf{0.8}$         & $\mathbf{0.69\pm0.02}$ & \textbf{[NeVI] emission line\tablefootmark{g}} \\
                   &         B  &          26 / 68  & $        104\pm11$    & $        0.996\pm0.004$   &  $1.65\pm0.17$          & $0.16\pm0.10$          &         [NeVI] emission line\tablefootmark{g} \\
  IRAS F10026+4949 & \textbf{A} & \textbf{ 19 / 88} & $\mathbf{149^{+8}_{-18}}$ & $\mathbf{>0.988}$     &  $\mathbf{0.8}$         & $\mathbf{0.63^{+0.06}_{-0.12}}$ & \textbf{...} \\
                   &         B  &          19 / 87  & $        150\pm60$    & $            >0.992$      &  $0.7^{+0.7}_{-0.5}$    & $0.6^{+0.3}_{-0.5}$    &         ...  \\
  IRAS F12509+3122 & \textbf{A} & \textbf{ 51 / 85} & $\mathbf{25.3^{+0.8}_{-0.3}}$ & $\mathbf{0.959^{+0.005}_{-0.003}}$  &  $\mathbf{0.8}$         & $\mathbf{<0.04}$       & \textbf{...} \\
                   &         B  &          51 / 84  & $        25\pm6$      & $0.962^{+0.008}_{-0.003}$ &  $0.87^{+0.02}_{-0.41}$ & $<0.2$                 &         ...  \\
  IRAS 12514+1027  & \textbf{A} & \textbf{116 / 73} & $\mathbf{57.4\pm1.0}$ & $\mathbf{0.958\pm0.003}$  &  $\mathbf{0.8}$         & $\mathbf{<0.02}$       & \textbf{[NeVI] emission line\tablefootmark{g}} \\
                   &         B  &         114 / 72  & $        57\pm7$      & $        0.956\pm0.004$   &  $0.77^{+0.01}_{-0.22}$ & $<0.12$                &         [NeVI] emission line\tablefootmark{g} \\
  IRAS 14026+4341  & \textbf{A} & \textbf{ 33 / 75} & $\mathbf{75\pm 2}$    & $\mathbf{0.977\pm0.004}$  &  $\mathbf{0.8}$         & $\mathbf{0.06\pm0.02}$ & \textbf{...} \\
                   &         B  &          31 / 74  & $70^{+13}_{-1}$       & $        0.978\pm0.004$   &  $0.93^{+0.02}_{-0.31}$ & $<0.18$                &         ...  \\
  IRAS F15307+3252 & \textbf{A} & \textbf{238 / 84} & $\mathbf{38\pm 2}$    & $\mathbf{0.981\pm0.004}$  &  $\mathbf{0.8}$         & $\mathbf{0.64\pm0.04}$ & \textbf{...} \\
                   &         B  &         173 / 83  & $21^{+6}_{-1}$        & $        0.975\pm0.006$   &  $>1.4$                 & $0.07^{+0.24}_{-0.03}$ &         ...  \\
  IRAS F16124+3241 & \textbf{A} & \textbf{ 38 / 78} & $\mathbf{6.6\pm1.3}$  & $\mathbf{0.87^{+0.03}_{-0.04}}$ & $\mathbf{0.8}$    & $\mathbf{1.3\pm0.2}$   & \textbf{...} \\
                   &         B  &          38 / 77  & $          7\pm 4$    & $  0.87^{+0.05}_{-0.11}$  &  $0.8^{+1.0}_{-0.6}$    & $1.3^{+0.5}_{-0.8}$    &         ...  \\
  ELAIS J1640+41   & \textbf{A} & \textbf{ 33 / 88} & $\mathbf{17.3\pm1.1}$ & $\mathbf{0.993\pm0.006}$   &  $\mathbf{0.8}$         & $\mathbf{0.13\pm0.06}$ & \textbf{...} \\
                   &         B  &          32 / 87  & $15^{+6}_{-1}$        & $        >0.988$          &  $1.06^{+0.02}_{-0.60}$ & $<0.4$                 &         ...  \\
  IRAS 18216+6418  &         A  &         117 / 77  & $136.1^{+2.2}_{-0.4}$ & $        >0.999$          &  $0.8$                  & $<0.001$               &         ...  \\
                   & \textbf{A} & \textbf{ 18 / 76} & $\mathbf{159^{+7}_{-19}}$ & $\mathbf{0.991^{+0.005}_{-0.003}}$ &  $\mathbf{<0.4}$        & $\mathbf{<0.17}$       & \textbf{...} \\
  \hline
  \end{tabular} \\ \\
  \textbf{Notes.} 
  \tablefoottext{a}{Sources where our model is not reliable are marked with *.}
  \tablefoottext{b}{HLIRG emission model 
                   (A: AGN power law slope fixed at 0.8;
                    B: AGN power law slope allowed to vary between 0.2 and 1.8).
                    See Sect.~\ref{sec:model} for a complete discussion of the adopted model.
                    The best-fit model is shown in boldface.}
  \tablefoottext{c}{Intrinsic flux at 6~\micr (corrected by dust absorption).}
  \tablefoottext{d}{Relative contribution of the AGN component at 6~\micr.}
  \tablefoottext{e}{Spectral index of the AGN power law.}
  \tablefoottext{f}{Optical depth at 6~\micr of the obscuring material.}
  \tablefoottext{g}{Emission line at 7.65~\micr. See Sect~\ref{sec:results} for a discussion about these spectral features.}
\end{minipage}
\end{table*}

\begin{table*}[ht]
\begin{minipage}[!ht]{0.8\textwidth}
\caption{SFR and AGN relative contributions to the IR output, and IR luminosities (in cgs units).}
\label{tab:luminIR}
 \renewcommand{\footnoterule}{}  
 \begin{tabular}{@{}lrrrr@{}}
  \hline \hline
   Sources \tablefootmark{a} & SFR \tablefootmark{b}       &  $\alpha_\mathrm{IR}$\tablefootmark{c} & $L_\mathrm{IR}$\tablefootmark{d}      & $L_\mathrm{IR}^\mathrm{SB}$\tablefootmark{e} \\
                             & [$\Msun \mathrm{yr^{-1}}$]  &                                        & [$\times10^{46}$]                     & [$\times10^{46}$]        \\
  \hline
   IRAS F00235+1024 & $1500^{+1100}_{-1000}$ & $0.68\pm0.12$          & $20\pm8$     & $6\pm4$ \\
   IRAS 07380-2342  & $<700$                 & $0.82^{+0.14}_{-0.16}$ & $14.2\pm0.6$ & $3\pm2$ \\
   IRAS 09104+4109* & $<600$                 & $0.92^{+0.08}_{-0.07}$ & $13.8\pm1.3$ & $1.1^{+1.1}_{-0.9}$ \\
   IRAS F10026+4949 & $<500$                 & $>0.8$                 & $140\pm40$   & $<20$ \\
   IRAS F12509+3122 & $2000\pm1400$          & $0.49^{+0.14}_{-0.13}$ & $36\pm13$    & $18\pm8$ \\
   IRAS 12514+1027  & $400\pm300$            & $0.53^{+0.17}_{-0.11}$ & $3.7\pm0.4$  & $1.7^{+0.7}_{-0.5}$ \\
   IRAS 14026+4341  & $400\pm300$            & $0.60^{+0.12}_{-0.10}$ & $6.2\pm0.5$  & $2.5^{+0.8}_{-0.7}$ \\
   IRAS F15307+3252 & $3000\pm2000$          & $0.59^{+0.13}_{-0.12}$ & $52\pm17$    & $21^{+10}_{-9}$ \\
   IRAS F16124+3241 & $1300^{+1700}_{-1100}$ & $0.25^{+0.13}_{-0.21}$ & $16\pm6$     & $12^{+5}_{-6}$ \\
   ELAIS J1640+41   & $<1200$                & $0.82^{+0.13}_{-0.11}$ & $<120$       & $<20$ \\
   IRAS 18216+6418  & $300^{+300}_{-200}$    & $0.80^{+0.12}_{-0.17}$ & $8.0\pm1.4$  & $1.6^{+1.0}_{-1.4}$ \\
  \hline
  \end{tabular} \\ \\
  \textbf{Notes.}
  \tablefoottext{a}{Sources where our model is not reliable are marked with *.}
  \tablefoottext{b}{Star formation rates estimated through PAH intensity (see Sect.\ref{sec:sfr}).}
  \tablefoottext{c}{AGN contribution to the total IR luminosity (see Sect.~\ref{sec:alfabol}).}
  \tablefoottext{d}{Total IR luminosity (8-1000~\micr), estimated through IRAS fluxes.}
  \tablefoottext{e}{IR luminosity (8-1000~\micr) of the SB component (see Sect.~\ref{sec:sfr}).}
 \end{minipage}
\end{table*}

\begin{table*}[ht]
\begin{minipage}[!ht]{0.8\textwidth}
\caption{AGN luminosities (in cgs units) and CF values.}
\label{tab:covfactor}
 \renewcommand{\footnoterule}{}  
 \begin{tabular}{@{}lrrrrr@{}}
  \hline \hline 
   Sources & $L_\mathrm{X}$ \tablefootmark{a}      &  $L_\mathrm{AGN}^\mathrm{DIR}$\tablefootmark{b} & $\lambda~L_\lambda(6~\mu\mathrm{m})$ \tablefootmark{c} & $L_\mathrm{AGN}^\mathrm{TH}$ \tablefootmark{d} & CF \tablefootmark{e} \\
           & [$\times10^{44}$]                     &  [$\times10^{46}$]                              & [$\times10^{46}$]                                      & [$\times10^{46}$]                              &    \\
   \hline
   IRAS F00235+1024\dag & $<1.6$              & $<0.6$                 & $6.1\pm1.2$   & $19^{+8}_{-6}$      & $>30$                  \\
   IRAS 07380-2342\dag  & $<2.0$              & $<0.8$                 & $3.0\pm0.3$   & $9^{+3}_{-2}$       & $>11$                  \\
   IRAS 09104+4109      & $20^{+26}_{-4}$     & $19^{+31}_{-4}$        & $6.8\pm0.3$   & $21^{+7}_{-5}$      & $1.1^{+1.9}_{-0.4}$    \\
   IRAS F12509+3122     & $1.8\pm0.2$         & $0.73\pm0.09$          & $3.9\pm0.3$   & $12^{+4}_{-3}$      & $16^{+6}_{-5}$         \\
   IRAS 12514+1027      & $0.2^{+4.5}_{-0.1}$ & $0.04^{+1.41}_{-0.03}$ & $0.88\pm0.06$ & $2.7^{+1.0}_{-0.7}$ & $70^{+3000}_{-60}$     \\
   IRAS 14026+4341\dag  & $3.0\pm1.8$         & $1.4^{+1.0}_{-0.9}$    & $1.27\pm0.07$ & $4.0^{+1.4}_{-1.0}$ & $2.8^{+2.2}_{-1.9}$    \\
   IRAS F15307+3252     & $31^{+7}_{-8}$      & $34^{+9}_{-10}$        & $8.1\pm0.7$   & $25^{+9}_{-7}$      & $0.7\pm0.3$            \\
   IRAS 18216+6418      & $37\pm4$            & $43\pm5$               & $2.11\pm0.10$ & $6.6^{+2.3}_{-1.7}$ & $0.15^{+0.06}_{-0.04}$ \\
  \hline
  \end{tabular} \\ \\
  \textbf{Notes.}
  \tablefoottext{a}{Absorption-corrected X-ray luminosity, from \citet{Ruiz07}.
A factor of 60 \citep{Panessa06} has been applied to
transform the observed X-ray luminosity to intrinsic X-ray luminosity to the sources marked with \dag.}
  \tablefoottext{b}{Bolometric AGN luminosity (i.e., intrinsic AGN emission above $\sim1$~\micr).}
  \tablefoottext{c}{Absorption-corrected AGN luminosity at 6~\micr.}
  \tablefoottext{d}{Reprocessed AGN luminosity (i.e., thermal emission due to heated dust).}
  \tablefoottext{e}{Covering factor.}
 \end{minipage}
\end{table*}

%

%

\end{document}